\documentclass[envcountsame,envcountchap]{svmono}

\input{hmm.sty} % Leave this first!

\begin{document}
\title{Bayesian Core:\\The Complete Solution Manual}
\author{
Christian P.\ Robert and Jean--Michel Marin\\% and Christian P.\ Robert\\
Universit\'e Paris Dauphine, and CREST, INSEE, Paris, \&~Institut de Math\'ematiques et Mod\'elisation de Montpellier,
Universit\'e Montpellier 2, and CREST, INSEE, Paris}
%Universit\'e Paris Dauphine \&\ CREST, INSEE, Paris}

\maketitle

\frontmatter%%%%%%%%%%%%%%%%%%%%%%%%%%%%%%%%%%%%%%%%%%%%%%%%%%%%%%

\addcontentsline{toc}{chapter}{\protect\numberline{}Preface}
\chapter*{Preface}

%\vglue 1truecm
\begin{quote}\begin{flushright}
The warning could not have been meant
for the place\\ where it could only be found after approach.\\
---{\bf Joseph Conrad}, \emph{\textbf{Heart of Darkness}}
\end{flushright}\end{quote}

\bigskip
This solution manual was initially intended for instructors using the book, so that they could entirely rely on the
exercises to provide graded homeworks. However, due to repeated criticisms and to requests from readers, as well as to the
fact that some exercises were stepping stones for following sections and to the lack of evidence that the exercises
were indeed used to set homeworks, we came to
realise (albeit late in the day!) that some solutions were needed by some (self-study or not) readers. From there, the move 
to make the {\em whole} set of solutions available to {\em all} readers was a rather natural step. Especially when contemplating 
the incoming revision of {\em Bayesian Core} towards a {\em Use R!} oriented version, with an attempt at reducing the math 
complexity, another reproach found in some of the published criticisms. Therefore, lo and behold!, by popular request,
the solution manual is now available for free use
and duplication by anyone, not only by instructors, on the book webpage as well as on Springer Verlag's website. 

However, there is a caveat to the opening of the manual to all: since this solution manual was first intended 
(and written) for instructors, some self-study readers may come to the realisation that the solutions provided 
here are too sketchy for them because the way we wrote those solutions assumes some minimal familiarity with the maths, 
the probability theory and with the statistics behind the arguments. There is unfortunately a limit to the time and 
to the efforts we can put in this solution manual and studying {\em Bayesian Core} requires some prerequisites in maths 
(such as matrix algebra and Riemann integrals), in probability theory (such as the use of joint and conditional densities) 
and some bases of statistics (such as the notions of inference, sufficiency and confidence sets) that we cannot cover here. 
Casella and Berger (2001) is a good reference in case a reader is lost with the ``basic" concepts or sketchy math
derivations. Indeed, we also came to realise that describing the book as``self-contained" was a dangerous add as readers were 
naturally inclined to always relate this term to their current state of knowledge, a bias resulting in inappropriate expectations. 
(For instance, some students unfortunately came to one of our short courses with no previous exposure to standard distributions 
like the $t$ or the gamma distributions.)

We obviously welcome comments and questions on possibly erroneous solutions, as well as suggestions for 
more elegant or more complete solutions: since this manual is distributed both freely and independently 
from the book, it can be updated and corrected [almost] in real time! Note however that the {\sf R} codes given in the following
pages are not optimised because we prefer to use simple and understandable codes, rather than condensed and 
efficient codes, both for time constraints (this manual took about
a whole week of August 2007 to complete) and for pedagogical purposes:
the readers must be able to grasp the meaning of the {\sf R} code with a minimum of effort since {\sf R}
programming is not supposed to be an obligatory entry to the book. In this respect, using {\sf R} replaces the pseudo-code
found in other books since it can be implemented as such but does not restrict understanding. Therefore, if you 
find better [meaning, more efficient/faster] codes than those provided along those pages, we would be glad to hear 
from you, but that does not mean that we will automatically substitute your {\sf R} code for the current one, 
because readability is also an important factor.

\bigskip
\begin{flushright}
{\bf Sceaux \&~Montpellier, France, \today\\
Christian P.~Robert \&~Jean-Michel Marin}
%Christian P.~Robert \&\ Jean-Michel Marin}
\end{flushright}
%\end{minipage}

\tableofcontents

\mainmatter%%%%%%%%%%%%%%%%%%%%%%%%%%%%%%%%%%%%%%%%%%%%%%%%%%%%%%%

\chapter{User's Manual}\label{ch:intro}

\begin{exoset} Given a function $g$ on $\mathbb{R}$, state the two basic conditions for $g$ to be
a probability density function (pdf) with respect to the Lebesgue measure. Recall 
the definition of the cumulative distribution function (cdf) associated
with $g$ and that of the quantile function of $g$.
\end{exoset}

\noindent If $g$ is integrable with respect to the Lebesgue measure, $g$ is a pdf if and only if
\begin{enumerate}
\item $g$ is non-negative, $g(x)\ge 0$
\item $g$ integrates to $1$,
$$
\int_\mathbb{R} g(x)\,\text{d}x = 1\,.
$$
\end{enumerate}

\begin{exoset}
If $(x_1,x_2)$ is a normal $\mathcal{N}_2((\mu_1,\mu_2),\Sigma)$ random vector, with 
$$
\Sigma=\left(\begin{matrix}
\sigma^2 &\omega\sigma\tau\cr \omega\sigma\tau &\tau^2\end{matrix}\right)\,,
$$
recall the conditions on
$(\omega,\sigma,\tau)$ for $\Sigma$ to be a (nonsingular) covariance matrix. Under those conditions,
derive the conditional distribution of $x_2$ given $x_1$.
\end{exoset}

\noindent The matrix $\Sigma$ is a covariance matrix if 
\begin{enumerate}
\item $\Sigma$ is symmetric and this is the case;
\item $\Sigma$ is semi-definite positive, i.e.~, for every $\bx\in\mathbb{R}^2$,
$\bx^\tee \Sigma \bx \ge 0$, or, for every $(x_1,x_2)$,
$$
\sigma^2 x_1^2 +2\omega\sigma\tau x_1 x_2 + \tau^2 x_2^2
= (\sigma x_1+\omega\tau x_2)^2 + \tau^2 x_2^2 (1-\omega^2) \ge 0\,.
$$
A necessary condition for $\Sigma$ to be positive semi-definite is that
$\text{det}(\Sigma) = \sigma^2\tau^2(1-\omega^2) \ge 0$, which is equivalent to $|\omega|\le 1$.
\end{enumerate}
In that case, $\bx^\tee \Sigma \bx \ge 0$.
The matrix $\Sigma$ is furthermore nonsingular if $\text{det}(\Sigma)>0$, which is equivalent to
$|\omega|< 1$.

Under those conditions, the conditional distribution of $x_2$ given $x_1$ is defined by
\begin{align*}
f(x_2|x_1) &\propto \exp\left\{ (x_1-\mu_1\quad x_2-\mu_2)\Sigma^{-1}{x_1-\mu_1\choose x_2-\mu_2}/2\right\}\\
&\propto \exp\left\{ (x_1-\mu_1\quad x_2-\mu_2) \left(\begin{matrix}
\tau^2 &-\omega\sigma\tau\cr -\omega\sigma\tau &\sigma^2\end{matrix}\right)
   {x_1-\mu_1\choose x_2-\mu_2}/2\text{det}(\Sigma)\right\}\\
&\propto \exp\left\{ \sigma^2(x_2-\mu_2)^2 
	-2\omega\sigma\tau(x_1-\mu_1)(x_2-\mu_2)/2\text{det}(\Sigma)\right\}\\
&\propto \exp\left\{ \frac{\sigma^2}{\sigma^2\tau^2(1-\omega^2)} 
\left( x_2-\mu_2-\omega\tau \frac{x_1-\mu_1}{\sigma}\right)^2/2 \right\}
\end{align*}
Therefore,
$$
x_2|x_1 \sim \mathscr{N}\left(\mu_2+\omega\tau \frac{x_1-\mu_1}{\sigma},\tau^2(1-\omega^2)\right)\,.
$$

\begin{exoset}
Test the \verb+help()+ command on the functions \verb+seq()+, \verb+sample()+, 
and \verb+order()+. ({\em Hint:} start with \verb+help()+.)
\end{exoset}

\noindent Just type\\

\begin{tabular}{ll}
\verb+> help()+                           & \\
\verb&> help(seq)&               & \\
\verb+> help(sample)+                          & \\
\verb+> help(order)+                         & \\
\end{tabular}\\

\noindent and try illustrations like\\

\begin{tabular}{ll}
\verb+> x=seq(1,500,10)+                  & \\
\verb&> y=sample(x,10,rep=T)&               & \\
\verb+> z=order(y)+                         & \\
\end{tabular}

\begin{exoset}
Study the properties of the {\sf R} function \verb+lm()+ using simulated
data as in
\begin{verbatim}
  > x=rnorm(20)
  > y=3*x+5+rnorm(20,sd=0.3)
  > reslm=lm(y~x)
  > summary(reslm)
\end{verbatim}
\end{exoset}

\noindent Generating the normal vectors $x$ and $y$ and calling the linear regression function 
{\sf lm} leads to
\begin{verbatim}
Call:
lm(formula = y ~ x)

Residuals:
    Min      1Q  Median      3Q     Max
-0.7459 -0.2216  0.1535  0.2130  0.8989

Coefficients:
            Estimate Std. Error t value Pr(>|t|)
(Intercept)  5.02530    0.10283   48.87   <2e-16 ***
x            2.98314    0.09628   30.98   <2e-16 ***
---
Sig. code:  0 `***' 0.001 `**' 0.01 `*' 0.05 `.' 0.1 ` ' 1

Residual standard error: 0.4098 on 18 degrees of freedom
Multiple R-Squared: 0.9816,     Adjusted R-squared: 0.9806
F-statistic:   960 on 1 and 18 DF,  p-value: < 2.2e-16
\end{verbatim}
Therefore, in this experiment, the regression coefficients $(\alpha,\beta)$
in $\mathbb{E}[y|x]=\alpha+\beta x$ are estimated by maximum likelihood as 
$\hat\alpha=2.98$ and $\hat\beta=5.03$, while they are $\alpha=3$ and 
$\beta=5$ in the simulated dataset.

\begin{exoset}
Of the {\sf R} functions you have met so far, check
which ones are written in {\sf R} by simply typing
their name without parentheses, as in \verb+mean+ or \verb+var+.
\end{exoset}

\noindent Since
\begin{verbatim}
> mean
function (x, ...)
UseMethod("mean")
<environment: namespace:base>
\end{verbatim}
and
\begin{verbatim}
> var
function (x, y = NULL, na.rm = FALSE, use)
{
    if (missing(use))
        use <- if (na.rm)
            "complete.obs"
        else "all.obs"
    na.method <- pmatch(use, c("all.obs", "complete.obs", 
		"pairwise.complete.obs"))
    if (is.data.frame(x))
        x <- as.matrix(x)
    else stopifnot(is.atomic(x))
    if (is.data.frame(y))
        y <- as.matrix(y)
    else stopifnot(is.atomic(y))
    .Internal(cov(x, y, na.method, FALSE))
}
<environment: namespace:stats>
\end{verbatim}
we can deduce that the first function is written in {\sf C}, while the second function is 
written in {\sf R}.

\chapter{Normal Models}\label{ch:norm}

\begin{exoset}
Check your current knowledge of the normal $\mathscr{N}(\mu,\sigma^2)$
distribution by writing down its density function and computing its
first four moments.
\end{exoset}

The density of the normal $\mathscr{N}(\mu,\sigma^2)$ distribution is given by
$$
\varphi(x|\mu,\sigma)=\frac{1}{\sqrt{2\pi}\sigma}\, \exp\left\{ -(x-\mu)^2/2\sigma^2 \right\}
$$
and, if $X\sim\mathscr{N}(\mu,\sigma^2)$,
\begin{eqnarray*}
\mathbb{E}[X] &=& \mu + \int_{-\infty}^{+\infty}
      \frac{x-\mu}{\sqrt{2\pi}\sigma} \exp\left\{ -(x-\mu)^2/2\sigma^2 \right\} \text{d}x\\
              &=& \mu + \frac{\sigma}{\sqrt{2\pi}} \int_{-\infty}^{+\infty} y \exp -y^2/2 \text{d}y\\
	      &=& \mu + \frac{\sigma}{\sqrt{2\pi}} [-\exp -y^2/2]_{y=-\infty}^{y=+\infty}\\
	      &=& \mu\,,
\end{eqnarray*}
then, using one integration by parts,
\begin{eqnarray*}
\mathbb{E}[(X-\mu)^2] &=& \int_{-\infty}^{+\infty} \frac{y^2}{\sqrt{2\pi}\sigma} 
		\exp -y^2/2\sigma^2 \text{d}y\\ 
      &=& \sigma^2 \int_{-\infty}^{+\infty} \frac{z^2}{\sqrt{2\pi}} \exp -z^2/2\,\text{d}z\\
      &=& \frac{\sigma^2 }{\sqrt{2\pi}} [-z\exp -z^2/2]_{z=-\infty}^{z=+\infty}
	+ \sigma^2 \, \int_{-\infty}^{+\infty} \frac{1}{\sqrt{2\pi}} \exp -z^2/2\,\text{d}z\\
      &=& \sigma^2\,,
\end{eqnarray*}
exploiting the fact that $(x-\mu)^3\,\exp\left\{ -(x-\mu)^2/2\sigma^2 \right\} $ is asymmetric wrt
the vertical axis $x=\mu$,
$$
\mathbb{E}[(X-\mu)^3] = \int_{-\infty}^{+\infty} \frac{y^3}{\sqrt{2\pi}\sigma}
                \exp -y^2/2\sigma^2 \text{d}y = 0
$$
and, using once again one integration by parts,
\begin{eqnarray*}
\mathbb{E}[(X-\mu)^4] &=& \frac{\sigma^4}{\sqrt{2\pi}} \int_{-\infty}^{+\infty} 
	z^4 \exp -z^2/2\,\text{d}z\\
    &=& \frac{\sigma^4}{\sqrt{2\pi}} [-z^3\exp -z^2/2]_{z=-\infty}^{z=+\infty}
      + \sigma^4 \, \int_{-\infty}^{+\infty} \frac{3z^2}{\sqrt{2\pi}} \exp -z^2/2\,\text{d}z\\
    &=& 3\sigma^4\,.
\end{eqnarray*}
Thus, the four first (centered) moments of the normal $\mathscr{N}(\mu,\sigma^2)$ distribution are
$\mu$, $\sigma^2$, $0$ and $3\sigma^4$.

\begin{exoset}
Before exiting to the next page, think of datasets that could be, or
could not be, classified as normal. In each case, describe your
reason for proposing or rejecting a normal modeling.
\end{exoset}

A good illustration of the opposition between normal and nonnormal modelings
can be found in insurrance claims: for minor damages, the histogram of the
data (or of its log-transform) is approximately normal, while, for the highest
claims, the tail is very heavy and cannot be modeled by a normal distribution
(but rather by an extreme value distribution). Take for instance 
\verb+http://www.statsci.org/data/general/carinsuk.html+

\begin{exoset}
Reproduce the histogram of Figure 2.1
and the subsequent analysis conducted in this chapter
for the relative changes in reported larcenies relative
to the 1995 figures, using the {\sf 90cntycr.wk1} file available
on the Webpage of the book.
\end{exoset}

A new datafile must be created out of the file {\sf 90cntycr.wk1}.
Then, plotting an histogram and doing inference on this
dataset follows from the directions provided within the chapter.

\begin{exoset}\label{exo:compaoumpa}
By creating random subwindows of the region plotted in Figure 2.2,
represent the histograms of these subsamples and examine whether they strongly differ
or not. Pay attention to the possible influence of the few ``bright spots" on the
image.
\end{exoset}

\begin{figure}[h]
\centerline{\includegraphics[height=5truecm,width=.7\textwidth]{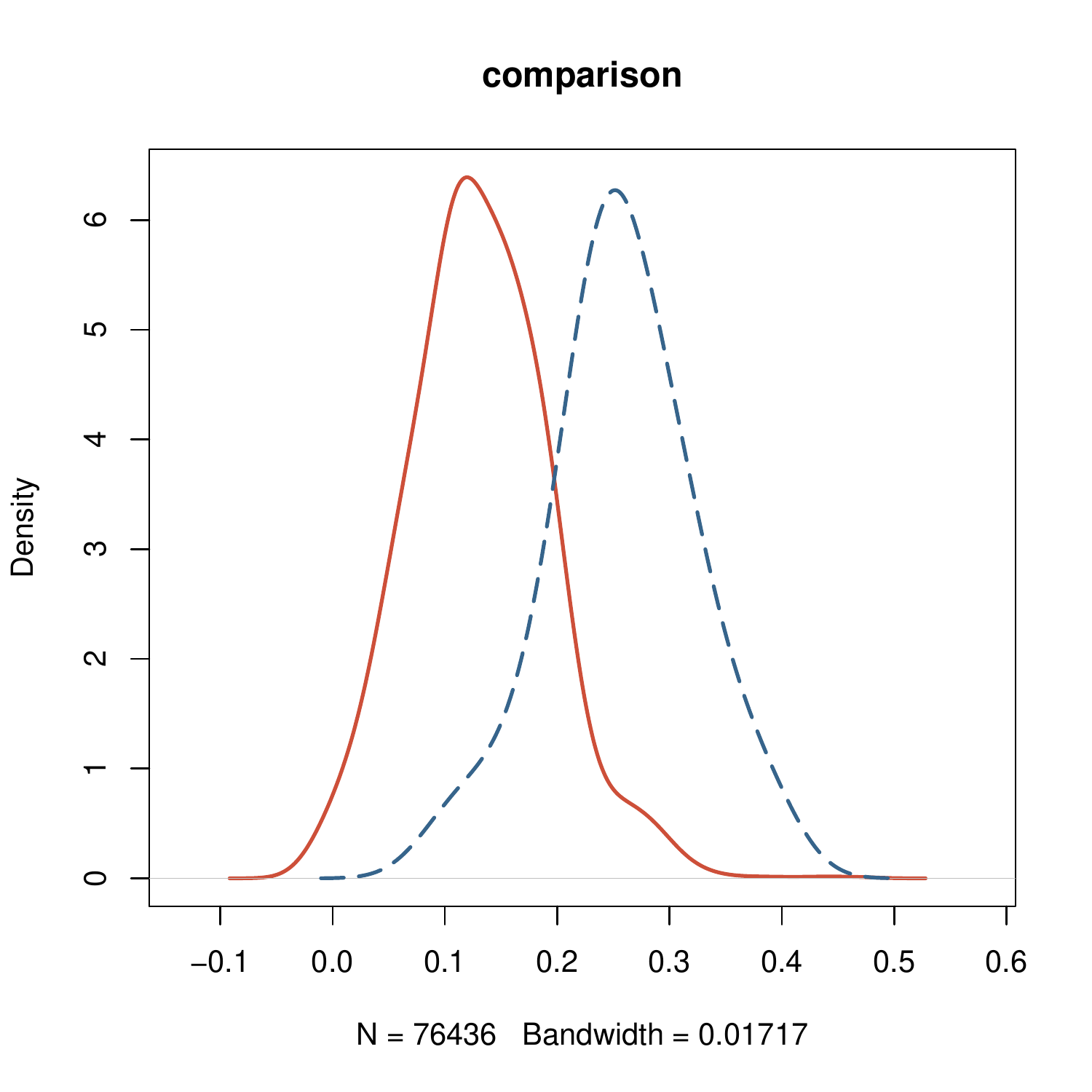}}
\caption{\label{fig:compa}
Comparison of two density estimates for two random subregions.}
\end{figure}

While a ``random subwindow" is not anything clearly defined, we can create a
$800\times 800$ matrix by\\
\verb+  > cmb=matrix(scan("CMBdata"),nrow=800)+\\
and define random subregions by
\begin{verbatim}
  > cmb1=cmb[sample(1:100,1):sample(101:300,1),
			sample(50:150,1):sample(401:600,1)]
  > cmb2=cmb[sample(701:750,1):sample(751:800,1),
			sample(650:750,1):sample(751:800,1)]
\end{verbatim}
Comparing the histograms can then be done as
\begin{verbatim}
  > hist(cmb1,proba=T,xlim=range(cmb))
  > par(new=T)
  > hist(cmb2,proba=T,xlim=range(cmb))
\end{verbatim}
or, more elaborately, through nonparametric density estimates
\begin{verbatim}
  > cnp1=density(cmb1,ad=3)  # Smooth out the bandwith
  > cnp2=density(cmb2,ad=3)  # Smooth out the bandwith
  > plot(cnp1,xlim=range(cmb),type="l",lwd=2,col="tomato3",
		main="comparison")
  > lines(cnp2,lty=5,col="steelblue4",lwd=2)
\end{verbatim}
which leads to Figure \ref{fig:compa}. In that case, both subsamples are roughly normal but with
different parameters.

\begin{exoset}
Show that (2.2) can be derived by first setting
$\theta$ as a random variable with density function $\pi$ and then $\mathscr{D}$
conditionally on $\theta$ as distributed from $\ell(\theta|\mathscr{D})$.
\end{exoset}

If $\pi(\theta)$ is the density of the {\em marginal} distribution of $\theta$ and then
$\ell(\theta|\mathscr{D})$ the density of the conditional distribution of $\mathscr{D}$
given $\theta$, the density of the joint distribution of $\theta$ and of $\mathscr{D}$
is given by
$$
\ell(\theta|\mathscr{D})\pi(\theta) \,.
$$
Therefore, Bayes's theorem simply is the derivation of the density of the conditional 
distribution of $\theta$ given $\mathscr{D}$ from this joint density.

\begin{exoset}
Show that the minimization (in $\hat\theta(\mathscr{D})$)
of the expectation $\mathbb{E}[\mbox{L} (\theta,\hat\theta))|
\mathscr{D}]$---that is, of the expectation of the quadratic loss function under the
distribution with density $\pi(\theta|\mathscr{D})$---produces the posterior
expectation as the solution in $\hat\theta$.
\end{exoset}

Since 
\begin{eqnarray*}
\mathbb{E}[\mbox{L} (\theta,\hat\theta))| \mathscr{D}]
&=& \mathbb{E}[||\theta-\hat\theta||^2|\mathscr{D}] \\
&=& \mathbb{E}[(\theta-\hat\theta)^\tee (\theta-\hat\theta)|\mathscr{D}]\\
&=& \mathbb{E}[||\theta||^2-2\theta^\tee\hat\theta+||\hat\theta||^2|\mathscr{D}]\\
&=& \mathbb{E}[||\theta||^2|\mathscr{D}] -2\hat\theta^\tee \mathbb{E}[\theta|\mathscr{D}]+||\hat\theta||^2\\
&=& \mathbb{E}[||\theta||^2|\mathscr{D}] -||\mathbb{E}[\theta|\mathscr{D}]||^2
+ ||\mathbb{E}[\theta|\mathscr{D}]-\hat\theta||^2\,,
\end{eqnarray*}
minimising $\mathbb{E}[\mbox{L} (\theta,\hat\theta))| \mathscr{D}]$ is equivalent
to minimising $||\mathbb{E}[\theta|\mathscr{D}]-\hat\theta||^2$ and hence the solution
is
$$
\hat\theta=\mathbb{E}[\theta|\mathscr{D}]\,.
$$

\begin{exoset}\label{exo:basex}
Show that the normal, binomial, geometric, Poisson, and exponential distributions are
all exponential families.
\end{exoset}

For each of those families of distributions, it is enough to achieve the standard form
of exponential families
\begin{equation}\label{eq:exp}
f_\theta(y) = h(y)\,\exp \left\{ \theta\cdot R(y) - \Psi(\theta) \right\}\,,
\end{equation}
as defined in the book. 

In the normal $\mathscr{N}(\theta,1)$ case,
$$
f_\theta(y) = \frac{1}{\sqrt{2\pi}}\,\exp\frac{1}{2}\,\left\{-y^2+2y\theta-\theta^2\right\}
$$
and so it fits the representation \eqref{eq:exp} with 
$R(y)=y$, $h(y)=\exp(-y^2/2)/\sqrt{2\pi}$ and $\Psi(\theta)=\theta^2/2$.

In the binomial $\mathscr{B}(n,p)$ case,
$$
f_p(y) = {n \choose y}\, \exp\left\{ y\log(p)+(n-y)\log(1-p) \right\}\,,
\quad y\in\{0,1,\ldots,n\}\,,
$$
and it also fits the representation \eqref{eq:exp} with $\theta=\log(p/(1-p))$,
$R(y)=y$, $h(y)={n \choose y}$ and $\Psi(\theta)=-n\log(1+e^\theta)$.

In the geometric $\mathscr{G}(p)$ case [corresponding to the number of 
failures before a success],
$$
f_p(y) = \exp\left\{ y\log(1-p)+\log(p)\right\}\,,
\quad y=0,1,\ldots,
$$
and it also fits the representation \eqref{eq:exp} with $\theta=\log(1-p)$,
$R(y)=y$, $h(y)=1$ and $\Psi(\theta)=-\log(1-e^\theta)$.

In the Poisson $\mathscr{P}(\lambda)$ case,
$$
f_\lambda (y) = \frac{1}{y!}\,\exp\left\{ y\log(\lambda)-\lambda \right\}
$$
and it also fits the representation \eqref{eq:exp} with $\theta=\log(\lambda)$,
$R(y)=y$, $h(y)=1/y!$ and $\Psi(\theta)=\exp(\theta)$.

In the exponential $\mathscr{E}xp(\lambda)$ case,
$$
f_\lambda (y) = \exp\left\{ -\lambda y +\log(\lambda)\right\}
$$
and it also fits the representation \eqref{eq:exp} with $\theta=\lambda$,
$R(y)=-y$, $h(y)=1$ and $\Psi(\theta)=-\log(\theta)$.

\begin{exoset}
Show that, for an exponential family, $\Psi(\theta)$ is defined by the constraint
that $f_\theta$ is a probability density and that the expectation of this distribution
can be written as $\partial\Psi(\theta)/\partial\theta$, the vector of the derivatives
of $\Psi(\theta)$ with respect to the components of $\theta$.
\end{exoset}

Using the representation \eqref{eq:exp},
$$
\int f_\theta(y)\,\text{d}y = 
\int h(y)\,\exp \left\{ \theta\cdot R(y) - \Psi(\theta) \right\} \,\text{d}y = 1
$$
implies that $\Psi(\theta)$ is uniquely defined by
$$
\int h(y)\,\exp \left\{ \theta\cdot R(y)\right\} \,\text{d}y = \int h(y)\,\text{d}y
\exp \left\{ \Psi(\theta) \right\}\,.
$$
When considering the expectation of $R(Y)$,
\begin{eqnarray*}
\mathbb{E}_\theta[R(Y)] &=&
    \int R(y)  h(y)\,\exp \left\{ \theta\cdot R(y) - \Psi(\theta) \right\} \,\text{d}y \\
&=& \int \frac{\partial}{\partial\theta} \left\{\theta\cdot R(y)\right\}
		h(y)\,\exp \left\{ \theta\cdot R(y) - \Psi(\theta) \right\} \,\text{d}y \\
&=& \int \frac{\partial}{\partial\theta} \left\{ \theta\cdot R(y)-\Psi(\theta)+\Psi(\theta) \right\}
                h(y)\,\exp \left\{ \theta\cdot R(y) - \Psi(\theta) \right\} \,\text{d}y \\
&=& \frac{\partial\Psi(\theta)}{\partial\theta} 
	\int h(y)\,\exp \left\{ \theta\cdot R(y) - \Psi(\theta) \right\} \,\text{d}y\\
&&+ \int h(y)\,\frac{\partial}{\partial\theta} \left[ \exp \left\{ \theta\cdot R(y) 
	- \Psi(\theta) \right\} \right] \,\text{d}y \\
&=& \frac{\partial\Psi(\theta)}{\partial\theta}\times 1 + \frac{\partial}{\partial\theta}
        \left\{ \int h(y)\,\exp \left\{ \theta\cdot R(y) - \Psi(\theta) \right\} \,\text{d}y\right\}\\
&=& \frac{\partial\Psi(\theta)}{\partial\theta}\,.
\end{eqnarray*}

\begin{exoset}\label{exo:update}
Show that the updated hyperparameters in (2.5) are given by
$$
\xi^\prime(y) = \xi + R(y)\,,\quad \lambda^\prime(y) = \lambda+1\,.
$$
Find the corresponding expressions for $\pi(\theta|\xi,\lambda,y_1,\ldots,y_n)$.
\end{exoset}

If we start with
$$
f_\theta(y) = h(y)\,\exp \left\{ \theta\cdot R(y) - \Psi(\theta) \right\}\,,
\quad\text{and}\quad
\pi(\theta|\xi,\lambda) \propto \exp \left\{
        \theta\cdot \xi - \lambda \Psi(\theta) \right\}\,,
$$
Bayes theorem implies that
\begin{eqnarray*}
\pi(\theta|\xi,\lambda,y)&\propto& f_\theta(y)\pi(\theta|\xi,\lambda)\\
&\propto& \exp \left\{ \theta\cdot R(y) - \Psi(\theta) \right\}\,
\exp \left\{\theta\cdot \xi - \lambda \Psi(\theta) \right\} \\
&=& \exp \left\{ \theta\cdot [R(y)+\xi] - (\lambda+1)\Psi(\theta) \right\}\,.
\end{eqnarray*}
Therefore, 
$$
\xi^\prime(y) = \xi + R(y)\,,\quad \lambda^\prime(y) = \lambda+1\,.
$$
Similarly,
\begin{eqnarray*}
\pi(\theta|\xi,\lambda,y_1,\ldots,y_n)&\propto& \prod_{i=1}^n f_\theta(y_i)\pi(\theta|\xi,\lambda)\\
&\propto& \exp \left\{ \sum_{i=1}^n [\theta\cdot R(y_i) - \Psi(\theta)] \right\}\,
\exp \left\{\theta\cdot \xi - \lambda \Psi(\theta) \right\} \\
&=& \exp \left\{ \theta\cdot [\sum_{i=1}^n R(y_i)+\xi] - (\lambda+n)\Psi(\theta) \right\}\,.
\end{eqnarray*}
Therefore,
$$
\xi^\prime(y_1,\ldots,y_n) = \xi + \sum_{i=1}^n R(y_i)\,,\quad \lambda^\prime(y_1,\ldots,y_n) = \lambda+n\,.
$$

\begin{exoset}
erive the posterior distribution for an iid sample $\mathscr{D}=(y_1,\ldots,y_n)$ from
$\mathscr{N}(\theta,1)$ and show that it only depends on the sufficient statistic
$\overline y=\sum_{i=1}^n y_i/n$.
\end{exoset}

Since (see Exercice \ref{exo:basex})
$$
f_\theta(y) = \frac{1}{\sqrt{2\pi}}\,\exp\frac{1}{2}\,\left\{-y^2+2y\theta-\theta^2\right\}
$$
fits the representation \eqref{eq:exp} with
$R(y)=y$, $h(y)=\exp(-y^2/2)/\sqrt{2\pi}$ and $\Psi(\theta)=\theta^2/2$,
a conjugate prior is 
$$
\pi(\theta|\xi,\lambda) \propto \exp\frac{1}{2}\,\left\{2\xi\theta-\lambda\theta^2\right\}\,,
$$
which is equivalent to a $\mathscr{N}(\xi/\lambda,1/\lambda)$ prior distribution. Following
the updating formula given in Exercice \ref{exo:update}, the posterior distribution is
a 
$$
\mathscr{N}(\xi^\prime(y_1,\ldots,y_n)/\lambda^\prime(y_1,\ldots,y_n),1/\lambda^\prime(y_1,\ldots,y_n))
$$ 
distribution, i.e.
$$
\mu|y_1,\ldots,y_n \sim \mathscr{N}\left( \frac{\xi 
	+ n \overline y}{\lambda+n}, \frac{1}{\lambda+n} \right)\,.
$$
It obviously only depends on the sufficient statistics $\overline y$.

\begin{exoset}
Give the range of values of the posterior mean (2.6)
as the pair $(\lambda,\lambda^{-1}\xi)$ varies over $\mathbb{R}^+\times\mathbb{R}$.
\end{exoset}

While 
$$
\frac{\lambda^{-1}}{1+\lambda^{-1}} \le 1\,,
$$
the fact that $\xi$ can take any value implies that this posterior mean has an unrestricted range,
which can be seen as a drawback of this conjugate modeling.

\begin{exoset}
A Weibull distribution $\mathscr{W}(\alpha,\beta,\gamma)$ is defined as the
power transform of a gamma $\mathscr{G}(\alpha,\beta)$ distribution: if
$X\sim\mathscr{W}(\alpha,\beta,\gamma)$, then $X^\gamma\sim
\mathscr{G}(\alpha,\beta)$. Show that, when $\gamma$ is known,
$\mathscr{W}(\alpha,\beta,\gamma)$ is an exponential family but that it is not
an exponential family when $\gamma$ is unknown.
\end{exoset}

The Weibull random variable $X$ has the density
\begin{equation*}
\frac{\gamma\alpha^{\beta}}{\Gamma(\beta)}\,x^{(\beta+1)\gamma-1}\,e^{-x^{\gamma}\alpha} \,,
\end{equation*}
since the Jacobian of the change of variables $y=x^{\gamma}$ is $\gamma x^{\gamma-1}$. 
So, checking the representation \eqref{eq:exp} leads to
$$
f(x|\alpha,\beta,\gamma) = \frac{\gamma\alpha^{\beta}}{\Gamma(\beta)}\,
\exp\left\{ [(\beta+1)\gamma-1]\log(x) - \alpha x^{\gamma} \right\}\,,
$$
with $R(x)=(\gamma\log(x),-x^{\gamma})$, $\theta=(\beta,\alpha)$ and $\Psi(\theta)=\log \Gamma(\beta)
-\log \gamma\alpha^{\beta}$.

If $\gamma$ is unknown, the term $x^{\gamma}\alpha$ in the exponential part makes it impossible to recover
the representation \eqref{eq:exp}.

\begin{exoset}
Show that, when the prior on $\theta=(\mu,\sigma^2)$ is
$\mathscr{N}(\xi,\sigma^2/\lambda_\mu)\times\mathscr{IG}(\lambda_\sigma,\alpha)$, the
marginal prior on $\mu$ is a Student's $t$ distribution
$\mathcal{T}(2\lambda_\sigma,\xi,\alpha/\lambda_\mu\lambda_\sigma)$ (see Example 2.3 below for the
definition of a Student's $t$ density).
Give the corresponding marginal prior on $\sigma^2$. For an iid sample $\mathscr{D}=(x_1,\ldots,x_n)$
from $\mathscr{N}(\mu,\sigma^2)$, derive the parameters of the posterior distribution of $(\mu,\sigma^2)$.
\end{exoset}

Since the joint prior distribution of $(\mu,\sigma^2)$ is
$$
\pi(\mu,\sigma^2) \propto (\sigma^2)^{-\lambda_\sigma-1-1/2}\,
\exp\frac{-1}{2\sigma^2}\left\{\lambda_\mu(\mu-\xi)^2 + 2\alpha \right\}
$$
(given that the Jacobian of the change of variable $\omega=\sigma^{-2}$ is $\omega^{-2}$),
integrating out $\sigma^2$ leads to
\begin{eqnarray*}
\pi(\mu) &\propto& \int_0^\infty (\sigma^2)^{-\lambda_\sigma-3/2}\,
\exp\frac{-1}{2\sigma^2}\left\{\lambda_\mu(\mu-\xi)^2 + 2\alpha \right\} \,\text{d}\sigma^2\\
&\propto& \int_0^\infty \omega^{\lambda_\sigma-1/2}\,\exp\frac{-\omega}{2}
\left\{\lambda_\mu(\mu-\xi)^2 + 2\alpha \right\} \,\text{d}\omega\\
&\propto& \left\{\lambda_\mu(\mu-\xi)^2 + 2\alpha \right\}^{-\lambda_\sigma-1/2}\\
&\propto& \left\{1 + \frac{ \lambda_\sigma\lambda_\mu(\mu-\xi)^2 }{2\lambda_\sigma\alpha} \right\}^{-\frac{2\lambda_\sigma+1}{2}}\,,
\end{eqnarray*}
which is the proper density of a Student's $t$ distribution
$\mathcal{T}(2\lambda_\sigma,\xi,\alpha/\lambda_\mu\lambda_\sigma)$.

By definition of the joint prior on $(\mu,\sigma^2)$, the marginal prior on $\sigma^2$ is a
inverse gamma $\mathscr{IG}(\lambda_\sigma,\alpha)$ distribution.

The joint posterior distribution of $(\mu,\sigma^2)$ is
$$
\pi((\mu,\sigma^2)|\mathscr{D}) 
\propto (\sigma^2)^{-\lambda_\sigma(\mathscr{D})}\exp\left\{-\left(\lambda_\mu(\mathscr{D})
                                  (\mu-\xi(\mathscr{D}))^2+\alpha(\mathscr{D})\right)/2\sigma^2\right\}\,,
$$
with
\begin{align*}
\lambda_\sigma(\mathscr{D}) &= \lambda_\sigma+3/2+n/2 \,,\\
\lambda_\mu(\mathscr{D}) &= \lambda_\mu + n\,,\\
\xi(\mathscr{D}) &= (\lambda_\mu\xi + n\overline x)/\lambda_\mu(\mathscr{D}) \,,\\
\alpha(\mathscr{D}) &= 2\alpha+\frac{\lambda_\mu(\mathscr{D})}{n\lambda_\mu}(\overline x - \xi)^2 +s^2(\mathscr{D})\,.
\end{align*}
This is the product of a marginal inverse gamma 
$$
\mathscr{IG}\left( \lambda_\sigma(\mathscr{D})-3/2,\alpha(\mathscr{D})/2 \right)
$$
distribution on $\sigma^2$ by a conditional normal 
$$
\mathscr{N}\left( \xi(\mathscr{D}), \sigma^2/\lambda_\mu(\mathscr{D}) \right)
$$
on $\mu$. (Hence, we do get a conjugate prior.) Integrating out $\sigma^2$ leads to
\begin{eqnarray*}
\pi(\mu|\mathscr{D}) &\propto&
	\int_0^\infty (\sigma^2)^{-\lambda_\sigma(\mathscr{D})}\,\exp\left\{-\left(\lambda_\mu(\mathscr{D})
       (\mu-\xi(\mathscr{D}))^2+\alpha(\mathscr{D})\right)/2\sigma^2\right\}\,\text{d}\sigma^2\\
&\propto& \int_0^\infty \omega^{\lambda_\sigma(\mathscr{D})-2}\,\exp\left\{-\left(\lambda_\mu(\mathscr{D})
       (\mu-\xi(\mathscr{D}))^2+\alpha(\mathscr{D})\right)\omega/2\right\}\,\text{d}\omega\\
&\propto& \left( \lambda_\mu(\mathscr{D}) (\mu-\xi(\mathscr{D}))^2+\alpha(\mathscr{D})\right)^{-(\lambda_\sigma(\mathscr{D})-1)}\,,
\end{eqnarray*}
which is the generic form of a Student's $t$ distribution.

\begin{exoset}
Show that, for location and scale models, Jeffreys' prior is given by
$\pi^J(\theta)=1$ and $\pi^J(\theta)=1/\theta$, respectively.
\end{exoset}

In the case of a location model, $f(y|\theta) = p(y-\theta)$, 
the Fisher information matrix of a location model is given by
\begin{eqnarray*}
I(\theta) &=& \mathbb{E}_\theta \left[ \frac{\partial \log p(Y-\theta)}{\partial\theta}^\tee
				       \frac{\partial \log p(Y-\theta)}{\partial\theta} \right]\\
          &=& \int \left[ \frac{\partial p(y-\theta)}{\partial\theta}\right]^\tee
	           \left[ \frac{\partial p(y-\theta)}{\partial\theta}\right] \big/ p(y-\theta)\,\text{d}y\\
          &=& \int \left[ \frac{\partial p(z)}{\partial z}\right]^\tee
                   \left[ \frac{\partial p(z)}{\partial z}\right] \big/ p(z)\,\text{d}z
\end{eqnarray*}
it is indeed constant in $\theta$. Therefore the determinant of $I(\theta)$ is also constant and Jeffreys' prior
can be chosen as $\pi^J(\theta)=1$ [or any other constant as long as the parameter space is not compact].

In the case of a scale model, if $y\sim f(y/\theta)/\theta$, a change of variable from $y$ to $z=\log(y)$ [if
$y>0$] implies that $\eta=\log(\theta)$ is a location parameter for $z$. Therefore, the Jacobian transform of
$\pi^J(\eta)=1$ is $\pi^J(\theta)=1/\theta$. When $y$ can take both negative and positive values, a transform of
$y$ into $z=\log(|y|)$ leads to the same result.

\begin{exoset}
In the case of an exponential family, derive Jeffreys' prior in terms of the
Hessian matrix of $\Psi(\theta)$, i.e.~the matrix of second derivatives of $\Psi(\theta)$.
\end{exoset}

Using the representation \eqref{eq:exp}
$$
\log f_\theta(y) = \log h(y) + \theta\cdot R(y) - \Psi(\theta)\,,
$$
we get
$$
 \frac{\partial^2}{\partial\theta\partial\theta^\tee} \log f_\theta(y) 
= - \frac{\partial^2 \Psi(\theta)}{\partial\theta\partial\theta^\tee}
$$
and therefore the Fisher information matrix is the Hessian matrix of $\Psi(\theta)$, $H(\theta)$. This implies
that $\pi^J(\theta) = \text{det} H(\theta)$.

\begin{exoset}
Show that, when $\pi(\theta)$ is a probability density, (2.8)
necessarily holds for all datasets $\mathscr{D}$.
\end{exoset}

Given that $\pi(\theta)$ is a (true) probability density and that the likelihood
$\ell(\theta|\mathscr{D})$ is also a (true) probability density in $\mathscr{D}$
that can be interpreted as a conditional density, the product 
$$
\pi(\theta)\ell(\theta|\mathscr{D})
$$
is a true joint probability density for $(\theta,\mathscr{D})$. The above integral
therefore defines the marginal density of $\mathscr{D}$, which is always defined. 

\begin{exoset}
Try to devise a parameterized model and an
improper prior such that, no matter the sample size, the posterior distribution
does not exist. (If you cannot find such a case, wait until Chapter 6.)
\end{exoset}

It is sufficient to find a function of the parameter $\theta$ that goes to infinity
faster than the likelihood goes to $0$, no matter what the sample size is. For instance,
take $\pi(\theta) \propto \exp \theta^2$ for a Cauchy $\mathscr{C}(\theta,1)$ model. Then, for
a sample of size $n$, the likelihood goes to $0$ as $\theta^{-2n}$ and it cannot beat the
exponential increase in the prior.

\begin{exoset}
Show that, under the loss $L_{a_0,a_1}$, the Bayes estimator associated with a prior $\pi$ is given by
$$
\delta^\pi(x)  =  \begin{cases} 1 & \hbox{if}\quad P^\pi(\theta \in \Theta_0|x)>a_1\big/{a_0+a_1}, \cr
                                0 & \hbox{otherwise.}\cr \end{cases}
$$
\end{exoset}

The posterior expected loss is
$$
\mathbb{E}\left[L_{a_0,a_1} (\theta ,d)|x\right] = 
\begin{cases} 
	      a_0\,P^\pi(\theta\in\Theta_0|x) & \hbox{if}\quad d=0\,, \cr
              a_1\,P^\pi(\theta\in\Theta_1|x) & \hbox{if}\quad d=1\,, \cr
\end{cases}
$$
thus the decision minimising this posterior loss is $d=1$ when 
$a_1\,P^\pi(\theta\in\Theta_1|x)<a_0\,P^\pi(\theta\in\Theta_0|x)$, i.e.
$$
a_1(1-P^\pi(\theta\in\Theta_0|x)) < a_0\,P^\pi(\theta\in\Theta_0|x)\,,
$$
and $d=0$ otherwise.

\begin{exoset}
When $\theta\in\{\theta_0,\theta_1\}$, show that the Bayesian procedure
only depends on the ratio $\varrho_0 f_{\theta_0}(x)/(1-\varrho_0)f_{\theta_1}(x)$,
where $\varrho_0$ is the prior weight on $\theta_0$.
\end{exoset}

In this special case, $\pi$ puts a point mass of $\varrho_0$ on $\theta_0$ and of
$(1-\varrho_0)$ on $\theta_1$. Therefore,
\begin{eqnarray*}
P^\pi(\theta = \theta_0|x) &=& \frac{\varrho_0\,f_{\theta_0}(x)}{\varrho_0\,f_{\theta_0}(x)
+(1-\varrho_0)\,f_{\theta_1}(x)} \\
&=&  \frac{1}{1+(1-\varrho_0)\,f_{\theta_1}(x)/(1-\varrho_0)\,f_{\theta_1}(x)}\,,
\end{eqnarray*}
which only depends on the ratio $\varrho_0 f_{\theta_0}(x)/(1-\varrho_0)f_{\theta_1}(x)$.

\begin{exoset}
Show that the limit of the posterior probability $P^\pi(\mu<0|x)$ when $\xi$ goes
to $0$ and $\tau$ goes to $\infty$ is $\Phi(-x/\sigma)$.
\end{exoset}

Since
\begin{eqnarray*}
P^\pi(\mu<0|x) &=& \Phi\left(-\xi(x)/ \omega \right)\\
     &=& \Phi\left( \frac{\sigma^2\xi+\tau^2 x}{\sigma^2 +\tau^2} 
	       \sqrt{\frac{\sigma^2+\tau^2}{\sigma^2\tau^2}} \right)\\
     &=& \Phi\left( \frac{\sigma^2\xi+\tau^2 x}{\sqrt{\sigma^2 +\tau^2}\sqrt{\sigma^2\tau^2}} \right)\,,
\end{eqnarray*}
when $\xi$ goes to $0$ and $\tau$ goes to $\infty$, the ratio 
$$
\frac{\sigma^2\xi+\tau^2 x}{\sqrt{\sigma^2 +\tau^2}\sqrt{\sigma^2\tau^2}}
$$
goes to
$$
\lim_{\tau\to\infty} \frac{\tau^2 x}{\sqrt{\sigma^2 +\tau^2}\sqrt{\sigma^2\tau^2}}
= \lim_{\tau\to\infty} \frac{\tau^2 x}{\tau^2\sigma} = \frac{x}{\sigma}\,.
$$

\begin{exoset}\label{ex:constu}
We recall that the normalizing constant for a Student's
$\mathcal{T}(\nu,\mu,\sigma^2)$ distribution is
$$
\frac{\Gamma((\nu+1)/2)/\Gamma(\nu/2)}{\sigma \sqrt{\nu\pi} }\,.
$$
Give the value of the integral in the denominator of $B^\pi_{10}$ above.
\end{exoset}

We have
$$
(\mu-\bar x)^2+(\mu-\bar y)^2 =
2 \left( \mu - \frac{\bar x+\bar y}{2} \right)^2 + \frac{(\bar x-\bar y)^2}{2}
$$
and thus
\begin{align*}
\int\,&\left[(\mu-\bar x)^2+(\mu-\bar y)^2+S^2\right]^{-n} \mathrm{d}\mu \\
&= 2^{-n}\int\,\left[\left( \mu - \frac{\bar x+\bar y}{2} \right)^2 
	+ \frac{(\bar x-\bar y)^2}{4}+\frac{S^2}{2}\right]^{-n} \mathrm{d}\mu\\
&= (2\sigma^2)^{-n}\int\,\left[1+\left( \mu - \frac{\bar x+\bar y}{2} \right)^2
	\big/\sigma^2\nu \right]^{-(\nu+1)/2} \mathrm{d}\mu\,,
\end{align*}
where $\nu=2n-1$ and 
$$
\sigma^2= \left[ \left( \frac{\bar x-\bar y}{2} \right)^{2} + \frac{S^2}{2} \right]\bigg/(2n-1)\,.
$$
Therefore,
\begin{align*}
\int\,&\left[(\mu-\bar x)^2+(\mu-\bar y)^2+S^2\right]^{-n} \mathrm{d}\mu \\
&= (2\sigma^2)^{-n}\,\frac{\sigma \sqrt{\nu\pi} }{\Gamma((\nu+1)/2)/\Gamma(\nu/2)}\\
&= \frac{\sqrt{\nu\pi} }{2^n \sigma^{2n-1} \Gamma((\nu+1)/2)/\Gamma(\nu/2)}\\
&= \frac{(2n-1)^{2n-1}\sqrt{\nu\pi} }{2^n 
\left[ \left( \frac{\bar x-\bar y}{2} \right)^{2} + \frac{S^2}{2} \right]^{2n-1} \Gamma((\nu+1)/2)/\Gamma(\nu/2)}\,.
\end{align*}
Note that this expression is used later in the simplified devivation of $B_{01}^\pi$ without
the term $(2n-1)^{2n-1}\sqrt{\nu\pi}/2^n \Gamma((\nu+1)/2)/\Gamma(\nu/2)$ because this term appears in both
the numerator and the denominator.

\begin{exoset}\label{exo:stew}
Approximate $B^\pi_{01}$ by a Monte Carlo experiment where $\xi$ is simulated from a Student's
$t$ distribution with mean $(\bar x+\bar y)/2$ and appropriate variance, and the integrand is
proportional to $\exp-\xi^2/2$. Compare the precision of the resulting estimator with the
above Monte Carlo approximation based on the normal simulation.
\end{exoset}

\begin{figure}[h]
\centerline{\includegraphics[height=5truecm,width=.8\textwidth]{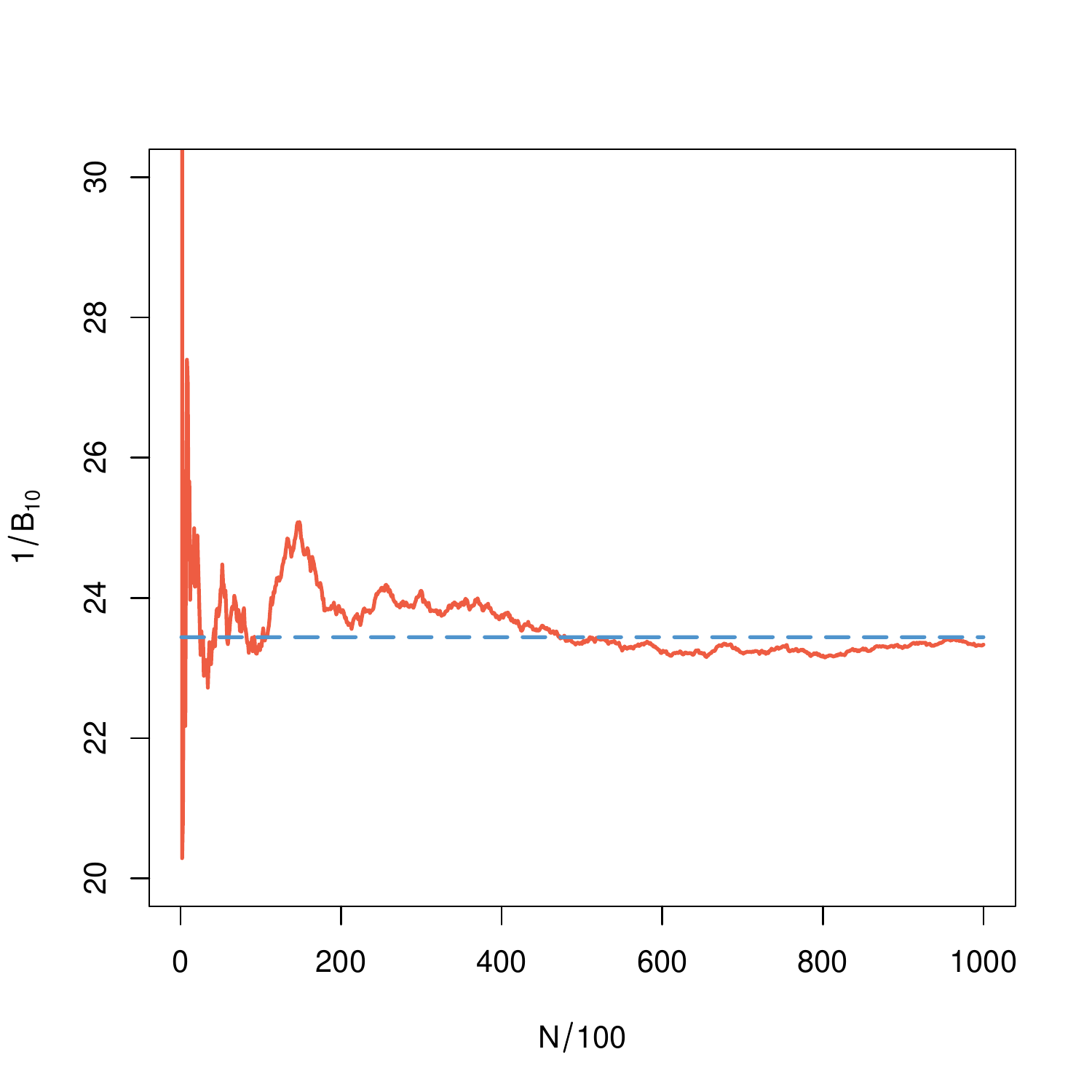}}
\caption{\label{fig:exo222}
Comparison of two approximations of the Bayes factor $B_{01}$ based on $10^6$ simulations.}
\end{figure}

The integral of interest in $B^\pi_{01}$ is
\begin{align*}
\int\,&\left[ (2\xi+\bar x-\bar y)^2/2+S^2 \right]^{-n+1/2} \,e^{-\xi^2/2}\,\hbox{d}\xi/\sqrt{2\pi}\\
&= \left(S^2\right)^{-n+1/2} \int\,\frac{\exp-\xi^2/2}{\sqrt{2\pi}}\, 
	\left[ \frac{4(n-1)(\xi-(\bar y-\bar x)/2)^2}{(2n-2)S^2}+1 \right]^{-n+1/2} \,\hbox{d}\xi\\
&= \left(S^2\right)^{-n+1/2} \int\,\frac{\exp-\xi^2/2}{\sqrt{2\pi}}\, 
	\left[ \frac{(\xi-(\bar y-\bar x)/2)^2}{\nu\sigma^2}+1 \right]^{-(\nu+1)/2} \,\hbox{d}\xi\\
&= C \int\,\frac{\exp-\xi^2/2}{\sqrt{2\pi}}\,\mathfrak{t}(\xi|\mu,\sigma,\nu)\,\hbox{d}\xi\,,
\end{align*}
where $\mathfrak{t}(\xi|\mu,\sigma,\nu)$ is the density of the Student's
$\mathcal{T}(\nu,\mu,\sigma^2)$ distribution with parameters $\nu=2n-2$, $\mu=(\bar y-\bar x)/2$,
and $\sigma^2 = S^2/4(n-1)$, and $C$ is the constant
$$
C=\left( S^2 \right)^{-n+1/2} \bigg/ \frac{\Gamma((\nu+1)/2)/\Gamma(\nu/2)}{\sigma \sqrt{\nu\pi} }\,.
$$
Therefore, we can simulate a sample $\xi_1,\ldots,\xi_n$ from the 
$\mathcal{T}(\nu,\mu,\sigma^2)$ distribution and approximate the above integral by 
the average
$$
\frac{C}{n} \sum_{i=1}^n \frac{\exp-\xi_i^2/2}{\sqrt{2\pi}}\,,
$$
using an {\sf R} program like
\begin{verbatim}
n=100
N=1000000
nu=2*n-2
barx=.088
bary=.1078
mu=.5*(bary-barx)
stwo=.00875
sigma=sqrt(.5*stwo/nu)
C=log(stwo)*(-n+.5)+log(sigma*sqrt(nu*pi))+
	lgamma(.5*nu)-lgamma(.5*nu+.5)

# T simulation
xis=rt(n=N,df=nu)*sigma + mu
B01=-log(cumsum(dnorm(xis))/(1:N))
B01=exp( (-n+.5)*log(.5*(barx-bary)^2+stwo)+B01-C )

# Normal simulation
xis=rnorm(N)
C01=cumsum((stwo+.5*(2*xis+barx-bary)^2)^(-n+.5))/(1:N) 
C01=((.5*(barx-bary)^2+stwo)^(-n+.5))/C01

# Comparison of the cumulated averages
plot(C01[seq(1,N,l=1000)],type="l",col="tomato2",lwd=2,
    ylim=c(20,30),xlab=expression(N/100),ylab=expression(1/B[10]))
lines(B01[seq(1,N,l=1000)],col="steelblue3",lwd=2,lty=5)

\end{verbatim}
As shown on Figure \ref{fig:exo222}, the precision of the estimator based
on the $\mathcal{T}(\nu,\mu,\sigma^2)$ simulation is immensely superior to
the precision of the estimator based on a normal sample: they both converge
to the same value $23.42$, but with very different variances.

\begin{exoset}
Discuss what happens to the importance sampling approximation when
the support of $g$ is larger than the support of $\gamma$.
\end{exoset}

If the support of $\gamma$, $\mathfrak{S}_\gamma$, is smaller than the
support of $g$, the representation
$$
\mathfrak{I} = \int\,\frac{h(x)g(x)}{\gamma(x)}\,\gamma(x)\,\hbox{d}x
$$
is not valid and the importance sampling approximation evaluates instead
the integral
$$
\int_{\mathfrak{S}_\gamma}\,\frac{h(x)g(x)}{\gamma(x)}\,\gamma(x)\,\hbox{d}x.
$$

\begin{exoset}\label{exo:poorcoco}
Show that the importance weights of Example 2.2 have infinite variance.
\end{exoset}

The importance weight is
$$
\exp\left\{(\theta-\mu)^2/2\right\}\,\prod_{i=1}^n [1+(x_i-\theta)^2]^{-1}
$$
with $\theta\sim\mathscr{N}(\mu,\sigma^2)$. While its expectation is finite---it would
be equal to $1$ were we to use the right normalising constants---, the expectation of its
square is not:
$$
\int \exp\left\{(\theta-\mu)^2/2\right\}\,\prod_{i=1}^n [1+(x_i-\theta)^2]^{-2} \,\text{d}\theta=\infty\,,
$$
due to the dominance of the exponential term over the polynomial term.

\begin{exoset}
Show that, when $\gamma$ is the normal {$\mathscr{N}(0,\nu/(\nu-2))$} density, the ratio
$${
{f_\nu^2(x) \over \gamma(x) } \propto
{e^{x^2(\nu-2)/2\nu}\over [1 + x^2/\nu]^{(\nu+1)}}
}$$
does not have a finite integral. What does this imply about the variance of the importance weights?
\end{exoset}

\begin{figure}[h]
\centerline{\includegraphics[height=7truecm,width=.95\textwidth]{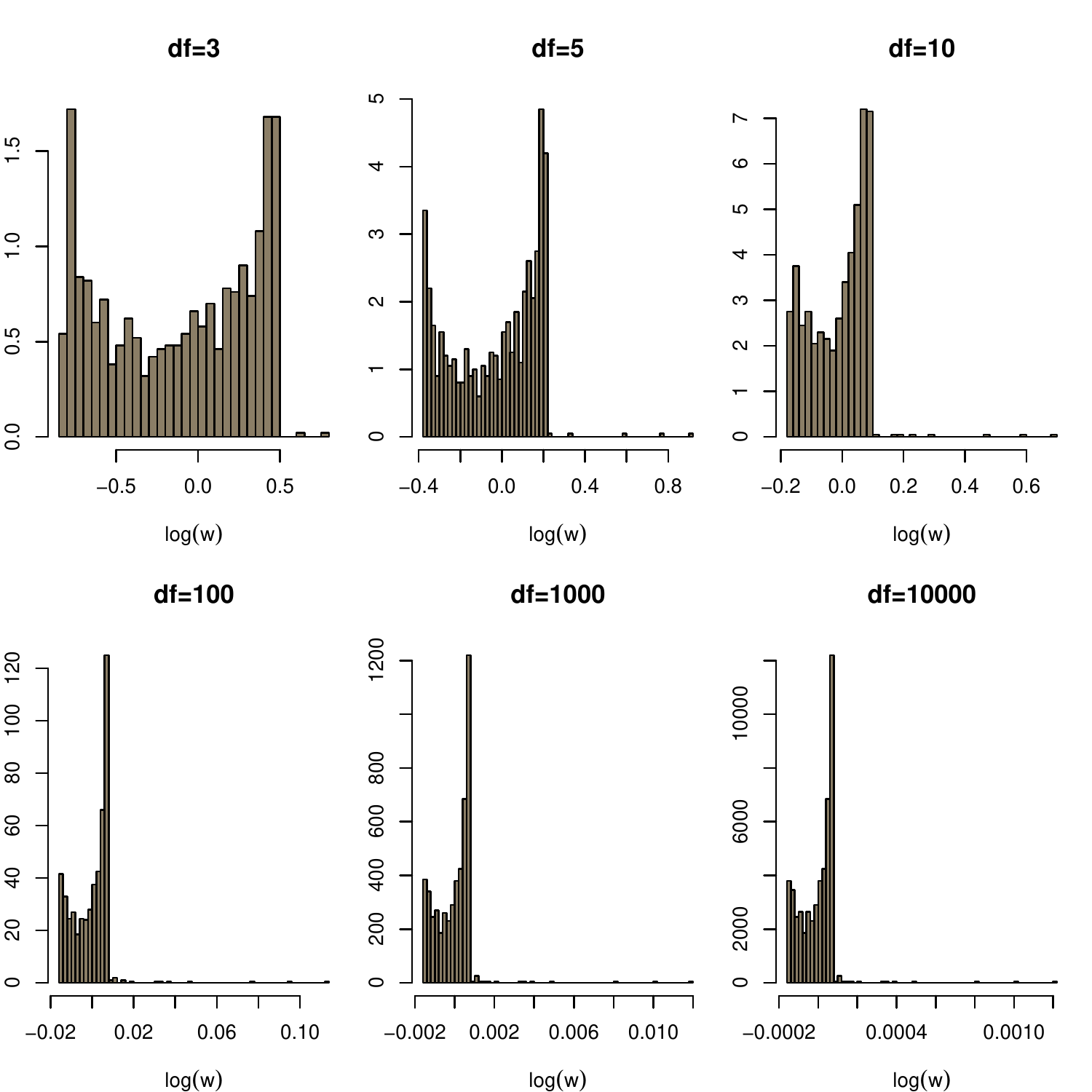}}
\caption{\label{fig:histowei}
Distributions of the log-importance weights for a normal importance distribution against
a Student's $t$ target for several values of $\nu$.}
\end{figure}

This is more or less a special case of Exercise \ref{exo:poorcoco}, with again the exponential
term dominating the polynomial term, no matter what the value of $\nu>2$ is. The importance weights
have no variance. When running an experiment like the following one
\begin{verbatim}
nu=c(3,5,10,100,1000,10000)
N=length(nu)
T=1000

nors=rnorm(T)
par(mfrow=c(2,N/2),mar=c(4,2,4,1))

for (nnu in nu){

   y=sqrt(nnu/(nnu-2))*nors
   isw=dt(y,df=nnu)/dnorm(y)

   hist(log(isw),prob=T,col="wheat4",nclass=T/20)
}
\end{verbatim}
the output in Figure \ref{fig:histowei} shows that the value of $\nu$ still matters very much in the
distribution of the weights. When $\nu$ is small, the probability to get very large weights is much higher
than with large $\nu$'s, and the dispersion decreases with $\nu$.

\begin{exoset}\label{ex:bridj}
Given two model densities $f_1(\mathscr{D}|\theta)$ and $f_2(\mathscr{D}|\theta)$
with the same parameter $\theta$ and corresponding priors
densities $\pi_1(\theta)$ and ${\pi}_2(\theta)$, denote
${\tilde\pi}_1(\theta|\mathscr{D})= f_1(\mathscr{D}|\theta){\pi}_1(\theta)$ and
${\tilde\pi}_2(\theta|\mathscr{D})= f_2(\mathscr{D}|\theta){\pi}_2(\theta)$, and
show that the Bayes factor corresponding to the comparison of both models satisfies
$$
B^\pi_{12} = \frac{ \displaystyle{ \int {\tilde\pi}_1
    (\theta|\mathscr{D}) \alpha(\theta) {\pi}_2(\theta|\mathscr{D}) \hbox{d}\theta } }
{ \displaystyle{ \int {\tilde\pi}_2(\theta|\mathscr{D}) \alpha(\theta)
     \pi_1(\theta|\mathscr{D}) \hbox{d}\theta }  }
$$
for every positive function $\alpha$ and deduce that
$$
   \displaystyle{ n_1 \sum_{i=1}^{n_2} {\tilde\pi}_1(\theta_{2i}|\mathscr{D})
   \alpha(\theta_{2i}) } \bigg/ \displaystyle{ n_2 \sum_{i=1}^{n_1}
   {\tilde\pi}_2(\theta_{1i}|\mathscr{D}) \alpha(\theta_{1i}) }
$$
is a convergent approximation of the Bayes factor $B^\pi_{12}$ when $\theta_{ji} \sim \pi_j(\theta|\mathscr{D})$
$(i=1,2,\,j=1,\ldots,n_j)$.
\end{exoset}

The missing normalising constants in ${\tilde\pi}_1(\theta|\mathscr{D})$ and ${\tilde\pi}_2(\theta|\mathscr{D})$
are the marginal densities $m_1(\mathscr{D})$ and $m_2(\mathscr{D})$, in the sense that $(i=1,2)$
$$
\pi_i(\theta|\mathscr{D})={\tilde\pi}_i(\theta|\mathscr{D})/m_i(\mathscr{D})\,.
$$
Therefore,
\begin{align*}
&\frac{ \displaystyle{ \int {\tilde\pi}_1
    (\theta|\mathscr{D}) \alpha(\theta) {\pi}_2(\theta|\mathscr{D}) \hbox{d}\theta } }
{ \displaystyle{ \int {\tilde\pi}_2(\theta|\mathscr{D}) \alpha(\theta)
     \pi_1(\theta|\mathscr{D}) \hbox{d}\theta }  }\\
&=  \frac{ \displaystyle{ \int m_1(\mathscr{D}) \pi_1(\theta|\mathscr{D}) \alpha(\theta) 
	{\pi}_2(\theta|\mathscr{D}) \hbox{d}\theta } }
{ \displaystyle{ \int m_2(\mathscr{D}) pi_1(\theta|\mathscr{D}) \alpha(\theta)
        {\pi}_2(\theta|\mathscr{D}) \hbox{d}\theta } }\\
&= \frac{m_1(\mathscr{D})}{m_2(\mathscr{D})} = B^\pi_{12} 
\end{align*}
and $\alpha$ is irrelevant for the computation of the ratio of integrals.

A Monte Carlo implementation of this remark is to represent each integral in the ratio
as an expectation under ${\pi}_2(\theta|\mathscr{D})$ and $\pi_1(\theta|\mathscr{D})$
respectively. Simulations $\theta_{ji}$'s from both posteriors then produce convergent
estimators of the corresponding integrals. This method is called {\em bridge sampling}
and the choice of $\alpha$ is relevant in the variance of the corresponding estimator. 

\begin{exoset}\label{ex:trueV}
Show that, when $n$ goes to infinity and when the prior has an
unlimited support, the predictive distribution converges to the exact (sampling) distribution
of $x_{n+1}$.
\end{exoset}

This property follows from the fact that the posterior distribution converges to a Dirac mass
at the true value $\theta^\star$ of the parameter when $n$ goes to infinity [under some regularity 
conditions on both $\pi$ and $f(x|\theta)$, as well as identifiability constraints]. Therefore,
$$
\int f(x_{n+1}|\theta) \pi(\theta|\mathscr{D}_n)\,\hbox{d}\theta
$$
converges to $f(x_{n+1}|\theta^\star)$.

\begin{exoset}\label{exo:invcdf}
Show that, when $X$ is distributed from an increasing and continuous
cdf $F$, $F(X)$ has a uniform distribution.
\end{exoset}

If $F$ is increasing and continuous, it is invertible and we have
$$
P(F(X)\le u) = P(X\le F^{-1}(u)) = F(F^{-1}(u)) = u\,,
$$
when $0\le u\le 1$. This demonstrates that $F(X)$ is uniformely distributed
and this property can be exploited for simulation purposes when $F$ is available
in closed form.

\chapter{Regression and Variable Selection}\label{ch:reg}

\begin{exoset}
Show that the matrix $X$ is of full rank if and only if the matrix $X^\tee X$ is invertible
(where $X^\tee$ denotes the transpose of the matrix $X$, which can produced in {\sf R} using 
the \verb+t(X)+ command). Deduce that this cannot happen when $k+1>n$.
\end{exoset}

The matrix $X$ is a $(n,k+1)$ matrix. It is of full rank if the $k+1$ columns of $X$ induce
a subspace of $\mathbb{R}^n$ of dimension $(k+1)$, or, in other words, if those columns are
linearly independent: there exists no solution to $X\gamma=\mathbf{0}_{n}$ other than $\gamma=\mathbf{0}_{n}$,
where $\mathbf{0}_{k+1}$ denotes the $(k+1)$-dimensional vector made of $0$'s. If $X^\tee X$ is invertible,
then $X\gamma=\mathbf{0}_{n}$ implies $X^\tee X \gamma=X^\tee\mathbf{0}_{n}=\mathbf{0}_{k+1}$ and thus
$\gamma=(X^\tee X)^{-1} \mathbf{0}_{k+1} = \mathbf{0}_{k+1}$, therefore $X$ is of full rank. If $X^\tee X$ is 
not invertible, there exist vectors $\beta$ and $\gamma\ne\beta$ such that $X^\tee X \beta = X^\tee X \gamma$,
i.e.~$X^\tee X (\beta-\gamma)=\mathbf{0}_{k+1}$. This implies that $||X(\beta-\gamma)||^2=0$ and hence
$X(\beta-\gamma)=\mathbf{0}_{n}$ for $\beta-\gamma\ne\mathbf{0}_{k+1}$, thus $X$ is not of full rank.

Obviously, the matrix $(k+1,k+1)$ matrix $X^\tee X$ cannot be invertible if $k+1>n$ since the columns of
$X$ are then necessarily linearly dependent.

\begin{exoset}
Show that solving the minimization program above requires solving
the system of equations $(X^\tee X)\beta=X^\tee \by$. Check that
this can be done via the {\sf R} command
\begin{verbatim}
   > solve(t(X)%*%(X),t(X)%*%y)
\end{verbatim}
\end{exoset}

If we decompose $(\by-X\beta)^\tee (\by-X\beta)$ as
$$
\by^\tee\by-2\by^\tee X\beta+\beta^\tee X^\tee X \beta
$$
and differentiate this expression in $\beta$, we obtain the equation
$$
-2 \by^\tee X + 2 \beta^\tee X^\tee X = \mathbf{0}_{k+1}\,,
$$
i.e.
$$
(X^\tee X)\beta=X^\tee \by
$$
by transposing the above.

As can be checked via \verb+help(solve)+, \verb+solve(A,b)+ is the {\sf R} function that solves the linear equation 
system $Ax=b$. Defining $X$ and $y$ from {\sf caterpillar}, we get
\begin{verbatim}
> solve(t(X)%*%X,t(X)%*%y)

                     [,1]
  rep(1, 33) 10.998412367
  V1         -0.004430805
  V2         -0.053830053
  V3          0.067939357
  V4         -1.293636435
  V5          0.231636755
  V6         -0.356799738
  V7         -0.237469094
  V8          0.181060170
  V9         -1.285316143
  V10        -0.433105521
\end{verbatim}
which [obviously] gives the same result as the call to the linear regression
function {\sf lm()}:
\begin{verbatim}
> lm(y~X-1)

Call:
lm(formula = y ~ X - 1)

Coefficients:
Xrep(1, 33)     XV1       XV2       XV3        XV4         XV5
 10.998412  -0.004431  -0.053830   0.067939  -1.29363   0.23163
      XV6       XV7       XV8       XV9        XV10
 -0.356800  -0.237469   0.181060  -1.285316  -0.43310
\end{verbatim}
Note the use of the {\sf -1} in the formula \verb+y~X-1+ that eliminates the intercept
already contained in $X$.

\begin{exoset}
Show that $\mathbb{V}(\hat\beta|\sigma^2,X)=\sigma^2(X^\tee X)^{-1}$.
\end{exoset}

Since $\hat\beta = (X^\tee X)^{-1} X^\tee \by$ is a linear transform of $\by\sim\mathscr{N}(X\beta,\sigma^2 I_n)$, we
have
$$
\hat\beta \sim \mathscr{N}\left(
(X^\tee X)^{-1} X^\tee X\beta,\sigma^2 
(X^\tee X)^{-1} X^\tee X (X^\tee X)^{-1} \right)\,,
$$
i.e.~
$$
\hat\beta \sim \mathscr{N}\left( \beta, \sigma^2 (X^\tee X)^{-1} \right)\,.
$$

\begin{exoset}\label{exo:matilde}
Taking advantage of the matrix identities
\begin{eqnarray*}
\left(M+X^\tee X\right)^{-1} & = & M^{-1}-M^{-1}\left(M^{-1}+(X^\tee X)^{-1}\right)^{-1}M^{-1} \\
                         & = & (X^\tee X)^{-1}-(X^\tee X)^{-1}\left(M^{-1}+(X^\tee X)^{-1}\right)^{-1}(X^\tee X)^{-1}
\end{eqnarray*}
and
\begin{eqnarray*}
X^\tee X(M+X^\tee X)^{-1}M & = & \left(M^{-1}(M+X^\tee X)(X^\tee X)^{-1}\right)^{-1} \\
                   & = & \left(M^{-1}+(X^\tee X)^{-1}\right)^{-1}\,,
\end{eqnarray*}
establish that (3.3) and (3.4) are the correct posterior distributions.
\end{exoset}

Starting from the prior distribution
$$
\beta|\sigma^2,X\sim\mathscr{N}_{k+1}(\tilde\beta,\sigma^2M^{-1})\,,\quad
\sigma^2|X\sim \mathscr{IG}(a,b)\,,
$$
the posterior distribution is
\begin{align*}
\pi(\beta,\sigma^2&|\hat\beta,s^2,X) \propto
\sigma^{-k-1-2a-2-n}\,\exp\frac{-1}{2\sigma^2}\left\{ 
(\beta-\tilde\beta)^\tee M (\beta-\tilde\beta) \right.\\
	&\quad\left. + (\beta-\hat\beta)^\tee (X^\tee X) (\beta-\hat\beta) +s^2 +2b \right\}\\
&=\sigma^{-k-n-2a-3}\,\exp\frac{-1}{2\sigma^2}\left\{
\beta^\tee (M+X^\tee X) \beta - 2 \beta^\tee (M\tilde\beta+X^\tee X\hat\beta)\right.\\
        &\quad\left. +\tilde\beta^\tee M\tilde\beta +\hat\beta^\tee (X^\tee X) \hat\beta +s^2 +2b \right\}\\
&=\sigma^{-k-n-2a-3}\,\exp\frac{-1}{2\sigma^2}\left\{
(\beta-\mathbb{E}[\beta|y,X])^\tee (M+X^\tee X) (\beta-\mathbb{E}[\beta|y,X])\right.\\
        &\quad\left. + \beta^\tee M\tilde\beta
+\hat\beta^\tee (X^\tee X) \hat\beta -\mathbb{E}[\beta|y,X]^\tee (M+X^\tee X) \mathbb{E}[\beta|y,X]+s^2 +2b \right\}
\end{align*}
with
$$
\mathbb{E}[\beta|y,X] = (M+X^\tee X)^{-1} (M\tilde\beta+X^\tee X\hat\beta)\,.
$$
Therefore, (3.3) is the conditional posterior distribution of $\beta$ given $\sigma^2$.
Integrating out $\beta$ leads to
\begin{align*}
\pi(\sigma^2&|\hat\beta,s^2,X) \propto \sigma^{-n-2a-2}\,\exp\frac{-1}{2\sigma^2}\left\{
\beta^\tee M\tilde\beta +\hat\beta^\tee (X^\tee X) \hat\beta \right.\\
        &\quad\left. -\mathbb{E}[\beta|y,X]^\tee (M+X^\tee X) \mathbb{E}[\beta|y,X]+s^2 +2b \right\}\\
&=\sigma^{-n-2a-2}\,\exp\frac{-1}{2\sigma^2}\left\{
\beta^\tee M\tilde\beta +\hat\beta^\tee (X^\tee X) \hat\beta +s^2 +2b \right.\\ 
	&\quad\left. -(M\tilde\beta+X^\tee X\hat\beta)^\tee (M+X^\tee X)^{-1} (M\tilde\beta+X^\tee X\hat\beta) \right\}
\end{align*}
Using the first matrix identity, we get that
\begin{align*}
(M\tilde\beta+&X^\tee X\hat\beta)^\tee
\left(M+X^\tee X\right)^{-1} (M\tilde\beta+X^\tee X\hat\beta) \\
& =  \tilde\beta^\tee M\tilde\beta-\tilde\beta^\tee\left(M^{-1}+(X^\tee X)^{-1}\right)^{-1}\tilde\beta \\
& + \hat\beta^\tee (X^\tee X)\hat\beta-\hat\beta^\tee\left(M^{-1}+(X^\tee X)^{-1}\right)^{-1}\hat\beta\\
& + 2\hat\beta^\tee (X^\tee X) \left(M+X^\tee X\right)^{-1} M \tilde\beta\\
& = \tilde\beta^\tee M\tilde\beta + \hat\beta^\tee (X^\tee X)\hat\beta\\
& - (\tilde\beta-\hat\beta)^\tee \left(M^{-1}+(X^\tee X)^{-1}\right)^{-1}(\tilde\beta-\hat\beta)
\end{align*}
by virtue of the second identity. Therefore,
\begin{align*}
\pi(\sigma^2|\hat\beta,s^2,X) &\propto \sigma^{-n-2a-2}\,\exp\frac{-1}{2\sigma^2}\left\{
(\tilde\beta-\hat\beta)^\tee \left(M^{-1}\right.\right.\\
        &\quad\left.\left.+(X^\tee X)^{-1}\right)^{-1}(\tilde\beta-\hat\beta) +s^2 +2b \right\}
\end{align*}
which is the distribution (3.4).

\begin{exoset}\label{exo:Hpd0}
Give a $(1-\alpha)$ HPD region on $\beta$ based on (3.6).
\end{exoset}

As indicated just before this exercise, 
$$
\beta|\by,X \sim \mathscr{T}_{k+1}\left(n+2a,\hat\mu,\hat\Sigma\right)\,.
$$
This means that
$$
\pi(\beta|\by,X) \propto \frac{1}{2}\,\left\{ 
1+\frac{(\beta-\hat\mu)^\tee\hat\Sigma^{-1}(\beta-\hat\mu) }{n+2a} \right\}^{(n+2a+k+1)}
$$
and therefore that an HPD region is of the form
$$
\mathfrak{H}_\alpha=\left\{ \beta;\,,(\beta-\hat\mu)^\tee\hat\Sigma^{-1}(\beta-\hat\mu) \le k_\alpha\right\}\,,
$$
where $k_\alpha$ is determined by the coverage probability $\alpha$.

Now, $(\beta-\hat\mu)^\tee\hat\Sigma^{-1}(\beta-\hat\mu)$ has the same distribution as $||z||^2$ when
$z\sim\mathscr{T}_{k+1}(n+2a,0,I_{k+1})$. This distribution is Fisher's $\mathcal{F}(k+1,n+2a)$
distribution, which means that the bound $k_\alpha$ is determined by the quantiles of this distribution.

\begin{exoset}\label{exo:linpred}
The regression model can also be used in a predictive sense: for a
given $(m,k+1)$ explanatory matrix $\tilde X$, the corresponding outcome
$\tilde \by$ can be inferred through the {\em predictive
distribution}\index{Distribution!predictive} $\pi(\tilde
\by|\sigma^2,\by,X,\tilde X)$. Show that $\pi(\tilde
\by|\sigma^2,\by,X,\tilde X)$ is a Gaussian density with mean
\begin{eqnarray*}
\mathbb{E}[\tilde \by|\sigma^2,\by,X,\tilde X]
& = & \mathbb{E}[\mathbb{E}(\tilde \by|\beta,\sigma^2,\by,X,\tilde X)|\sigma^2,\by,X,\tilde X] \\
                                  & = & \mathbb{E}[\tilde X\beta|\sigma^2,\by,X,\tilde X] \\
                                  & = & \tilde X(M+X^\tee X)^{-1}(X^\tee X\hat\beta+M\tilde\beta)
\end{eqnarray*}
and covariance matrix
\begin{eqnarray*}
\mathbb{V}(\tilde \by|\sigma^2,\by,\tilde X)
  x&=& \mathbb{E}[\mathbb{V}(\tilde \by|\beta,\sigma^2,\by,X,\tilde X)|\sigma^2,\by,X,\tilde X] \\
  &&  \quad+\mathbb{V}(\mathbb{E}[\tilde \by|\beta,\sigma^2,\by,X,\tilde X]|\sigma^2,\by,X,\tilde X) \\
  &=& \mathbb{E}[\sigma^2I_m|\sigma^2,\by,X,\tilde X]+\mathbb{V}(\tilde X\beta|\sigma^2,\by,X,\tilde X) \\
  &=& \sigma^2(I_m+\tilde X(M+X^\tee X)^{-1}\tilde X^\tee ) \,.
\end{eqnarray*}
Deduce that
\begin{eqnarray*}
\tilde \by|\by,X,\tilde X & \sim & \mathscr{T}_m\left(n+2a,\tilde
X(M+X^\tee X)^{-1}(X^\tee
    X\hat\beta+M\tilde\beta),\right. \\
&&\quad\frac{2b+s^2+(\tilde\beta-\hat\beta)^\tee \left(M^{-1}+(X^\tee X)^{-1}\right)^{-1}
    (\tilde\beta-\hat\beta)}{n+2a} \\
&&\quad\times\left.\left\{ I_m+\tilde X(M+X^\tee X)^{-1}\tilde X^\tee \right\}\right).
\end{eqnarray*}
\end{exoset}

Since 
$$
\tilde \by|\tilde X,\beta,\sigma \sim \mathscr{N}\left(\tilde X\beta,\sigma^2 I_m \right)
$$
and since the posterior distribution of $\beta$ conditional on $\sigma$ is given by (3.3),
we have that
$$
\tilde \by|\by,X,\tilde X,\sigma \sim \mathscr{N}\left(
\tilde X\mathbb{E}[\beta|\sigma^2,\by,X],\sigma^2 I_m+\tilde X\text{var}(\beta|\by,X,\sigma) \tilde X^\tee \right)\,,
$$
with mean $\tilde X(M+X^\tee X)^{-1}(X^\tee X\hat\beta+M\tilde\beta)$,
as shown by the derivation of Exercise \ref{exo:matilde}.
The variance is equal to $\sigma^2\left(I_m+\tilde X(M+X^\tee X)^{-1} \tilde X^\tee \right)$.

Integrating $\sigma^2$ against the posterior distribution (3.4) means that
\begin{align*}
\pi(\tilde \by&|\by,X,\tilde X) \propto 
\int_0^\infty \sigma^{-m-n-2a-1}\,\exp\frac{-1}{2\sigma^2}\left\{
2b+s^2+ \right.\\
&\quad\left.(\tilde\beta-\hat\beta)^\tee \left(M^{-1}+(X^\tee X)^{-1}\right)^{-1} (\tilde\beta-\hat\beta)
+(\tilde \by-\mathbb{E}[\tilde \by|\sigma^2,\by,X,\tilde X])^\tee (I_m \right.\\ 
&\quad\left. +\tilde X(M+X^\tee X)^{-1}\tilde X^\tee )^{-1} (\tilde\beta-\hat\beta)
(\tilde \by-\mathbb{E}[\tilde \by|\sigma^2,\by,X,\tilde X])^\tee \right\}\,\text{d}\sigma^2\\
&\propto \left\{2b+s^2+(\tilde\beta-\hat\beta)^\tee \left(M^{-1}+(X^\tee X)^{-1}\right)^{-1} (\tilde\beta-\hat\beta)
+(\tilde \by \right.\\
&\quad\left.-\mathbb{E}[\tilde \by|\sigma^2,\by,X,\tilde X])^\tee (I_m 
+\tilde X(M+X^\tee X)^{-1}\tilde X^\tee )^{-1} (\tilde\beta-\hat\beta)
(\tilde \by\right.\\
&\quad\left.-\mathbb{E}[\tilde \by|\sigma^2,\by,X,\tilde X])^\tee \right\}^{-(m+n+2a)/2}
\end{align*}
which corresponds to a Student's $\mathcal{T}$ distribution with $(n+2a)$ degrees of freedom, a location
parameter equal to $\mathbb{E}[\tilde \by|\sigma^2,\by,X,\tilde X]$ [that does not depend on $\sigma$] and a scale 
parameter equal to
\begin{align*}
&\left\{2b+s^2+(\tilde\beta-\hat\beta)^\tee \left(M^{-1}+(X^\tee X)^{-1}\right)^{-1} (\tilde\beta-\hat\beta)\right\}\\
&\quad \times \left[ I_m+\tilde X(M+X^\tee X)^{-1}\tilde X^\tee \right] / (n+2a) \,.
\end{align*}

\begin{exoset}\label{exo:straightA}
Show that the marginal distribution of $\by$ associated with
(3.3) and (3.4) is given by
$$
\by|X\sim\mathscr{T}_n\left(2a,X\tilde\beta,\frac{b}{a}(I_n+XM^{-1}X^\tee)
\right).
$$
\end{exoset}

This is a direct consequence of Exercise \ref{exo:linpred} when replacing $(\tilde \by,\tilde X)$ with $(\by, X)$
and $(\by, X)$ with the empty set. This is indeed equivalent to take $m=n$, $n=0$, $X=0$, $s^2=0$ and
$$
(\tilde\beta-\hat\beta)^\tee \left(M^{-1}+(X^\tee X)^{-1}\right)^{-1} (\tilde\beta-\hat\beta)=0
$$
in the previous exercice.

\begin{exoset}\label{exo:nullin}
Given the null hypothesis $H_0:R\beta=0$, where $R$ is a $(q,p)$
matrix of rank $q$, show that the restricted model on $\by$ given
$X$ can be represented as
$$
\by|\beta_0,\sigma_0^2,X_0\stackrel{H_0}{\sim}\mathscr{N}_n\left(X_0\beta_0,\sigma_0^2I_n\right)
$$
where $X_0$ is a $(n,k-q)$ matrix and $\beta_0$ is a $(k-q)$ dimensional vector. ({\em Hint:}
Give the form of $X_0$ and $\beta_0$ in terms of $X$ and $\beta$.) Under the hypothesis specific prior
$\beta_0|H_0,\sigma_0^2\sim\mathscr{N}_{k-q}\left(\tilde\beta_0,\sigma^2(M_0)^{-1}\right)$
and $\sigma_0^2|H_0\sim \mathscr{IG}(a_0,b_0)$, construct the Bayes factor associated with the test of $H_0$.
\end{exoset}

When $R\beta=0$, $\beta$ satisfies $q$ independent linear constraints, which means that $q$ coordinates of
$\beta$ can be represented as linear combinations of the $(k-q)$ others, denoted by $\beta_0$, e.g.
$$
\beta_{i_1} = \mathfrak{s}_{i_1}^\tee \beta_0,\ldots,\beta_{i_q} = \mathfrak{s}_{i_q}^\tee \beta_0\,,
$$
where $i_1<\cdots<i-q$ are the above coordinates. This implies that
$$
X\beta = X \left( \begin{matrix} \mathfrak{s}_1^\tee\cr \cdots \cr \mathfrak{s}_q^\tee\cr \end{matrix} \right)\beta_0
       = X_0 \beta_0\,,
$$
where $\mathfrak{s}_i$ is either one of the above linear coefficients or contains $0$ except for a single $1$.
Therefore, when $H_0$ holds, the expectation of $\by$ conditional on $X$ can be written as $X_0\beta_0$ 
where $X_0$ is a $(n,k-q)$ matrix and $\beta_0$ is a $(k-q)$ dimensional vector. (Actually, there is an infinite
number of ways to write $\mathbb{E}[\by|X]$ in this format.) The change from $\sigma$ to $\sigma_0$ is purely notational
to indicate that the variance $\sigma_0^2$ is associated with another model.

If we set a conjugate prior on $(\beta_0,\sigma_0)$, the result of Exercise \ref{exo:straightA} also applies for
this (sub-)model, in the sense that the marginal distribution of $\by$ for this model is
$$
\by|X_0\sim\mathscr{T}_n\left(2a_0,X_0\tilde\beta_0,\frac{b_0}{a_0}(I_n+X_0M_0^{-1}X_0^\tee) \right).
$$
Therefore, the Bayes factor for testing $H_0$ can be written in closed form as
\begin{align*}
B_{01} &= \frac{
\Gamma((2a_0+n)/2)/\Gamma(2a_0/2) \big/ \sigma \sqrt{2a_0}(b_0/a_0)^{n/2}|I_n+X_0M_0^{-1}X_0^\tee|^{1/2}
}{
\Gamma((2a+n)/2)/\Gamma(2a/2) \big/ \sigma \sqrt{2a}(b/a)^{n/2}|I_n+XM^{-1}X^\tee|^{1/2} }\\
&\quad\times \frac{
\left\{ 1+(\by-X_0\tilde\beta_0)^\tee (I_n+X_0M_0^{-1}X_0^\tee)^{-1} 
(\by-X_0\tilde\beta_0)/2b_0 \right\}^{-(2a_0+n)/2}
}{
\left\{ 1+(\by-X\tilde\beta)^\tee (I_n+XM^{-1}X^\tee)^{-1}
(\by-X\tilde\beta)/2b \right\}^{-(2a+n)/2}
}
\end{align*}
% \frac{\Gamma((\nu+1)/2)/\Gamma(\nu/2)}{\sigma \sqrt{\nu\pi} }
Note that, in this case, the normalising constants matter because they differ under $H_0$ and under
the alternative.

\begin{exoset}\label{exo:betastud}
Show that
\begin{eqnarray*}
\beta|\by,X &\sim& \mathscr{T}_{k+1}\left(n,\frac{c}{c+1}\left(\frac{\tilde\beta}{c}+\hat\beta\right),\right. \\
    && \left.\frac{c(s^2+(\tilde\beta-\hat\beta)^\tee X^\tee
    X(\tilde\beta-\hat\beta))}{n(c+1)}(X^\tee X)^{-1}\right).
\end{eqnarray*}
\end{exoset}

Since
\begin{eqnarray*}
\beta|\sigma^2,\by,X&\sim&\mathscr{N}_{k+1}\left(\frac{c}{c+1}(\tilde\beta/c+\hat\beta),
\frac{\sigma^2c}{c+1}(X^\tee X)^{-1}\right),\\
\sigma^2|\by,X&\sim&\mathcal{IG}\left(\frac{n}{2},\frac{s^2}{2}+\frac{1}{2(c+1)}
(\tilde\beta-\hat\beta)^\tee X^\tee
X(\tilde\beta-\hat\beta)\right)\,,
\end{eqnarray*}
we have that
$$
\sqrt{\frac{c+1}{c}}\,\left[X^\tee X\right]^{1/2}\,\left\{ \beta - \frac{c}{c+1}(\tilde\beta/c+\hat\beta) \right\}
\sim \mathscr{N}_{k+1}\left(0,\sigma^2 I_n \right)\,,
$$
with 
$$
\left[s^2+\frac{1}{(c+1)} (\tilde\beta-\hat\beta)^\tee X^\tee
X(\tilde\beta-\hat\beta) \right] / \sigma^2 \sim \chi^2_n\,,
$$
which is the definition of the Student's 
$$
\mathscr{T}_{k+1}\left(n,\frac{c}{c+1}\left(\frac{\tilde\beta}{c}+\hat\beta\right),
\frac{c(s^2+(\tilde\beta-\hat\beta)^\tee X^\tee
    X(\tilde\beta-\hat\beta)/(c+1))}{n(c+1)}(X^\tee X)^{-1}\right)
$$
distribution.

\begin{exoset}
Show that $\pi(\tilde \by|\sigma^2,\by,X,\tilde X)$ is a Gaussian density.
\end{exoset}

Conditional on $\sigma^2$, there is no difference with the setting of Exercise \ref{exo:linpred}
since the only difference in using Zellner's $G$-prior compared with the conjugate priors is in
the use of the noninformative prior $\pi(\sigma^2|X)\propto\sigma^{-2}$. Therefore, this is a
consequence of Exercise \ref{exo:linpred}.

\begin{exoset}\label{exo:Ob10}
The posterior predictive distribution is obtained by
integration over the marginal posterior distribution of $\sigma^2$.
Derive $\pi(\tilde \by|\by,X,\tilde X)$.
\end{exoset}

Once more, integrating the normal distribution over the inverse gamma random variable $\sigma^2$
produces a Student's $\mathscr{T}$ distribution. Since 
$$
\sigma^2|\by,X \sim \mathcal{IG}\left(\frac{n}{2},\frac{s^2}{2}+\frac{1}{2(c+1)}
(\tilde\beta-\hat\beta)^\tee X^\tee X(\tilde\beta-\hat\beta)\right)
$$
under Zellner's $G$-prior, the predictive distribution is a 
\begin{eqnarray*}
\tilde\by|\by,X,\tilde X &\sim& \mathscr{T}_{k+1}\left(n, \tilde X\frac{\tilde\beta+c\hat\beta}{c+1},
   \frac{c(s^2+(\tilde\beta-\hat\beta)^\tee X^\tee X(\tilde\beta-\hat\beta)/(c+1))}{n(c+1)}\right. \\
    && \left.  \times\left\{I_m+\frac{c}{c+1}\tilde X(X^\tee X)^{-1}\tilde X^\tee \right\}\right)
\end{eqnarray*}
distribution.

\begin{exoset}
Give a joint $(1-\alpha)$ HPD region on $\beta$.
\end{exoset}

Since we have (Exercise \ref{exo:betastud})
\begin{eqnarray*}
\beta|\by,X &\sim& \mathscr{T}_{k+1}\left(n,\frac{c}{c+1}\left(\frac{\tilde\beta}{c}+\hat\beta\right),\right. \\
    && \left.\frac{c(s^2+(\tilde\beta-\hat\beta)^\tee X^\tee
    X(\tilde\beta-\hat\beta))}{n(c+1)}(X^\tee X)^{-1}\right)\,,
\end{eqnarray*}
with 
$$
\hat\Sigma = \frac{c(s^2+(\tilde\beta-\hat\beta)^\tee X^\tee
    X(\tilde\beta-\hat\beta))}{n(c+1)}(X^\tee X)^{-1} \,,
$$
an HPD region is of the form
$$
\mathfrak{H}_\alpha=\left\{ \beta;\,,(\beta-\hat\mu)^\tee\hat\Sigma^{-1}(\beta-\hat\mu) \le k_\alpha\right\}\,,
$$
where $k_\alpha$ is determined by the coverage probability $\alpha$ and
$$
\hat\mu = \frac{c}{c+1}\left(\frac{\tilde\beta}{c}+\hat\beta\right)\,,\quad
\hat\Sigma = \frac{c(s^2+(\tilde\beta-\hat\beta)^\tee X^\tee
    X(\tilde\beta-\hat\beta))}{n(c+1)}(X^\tee X)^{-1}\,.
$$
As in Exercise \ref{exo:Hpd0}, the distribution of $(\beta-\hat\mu)^\tee\hat\Sigma^{-1}(\beta-\hat\mu)$
is a Fisher's $\mathcal{F}(k+1,n)$ distribution.

\begin{exoset}
Show that the matrix $(I_n+cX(X^\tee X)^{-1}X^\tee )$ has $1$ and $c+1$ as eigenvalues.
({\em Hint:} Show that the eigenvectors associated with $c+1$ are of the form $X\beta$
and that the eigenvectors associated with $1$ are those orthogonal to $X$, i.e. $z$'s
such that $X^\tee z =0$.) Deduce that the determinant of the matrix $(I_n+cX(X^\tee X)^{-1}X^\tee )$
is indeed $(c+1)^{(k+1)/2}$.
\end{exoset}

Given the hint, this is somehow obvious: 
\begin{eqnarray*}
(I_n+cX(X^\tee X)^{-1}X^\tee )X\beta &=& X\beta + cX(X^\tee X)^{-1}X^\tee X\beta=(c+1)X\beta\\
(I_n+cX(X^\tee X)^{-1}X^\tee )z&=& z + cX(X^\tee X)^{-1}X^\tee z = z
\end{eqnarray*}
for all $\beta$'s in $\mathbb{R}^{k+1}$ and all $z$'s orthogonal to $X$. Since the addition of those
two subspaces generates a vector space of dimension $n$, this defines the whole set of eigenvectors
for both eigenvalues. And since the vector subspace generated by $X$ is of dimension $(k+1)$, this
means that the determinant of $(I_n+cX(X^\tee X)^{-1}X^\tee )$ is $(c+1)^{k+1}\times 1^{n-k-1}$.

\begin{exoset}\label{exo:jefflin}
Derive the marginal posterior distribution of $\beta$ for this model.
\end{exoset}

The joint posterior is given by
\begin{eqnarray*}
\beta|\sigma^2,\by,X&\sim&\mathscr{N}_{k+1}\left(\hat\beta,\sigma^2(X^\tee X)^{-1}\right), \\
\sigma^2|\by,X &\sim&\mathscr{IG}((n-k-1)/2,s^2/2).  
\end{eqnarray*}
Therefore, 
$$
\beta|\by,X\sim\mathscr{T}_{k+1}\left(n-k-1,\hat\beta,\frac{s^2}{n-k-1}(X^\tee X)^{-1}\right)
$$
by the same argument as in the previous exercises.

\begin{exoset}\label{exo:teemar}
Show that the marginal posterior distribution of $\beta_i$ $(1\le i\le k)$ is a
$\mathscr{T}_1(n-k-1,\hat\beta_i, \omega_{(i,i)}s^2/(n-k-1))$ distribution.
({\em Hint:} Recall that $\omega_{(i,i)}=(X^\tee X)_{(i,i)}^{-1}$.)
\end{exoset}

The argument is straightforward: since $\beta|\sigma^2,\by,X\sim\mathscr{N}_{k+1}\left(\hat\beta,
\sigma^2(X^\tee X)^{-1}\right)$, $\beta_i|\sigma^2,\by,X\sim\mathscr{N}\left(\hat\beta_i,
\sigma^2\omega_{(i,i)}\right)$. Integrating out $\sigma^2$ as in the previous exercise leads
to 
$$
\beta_i|\sigma^2,\by,X\sim\mathscr{T}_1(n-k-1,\hat\beta_i, \omega_{(i,i)}s^2/(n-k-1))\,.
$$

\begin{exoset}
Give the predictive distribution of $\tilde \by$, the $m$ dimensional
vector corresponding to the $(m,k)$ matrix of explanatory variables $\tilde X$.
\end{exoset}

This predictive can be derived from Exercise \ref{exo:linpred}. Indeed, Jeffreys' prior is
nothing but a special case of conjugate prior with $a=b=0$. Therefore, Exercise \ref{exo:linpred}
implies that, in this limiting case, 
\begin{eqnarray*}
\tilde \by|\by,X,\tilde X & \sim & \mathscr{T}_m\left(n,\tilde
X(M+X^\tee X)^{-1}(X^\tee
    X\hat\beta+M\tilde\beta),\right. \\
&&\quad\frac{s^2+(\tilde\beta-\hat\beta)^\tee \left(M^{-1}+(X^\tee X)^{-1}\right)^{-1}
    (\tilde\beta-\hat\beta)}{n} \\
&&\quad\times\left.\left\{ I_m+\tilde X(M+X^\tee X)^{-1}\tilde X^\tee \right\}\right).
\end{eqnarray*}

\begin{exoset}
When using the prior distribution $\pi(c)=1/c^2$, compare the results with Table 3.6.
\end{exoset}

In the file {\sf {\#}3.txt} provided on the Webpage, it suffices to replace \verb+ cc^(-1)+ with \verb+ cc^(-2)+ :
for instance, the point estimate of $\beta$ is now
\begin{verbatim}
> facto=sum(cc/(cc+1)*cc^(-2)*(cc+1)^(-11/2)*
+ (t(y)%*%y-cc/(cc+1)*t(y)%*%P%*%y)^(-33/2))/
+ sum(cc^(-2)*(cc+1)^(-11/2)*(t(y)%*%y-cc/
+ (cc+1)*t(y)%*%P%*%y)^(-33/2))
> facto*betahat 
                [,1]
   [1,]  8.506662193
   [2,] -0.003426982
   [3,] -0.041634562
   [4,]  0.052547326
   [5,] -1.000556061
   [6,]  0.179158187
   [7,] -0.275964816
   [8,] -0.183669178
   [9,]  0.140040003
  [10,] -0.994120776
  [11,] -0.334983109

\end{verbatim}

\begin{exoset}
Show that both series (3.10) and (3.11) converge.
\end{exoset}

Given that
\begin{eqnarray*}
f(\by|X,c) &\propto& (c+1)^{-(k+1)/2} \left[\by^\tee
\by-\frac{c}{c+1}\by^\tee X(X^\tee X)^{-1}X^\tee \by\right]^{-n/2}\\
&\approx&
c^{-(k+1)/2} \left[\by^\tee \by-\by^\tee X(X^\tee X)^{-1}X^\tee \by\right]^{-n/2}
\end{eqnarray*}
when $c$ goes to $\infty$, the main term in the series goes to $0$ as
a $\text{o}(c^{-(k+3)/2})$ and the series converges.

Obviously, if the first series converges, then so does the second series.

\begin{exoset}
Give the predictive distribution of $\tilde \by$, the $m$-dimensional
vector corresponding to the $(m,k)$ matrix of explanatory variables $\tilde X$.
\end{exoset}

The predictive of $\tilde \by$ given $\by,X,\tilde X$ is then the
weighted average of the predictives given $\by,X,\tilde X$ and $c$:
$$
\pi(\tilde \by|\by,X,\tilde X) \propto
\sum_{c=1}^\infty \pi(\tilde\by|\by,X,\tilde X,c)f(\by|X,c)\,c^{-1}
$$
where $\pi(\tilde\by|\by,X,\tilde X,c)$ is the Student's $\mathscr{T}$ 
distribution obtained in Exercise \ref{exo:Ob10}.

\begin{exoset}\label{exo:notire}
If $(x_1,x_2)$ is distributed from the uniform distribution on
$$
\left\{(x_1,x_2);\,(x_1-1)^2+(x_2-1)^2\le 1\right\}\cup
\left\{(x_1,x_2);\,(x_1+1)^2+(x_2+1)^2\le 1\right\}\,,
$$
show that the Gibbs sampler does not produce an irreducible chain.
For this distribution, find an alternative Gibbs sampler that works.
({\em Hint:} Consider a rotation of the coordinate axes.)\end{exoset}

\begin{figure}[bt]
\begin{center}
\includegraphics[width=5cm,height=6cm]{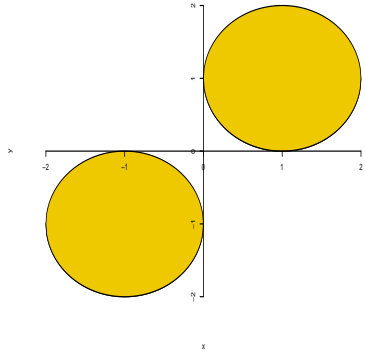}
\caption{\label{fig:nonconnect}Support of the uniform distribution.}
\end{center}
\end{figure}

The support of this uniform distribution is made of two disks with
respective centers $(-1,-1)$ and $(1,1)$, and with radius $1$. This
support is not connected (see Figure \ref{fig:nonconnect}) and 
conditioning on $x_1<0$ means that the conditional distribution of $x_2$ is 
$\mathscr{U}(-1-\sqrt{1-x_1^2},-1+\sqrt{1-x_1^2}$,
thus cannot produce a value in $[0,1]$. Similarly, when simulating the
next value of $x_1$, it necessarily remains negative. The Gibbs sampler
thus produces two types of chains, depending on whether or not it is started
from the negative disk.
If we now consider the Gibbs sampler for the new parameterisation
$$
y_1 = x_1+x_2,\quad y_2=x_2-x_1\,,
$$
conditioning on $y_1$ produces a uniform distribution on the union
of a negative and of a positive interval. Therefore, one iteration
of the Gibbs sampler is sufficient to jump [with positive probability]
from one disk to the other one.

\begin{exoset}\label{exo:babyHaC}
If a joint density $g(y_1,y_2)$ corresponds to the conditional
distributions $g_1(y_1|y_2)$ and $g_2(y_2|y_1)$, show that it is
given by
$$
g(y_1,y_2) = {g_2(y_2|y_1) \over \int \;
   g_2(v|y_1)/g_1(y_1|v) \;\hbox{d}v}.
$$
\end{exoset}

If the joint density $g(y_1,y_2)$ exists, then
\begin{align*}
g(y_1,y_2) &= g^1(y_1) g_2(y_2|y_1) \\
           &= g^2(y_2) g_1(y_1|y_2)
\end{align*}
where $g^1$ and $g^2$ denote the densities of the marginal distributions of $y_1$ and $y_2$, respectively.
Thus,
\begin{align*}
g^1(y_1) &= \frac{g_1(y_1|y_2)}{g_2(y_2|y_1)} g^2(y_2)\\
         &\propto \frac{g_1(y_1|y_2)}{g_2(y_2|y_1)}\,,
\end{align*}
as a function of $y_1$ [$g^2(y_2)$ is irrelevant]. Since $g^1$ is a density,
$$
g^1(y_1) = \frac{g_1(y_1|y_2)}{g_2(y_2|y_1)} \bigg/ \int \frac{g_1(u|y_2)}{g_2(y_2|u)} \text{d}u
$$
and
$$
g(y_1,y_2) = g_1(y_1|y_2) \bigg/ \int \frac{g_1(u|y_2)}{g_2(y_2|u)} \text{d}u\,.
$$
Since $y_1$ and $y_2$ play symmetric roles in this derivation, the symmetric version
also holds.

\begin{exoset}
Check that the starting value of $\mu$ in the setting of Example 3.2
has no influence on the output of the above Gibbs sampler after $N=1000$ iterations.
\end{exoset}

The core of the Gibbs program is
\begin{verbatim}
   > mu = rnorm(1,sum(x*omega)/sum(omega+.05),
   + sqrt(1/(.05+2*sum(omega)))
   > omega = rexp(2,1+(x-mu)^2)
\end{verbatim}
which needs to be iterated $N=1000$ times to produce a Gibbs $N$-sample from $\pi(\mu|\mathcal{D})$.
A {\sf R} program evaluating the [lack of] influence of the starting value $\mu^{(0)}$ will thus need 
to compare histograms of $\mu^{(1000)}$'s for different starting values. It therefore requires three
loops:
\begin{verbatim}
x=c(3.2,-1.5)	# observations
mu0=seq(-10,10,length=20) # starting values
muk=rep(0,250)

par(mfrow=c(5,4),mar=c(4,2,4,1)) # multiple histograms

for (i in 1:20){

  for (t in 1:250){

    mu=mu0[i]
    for (iter in 1:1000){

      omega = rexp(2,1+(x-mu)^2)
      mu = rnorm(1,sum(x*omega)/sum(omega+.05),
           sqrt(1/(.05+2*sum(omega))))
    }
    muk[t]=mu
  }
  hist(muk,proba=T,col="wheat",main=paste(mu0[i]))

}
\end{verbatim}
Be warned that the use of this triple loop induces a long wait on most machines!

\begin{exoset}
In the setup of Section 3.5.3, show that
\begin{eqnarray*}
\pi(\gamma|\by,X)&\propto&\sum_{c=1}^\infty c^{-1}(c+1)^{-(q_\gamma+1)/2}\left[\by^\tee \by-\right.\nonumber\\
&&\quad \left.\frac{c}{c+1}\by^\tee X_{\gamma}\left(X_{\gamma}^\tee
X_{\gamma}\right)^{-1}X_{\gamma}^\tee \by\right]^{-n/2}
\end{eqnarray*}
and that the series converges. If $\pi(c)\propto c^{-\alpha}$, find which values of $\alpha$ lead to a proper posterior.
\end{exoset}

We can take advantage of Section 3.5.2: when $c$ is fixed in Zellner's informative $G$-prior and
$\tilde\beta_{\gamma}=\mathbf{0}_{q_\gamma+1}$ for all $\gamma$'s, 
$$
\pi(\gamma|\by,X,c) \propto (c+1)^{-(q_\gamma+1)/2}\left[\by^\tee \by-\frac{c}{c+1}\by^\tee 
X_{\gamma}\left(X_{\gamma}^\tee X_{\gamma}\right)^{-1}X_{\gamma}^\tee \by\ \right]^{-n/2}\,,
$$
thus
$$
\pi(\gamma,c|\by,X,c) \propto c^{-1}(c+1)^{-(q_\gamma+1)/2}\left[\by^\tee \by-\frac{c}{c+1}\by^\tee
X_{\gamma}\left(X_{\gamma}^\tee X_{\gamma}\right)^{-1}X_{\gamma}^\tee \by\ \right]^{-n/2}\,.
$$
and
\begin{eqnarray*}
\pi(\gamma|\by,X)&=&\sum_{c=1}^\infty \pi(\gamma,c|\by,X,c)\\
&\propto& \sum_{c=1}^\infty c^{-1}(c+1)^{-(q_\gamma+1)/2}\left[\by^\tee \by-\frac{c}{c+1}\by^\tee
X_{\gamma}\left(X_{\gamma}^\tee X_{\gamma}\right)^{-1}X_{\gamma}^\tee \by\ \right]^{-n/2}\,.
\end{eqnarray*}
For $\pi(c)\propto c^{-\alpha}$, the series
$$
\sum_{c=1}^\infty c^{-\alpha}(c+1)^{-(q_\gamma+1)/2}\left[\by^\tee \by-\frac{c}{c+1}\by^\tee
X_{\gamma}\left(X_{\gamma}^\tee X_{\gamma}\right)^{-1}X_{\gamma}^\tee \by\ \right]^{-n/2}
$$
converges if and only if, for all $\gamma$'s, 
$$
\alpha + \frac{q_\gamma+1}{2} > 1\,,
$$
which is equivalent to $2\alpha+q_\gamma>1$. Since $\min(q_\gamma)=0$, the constraint for
propriety of the posterior is 
$$
\alpha > 1/2\,.
$$

\chapter{Generalized Linear Models}\label{ch:glm}

\begin{exoset}
For {\sf bank}, derive the maximum likelihood estimates of $\beta_0$ and $\beta_1$
found in the previous analysis. Using Jeffreys prior on the parameters
$(\beta_0,\beta_1,\sigma^2)$ of the linear regression model, compute the
corresponding posterior expectation of $(\beta_0,\beta_1)$.
\end{exoset}

The code is provided in the file \verb+ #4.txt+ on the Webpage. If the \verb+bank+ dataset is not
available, it can be downloaded from the Webpage and the following code can be used:
\begin{verbatim}
  bank=matrix(scan("bank"),byrow=T,ncol=5)
  y=as.vector(bank[,5])
  X=cbind(rep(1,200),as.vector(bank[,1]),as.vector(bank[,2]),
       as.vector(bank[,3]),as.vector(bank[,4]))
  summary(lm(y~X[,5]))
\end{verbatim}
which produces the output [leading to eqn.~(4.1) in the book]:
\begin{verbatim}
> summary(lm(y~X[,5]))

Call:
lm(formula = y ~ X[, 5])

Residuals:
     Min       1Q   Median       3Q      Max
-0.76320 -0.21860 -0.06228  0.18322  1.04046

Coefficients:
            Estimate Std. Error t value Pr(>|t|)
(Intercept) -2.02282    0.14932  -13.55   <2e-16 ***
X[, 5]       0.26789    0.01567   17.09   <2e-16 ***
---
Sig. codes: 0 `***' .001 `**' .01 `*' .05 `.' .1 ` ' 1

Residual standard error: 0.3194 on 198 degrees of freedom
Multiple R-Squared: 0.596,      Adjusted R-squared: 0.594
F-statistic: 292.2 on 1 and 198 DF,  p-value: < 2.2e-16

\end{verbatim}

As shown in Exercise \ref{exo:jefflin}, the [marginal] posterior in $\beta$ associated 
with the Jeffreys prior is
$$
\beta|\by,X\sim\mathscr{T}_{k+1}\left(n-k-1,\hat\beta,\frac{s^2}{n-k-1}(X^\tee X)^{-1}\right)
$$
so the posterior expectation of $(\beta_0,\beta_1)$ is again $\hat\beta$.

\begin{exoset} Show that, in the setting of Example 4.1, the statistic $\sum_{i=1}^n y_i\,\bx^{i}$ is
sufficient when conditioning on the $\bx^i$'s $(1\le i\le n)$ and
give the corresponding family of conjugate priors.
\end{exoset}

Since the likelihood is
\begin{align*}
\exp &\left\{\sum_{i=1}^n y_i\,\bx^{i\tee}\beta \right\}
\bigg/ \prod_{i=1}^n \left[ 1+\exp(\bx^{i\tee}\beta )\right]\\
&= \exp \left\{\sum_{i=1}^n \left[y_i\,\bx^i\right]^\tee \beta \right\}
\bigg/ \prod_{i=1}^n \left[ 1+\exp(\bx^{i\tee}\beta )\right]\,,
\end{align*}
it depends on the observations $(y_1,\ldots,y_n)$ only through the sum $\sum_{i=1}^n y_i\,\bx^{i}$
which is thus a sufficient statistic in this conditional sense.

The family of priors
$(\xi\in\mathbb{R}^k\,,\lambda>0)$
$$
\pi(\beta|\xi,\lambda) \propto \exp \left\{\xi^\tee \beta \right\}
\bigg/ \prod_{i=1}^n \left[ 1+\exp(\bx^{i\tee}\beta )\right]^\lambda\,,
$$
is obviously conjugate. The corresponding posterior is 
$$
\pi\left( \beta\left|\xi+\sum_{i=1}^n y_i\,\bx^{i},\lambda+1\right.\right)\,,
$$
whose drawback is to have $\lambda$ updated in $\lambda+1$ rather than $\lambda+n$ as in other
conjugate settings. This is due to the fact that the prior is itself conditional on $X$ and therefore on $n$.

\begin{exoset}
Show that the logarithmic link is the canonical link function in the case of the Poisson regression model.
\end{exoset}

The likelihood of the Poisson regression model is
\begin{align*}
\ell(\beta|\by,X)&=\prod_{i=1}^n\left(\frac{1}{y_i!}\right)\exp\left\{
y_i\,\bx^{i\tee}\beta-\exp(\bx^{i\tee}\beta)\right\}\\
&= \prod_{i=1}^n\frac{1}{y_i!}\exp\left\{
y_i\,\log(\mu_i)-\mu_i\right\}\,,
\end{align*}
so $\log(\mu_i)=\bx^{i\tee}\beta$ and the logarithmic link is indeed the canonical link function.

\begin{exoset}
Suppose $y_1,\ldots,y_k$ are independent Poisson
$\mathcal{P}(\mu_i)$ random variables. Show that, conditional on
$n=\sum_{i=1}^k y_i$,
$$
\by = (y_1,\ldots,y_k) \sim
\mathcal{M}_k(n;\alpha_1,\ldots,\alpha_k)
$$
and determine the $\alpha_i$'s.
\end{exoset}

The joint distribution of $\by$ is
$$
f(\by|\mu_1,\ldots,\mu_k) = \prod_{i=1}^k\left(\frac{\mu_i^{y_i}}{y_i!}\right)\,\exp\left\{-\sum_{i=1}^k\mu_i\right\}\,,
$$
while $n=\sum_{i=1}^k y_i\sim \mathcal{P}(\sum_{i=1}^k\mu_i)$ [which can be established using the moment generating
function of the $\mathcal{P}(\mu)$ distribution]. Therefore, the conditional distribution of $\by$ given $n$ is
\begin{align*}
f(\by|\mu_1,\ldots,\mu_k,n) &= \frac{
\prod_{i=1}^k\left(\frac{\mu_i^{y_i}}{y_i!}\right)\,\exp\left\{-\sum_{i=1}^k\mu_i\right\}
}{ \frac{[\sum_{i=1}^k\mu_i]^n}{n!} \exp\left\{-\sum_{i=1}^k\mu_i\right\} 
}\,\mathbb{I}_n \left(\sum_{i=1}^k y_i\right)\\
&= \frac{n!}{\prod_{i=1}^k y_i!}\,\prod_{i=1}^k \left( \frac{\mu_i}{\sum_{i=1}^k\mu_i} \right)^{y_i}
\,\mathbb{I}_n \left(\sum_{i=1}^k y_i\right)\,,
\end{align*}
which is the pdf of the $\mathcal{M}_k(n;\alpha_1,\ldots,\alpha_k)$ distribution, with
$$
\alpha_i = \frac{\mu_i}{\sum_{j=1}^k\mu_j}\,,\qquad i=1,\ldots,k\,.
$$

This conditional representation is a standard property used in the statistical analysis 
of contingency tables (Section 4.5): when the margins are random, the cells are Poisson
while, when the margins are fixed, the cells are multinomial.

\begin{exoset}
Show that the detailed balance equation also holds for the {\em Boltzmann} acceptance probability
$$
\rho(x,y) = \frac{\pi(y)q(y,x)}{\pi(y)q(y,x)+\pi(x) q(x,y)}\,.
$$
\end{exoset}

The detailed balance equation is
$$
\pi(x) q(x,y) \rho(x,y) = \pi(y) q(y,x) \rho(y,x)\,.
$$
Therefore, in the case of the Boltzmann acceptance probability
\begin{align*}
\pi(x) q(x,y) \rho(x,y) &=
\pi(x) q(x,y) \frac{\pi(y)q(y,x)}{\pi(y)q(y,x)+\pi(x) q(x,y)}\\
&= \frac{\pi(x) q(x,y) \pi(y)q(y,x)}{\pi(y)q(y,x)+\pi(x) q(x,y)}\\
&= \pi(y)q(y,x) \frac{\pi(x) q(x,y)}{\pi(y)q(y,x)+\pi(x) q(x,y)}\\
&= \pi(y) q(y,x) \rho(y,x)\,.
\end{align*}
Note that this property also holds for the generalized Boltzmann acceptance probability
$$
\rho(x,y) = \frac{\pi(y)q(y,x)\alpha(x,y)}{\pi(y)q(y,x)\alpha(x,y)+\pi(x) q(x,y)\alpha(y,x)}\,,
$$
where $\alpha(x,y)$ is an arbitrary positive function.

\begin{exoset}
For $\pi$ the density of an inverse normal distribution with parameters $\theta_1=3/2$ and $\theta_2=2$,
$$
\pi(x)\propto x^{-3/2}\exp(-3/2x-2/x)\mathbb{I}_{x>0},
$$
write down and implement an independence MH sampler with a Gamma proposal with
parameters $(\alpha,\beta)=(4/3,1)$ and
$(\alpha,\beta)=(0.5\sqrt{4/3},0.5)$.
\end{exoset}

\begin{figure}
\begin{center}
\includegraphics[width=\textwidth,height=6cm]{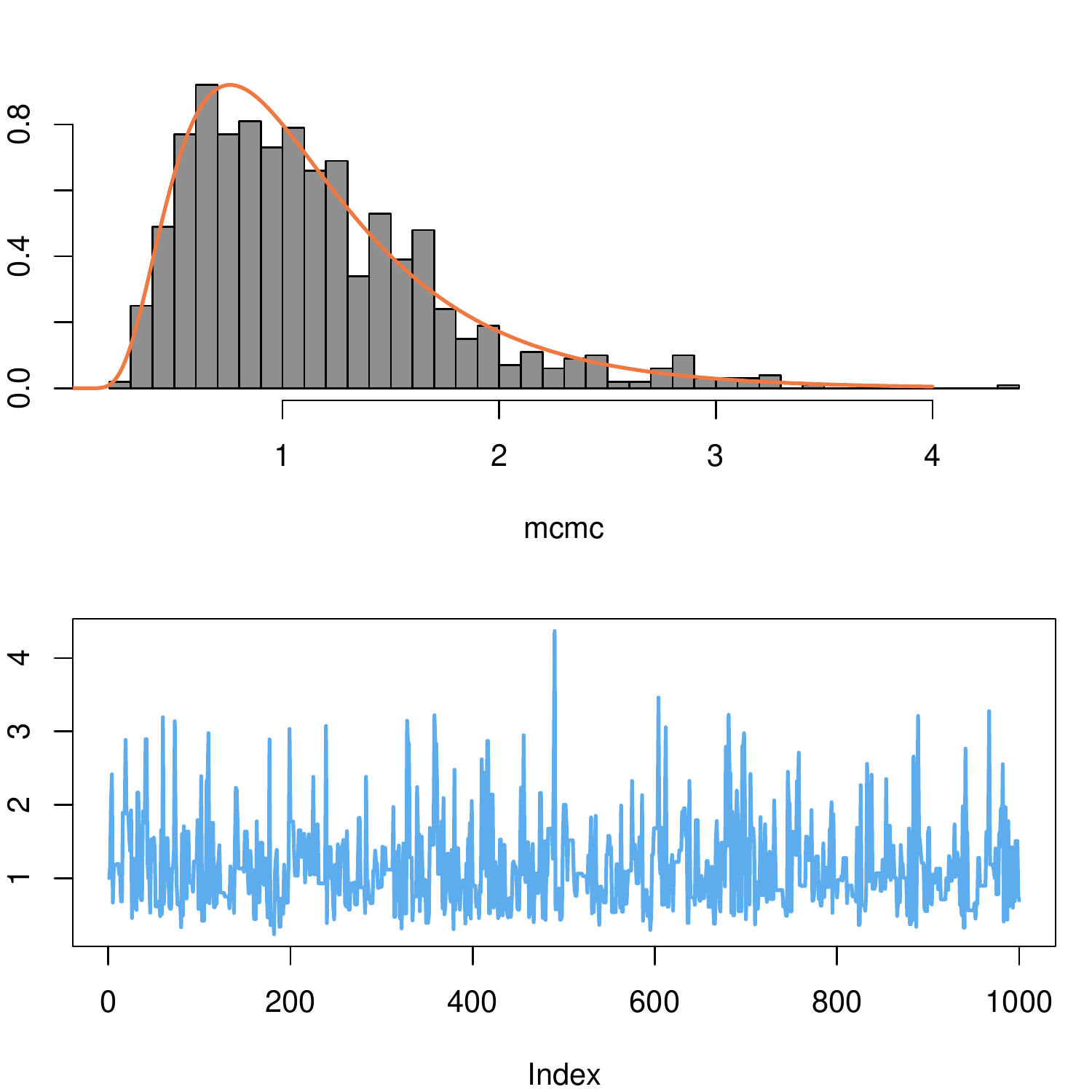}
\caption{\label{fig:exo406}Output of an MCMC simulation of the inverse
normal distribution.}
\end{center}
\end{figure}

A {\sf R} possible code for running an independence Metropolis--Hastings sampler in this setting is
as follows:
\begin{verbatim}
# target density
target=function(x,the1=1.5,the2=2){ 
  x^(-the1)*exp(-the1*x-the2/x)
  }

al=4/3
bet=1

# initial value
mcmc=rep(1,1000)

for (t in 2:1000){

  y = rgamma(1,shape=al,rate=bet)
  if (runif(1)<target(y)*dgamma(mcmc[t-1],shape=al,rate=bet)/
	(target(mcmc[t-1])*dgamma(y,shape=al,rate=bet)))
    mcmc[t]=y
    else
      mcmc[t]=mcmc[t-1]
  }

# plots
par(mfrow=c(2,1),mar=c(4,2,2,1))
res=hist(mcmc,freq=F,nclass=55,prob=T,col="grey56",
  ylab="",main="")
lines(seq(0.01,4,length=500),valpi*max(res$int)/max(valpi),
  lwd=2,col="sienna2")
plot(mcmc,type="l",col="steelblue2",lwd=2)

\end{verbatim}
The output of this code is illustrated on Figure \ref{fig:exo406} and shows a reasonable fit of
the target by the histogram and a proper mixing behaviour. Out of the $1000$ iterations
in this example, $600$ corresponded to an acceptance of the Gamma random variable. (Note
that to plot the density on the same scale as the histogram, we resorted to a trick on the
maxima of the histogram and of the density.)

\begin{exoset}
Estimate the mean of a $\mathscr{G}a(4.3,6.2)$ random variable using
\begin{enumerate}
\item direct sampling from the distribution via {\sf R} command\\
\verb+  > x=rgamma(n,4.3,rate=6.2)+ 
\item Metropolis--Hastings with a $\mathscr{G}a(4,7)$ proposal distribution;
\item Metropolis--Hastings with a $\mathscr{G}a(5,6)$ proposal distribution.
\end{enumerate}
In each case, monitor the convergence of the cumulated average.
\end{exoset}

\begin{figure}
\begin{center}
\includegraphics[width=\textwidth,height=6cm]{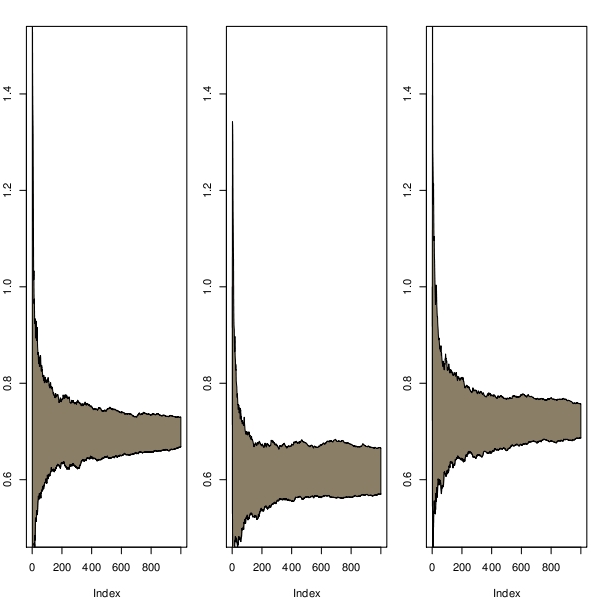}
\caption{\label{fig:ranj407}Range of three samplers for the approximation
of the $\mathscr{G}a(4.3,6.2)$ mean: {\em (left)} iid; {\em (center)} $\mathscr{G}a(4,7)$ proposal;
{\em (right)} $\mathscr{G}a(5,6)$ proposal.}
\end{center}
\end{figure}

Both independence Metropolis--Hastings samplers can be implemented via an {\sf R}
code like
\begin{verbatim}
al=4.3
bet=6.2

mcmc=rep(1,1000)
for (t in 2:1000){

  mcmc[,t]=mcmc[,t-1]
  y = rgamma(500,4,rate=7)
  if (runif(1)< dgamma(y,al,rate=bet)*dgamma(mcmc[t-1],4,rate=7)/
	(dgamma(mcmc[t-1],al,rate=bet)*dgamma(y,4,rate=7))){
    mcmc[t]=y
    }
}
aver=cumsum(mcmc)/1:1000

\end{verbatim}
When comparing those samplers, their variability can only be evaluated through repeated calls
to the above code, in order to produce a range of outputs for the three methods. For instance,
one can define a matrix of cumulated averages \verb+aver=matrix(0,250,1000)+ and take the range
of the cumulated averages over the $250$ repetitions as in \verb+ranj=apply(aver,1,range)+, leading
to something similar to Figure \ref{fig:ranj407}. The complete code for one of the ranges is
\begin{verbatim}
al=4.3
bet=6.2

mcmc=matrix(1,ncol=1000,nrow=500)
for (t in 2:1000){
  mcmc[,t]=mcmc[,t-1]
  y = rgamma(500,4,rate=7)
  valid=(runif(500)<dgamma(y,al,rate=bet)*
    dgamma(mcmc[i,t-1],4,rate=7)/(dgamma(mcmc[,t-1],al,rate=bet)*
    dgamma(y,4,rate=7)))
  mcmc[valid,t]=y[valid]
  }
aver2=apply(mcmc,1,cumsum)
aver2=t(aver2/(1:1000))
ranj2=apply(aver2,2,range)
plot(ranj2[1,],type="l",ylim=range(ranj2),ylab="")
polygon(c(1:1000,1000:1),c(ranj2[2,],rev(ranj2[1,])))

\end{verbatim}
which removes the Monte Carlo loop over the $500$ replications by running the simulations in parallel.
We can notice on Figure \ref{fig:ranj407} that, while the output from the third sampler is quite similar 
with the output from the iid sampler [since we use the same scale on the $y$ axis], 
the Metropolis--Hastings algorithm based on the $\mathscr{G}a(4,7)$ proposal is rather biased,
which may indicate a difficulty in converging to the stationary distribution. This is somehow an
expected problem, in the sense that the ratio target-over-proposal is proportional to $x^{0.3}\,\exp(0.8x)$,
which is explosive at both $x=0$ and $x=\infty$.

\begin{exoset}
Consider $x_1$, $x_2$ and $x_3$ iid $\mathscr{C}(\theta,1)$, and $\pi(\theta)\propto \exp(-\theta^2/100)$.
Show that the posterior distribution of $\theta$, $\pi(\theta|x_1,x_2,x_3)$, is proportional to
\begin{equation}\label{eq:trimoco}
\exp(-\theta^2/100)[(1+(\theta-x_1)^2)(1+(\theta-x_2)^2)(1+(\theta-x_3)^2)]^{-1}
\end{equation}
and that it is trimodal when  $x_1=0$, $x_2=5$ and $x_3=9$.
Using a random walk based on the Cauchy distribution $\mathscr{C}(0,\sigma^2)$, estimate the
posterior mean of $\theta$ using different values of $\sigma^2$. In each case, monitor the convergence.
\end{exoset}

The function \eqref{eq:trimoco} appears as the product of the prior by the three densities $f(x_i|\theta)$.
The trimodality of the posterior can be checked on a graph when plotting the function \eqref{eq:trimoco}.

\begin{figure}
\begin{center}
\includegraphics[width=\textwidth,height=6cm]{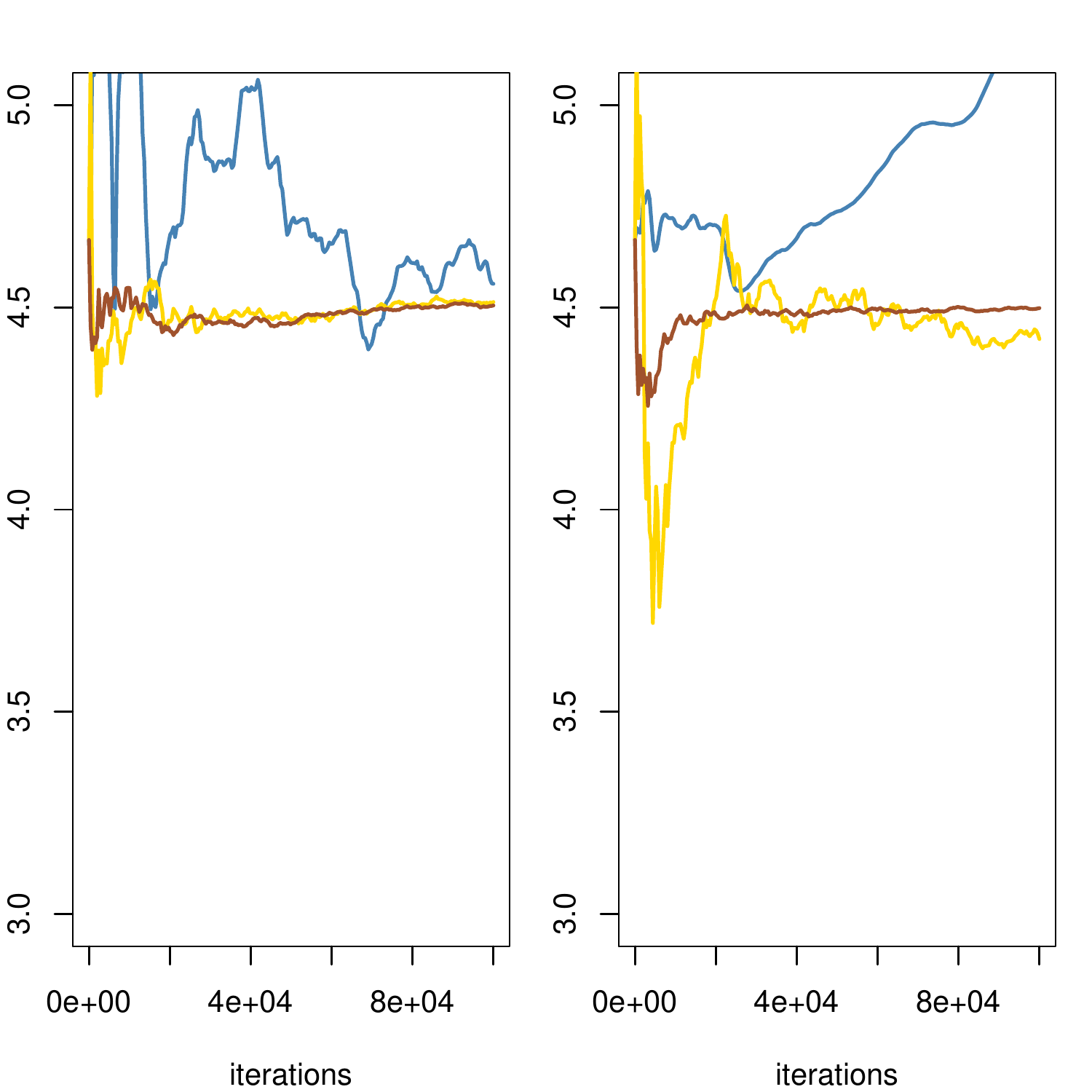}
\caption{\label{fig:trimod}Comparison of the three scale factors $\sigma=.009$ (blue),
$\sigma=.45$ (gold) and $\sigma=9$ (brown),
when using a Cauchy noise {\em (left)} and a normal noise {\em (right)}.}
\end{center}
\end{figure}

A random walk Metropolis--Hastings algorithm can be coded as follows
\begin{verbatim}
x=c(0,5,9)
# target
targ=function(y){
  dnorm(y,sd=sqrt(50))*dt(y-x[1],df=1)*
  dt(y-x[2],df=1)*dt(y-x[3],df=1)
}

# Checking trimodality
plot(seq(-2,15,length=250),
  targ(seq(-2,15,length=250)),type="l")

sigma=c(.001,.05,1)*9 # different scales
N=100000 # number of mcmc iterations

mcmc=matrix(mean(x),ncol=3,nrow=N)
for (t in 2:N){

   mcmc[t,]=mcmc[t-1,]
   y=mcmc[t,]+sigma*rt(3,1) # rnorm(3)
   valid=(runif(3)<targ(y)/targ(mcmc[t-1,]))
   mcmc[t,valid]=y[valid]
   }
\end{verbatim}
The comparison of the three cumulated averages is given in Figure \ref{fig:trimod}
and shows that, for the Cauchy noise, both large scales are acceptable while the
smallest scale slows down the convergence properties of the chain. For the normal
noise, these features are exacerbated in the sense that the smallest scale does 
not produce convergence for the number of iterations under study [the blue curve leaves
the window of observation], the medium scale
induces some variability and it is only the largest scale that gives an
acceptable approximation to the mean of the distribution \eqref{eq:trimoco}.

\begin{exoset}
Rerun the experiment of Example  4.4 using instead a
mixture of five random walks with variances
$\sigma=0.01,0.1,1,10,100$, and equal weights, 
and compare its output with the output of Figure 4.3.
\end{exoset}

The original code is provided in the files\verb+ #4.R+ and\verb+ #4.txt+ on the
webpage. The modification of the \verb+hm1+ function is as follows:
\begin{verbatim}
hmi=function(n,x0,sigma2)
{
x=rep(x0,n)

for (i in 2:n){

  x[i]=x[i-1]
  y=rnorm(1,x[i-1],sqrt(sample(sigma2,1)))
  if (runif(1)<dnorm(y)/dnorm(x[i-1]))
    x[i]=y 
  }
x
}
\end{verbatim}
Note that picking the variance at random does not modify the random walk structure of the
proposal, which is then a mixture of normal distributions all centered in $x^{(t-1)}$.
The output compares with Figure 4.3 [from the book] but the histogram is not as smooth
and the autocorrelations are higher, which can be easily explained by the fact that using
a whole range of scales induces inefficiencies in that the poor scales are chosen 
for ``nothing" from time to time.

\begin{figure}
\begin{center}
\includegraphics[width=.8\textwidth,height=6cm]{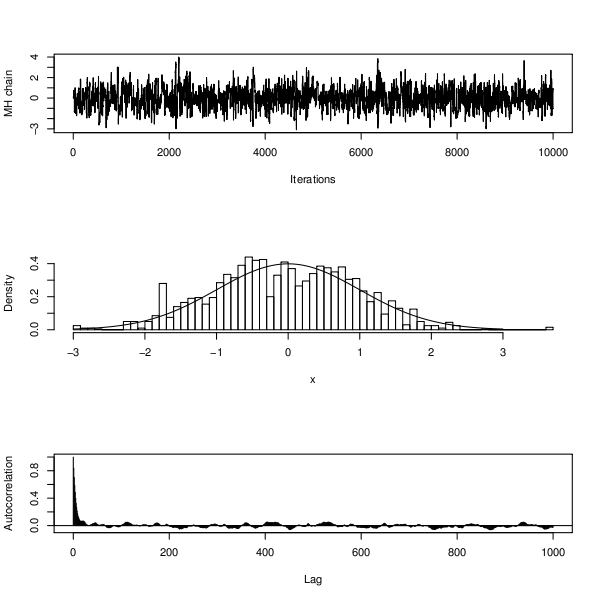}
\caption{\label{fig:mixrw}
Simulation of a $\mathscr{N}(0,1)$ target with a normal mixture:
{\em top}: sequence of $10,000$ iterations subsampled at every $10$-th iteration;
{\em middle}: histogram of the $2,000$ last iterations compared with the target density;
{\em bottom}: empirical autocorrelations using {\sf R} function {\sf plot.acf}.}
\end{center}
\end{figure}

\begin{exoset}\label{exo:proprobi}
Find conditions on the observed pairs $(\bx^i,y_i)$ for the posterior distribution 
above to be proper.
\end{exoset}

This distribution is proper (i.e.~well-defined) if the integral
$$
\mathfrak{I} = \int \prod_{i=1}^n \Phi(
\bx^{i\tee}\beta)^{y_i}\left[1-\Phi(
\bx^{i\tee}\beta)\right]^{1-y_i}\,\text{d}\beta
$$
is finite. If we introduce the latent variable behind $\Phi(\bx^{i\tee}\beta)$, we get by
Fubini that
$$
\mathfrak{I} = \int \prod_{i=1}^n \varphi(z_i)
\int_{\left\{ \beta\,; \bx^{i\tee}\beta) \gtrless z_i\,,\ i=1,\ldots,n \right\}}\,
\text{d}\beta\,\text{d}z_1\cdots \text{d}z_n\,,
$$
where $\bx^{i\tee}\beta \gtrless z_i$ means that the inequality is $\bx^{i\tee}\beta < z_i$ if $y_i=1$ and
$\bx^{i\tee}\beta < z_i$ otherwise.
Therefore, the inner integral is finite if and only if the set 
$$
\mathfrak{P}=\left\{ \beta\,; \bx^{i\tee}\beta \gtrless z_i\,,\ i=1,\ldots,n \right\}
$$
is compact. The fact that the whole integral $\mathfrak{I}$ is finite follows from the fact that the
volume of the polyhedron defined by $\mathfrak{P}$ grows like $|z_i|^k$ when $z_i$ goes to
infinity. This is however a rather less than explicit constraint on the $(\bx^i,y_i)$'s!

\begin{exoset}
Include an intercept in the probit analysis of
{\sf bank} and run the corresponding version of Algorithm 4.2
to discuss whether or not the posterior variance of the intercept is high
\end{exoset}

\begin{figure}
\begin{center}
\includegraphics[width=\textwidth,height=12cm]{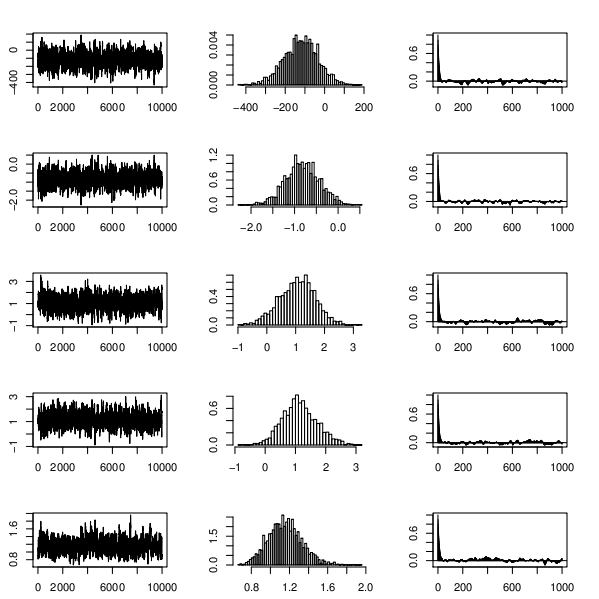}
\caption{\label{fig:bankincpt}
{\sf bank}: estimation of the probit coefficients [including one intercept $\beta_0$] 
via Algorithm 4.2 and a flat prior.
{\em Left:} $\beta_i$'s ($i=0,\ldots,4$); {\em center:} histogram
over the last $9,000$ iterations; {\em right:} auto-correlation over the last $9,000$ iterations.
}
\end{center}
\end{figure}

We simply need to add a column of $1$'s to the matrix $X$, as for instance in\\
\begin{verbatim} 
	> X=as.matrix(cbind(rep(1,dim(X)[1]),X)) 
\end{verbatim}
and then use the code provided in the file \verb+#4.txt+, i.e.
\begin{verbatim}
flatprobit=hmflatprobit(10000,y,X,1)
par(mfrow=c(5,3),mar=1+c(1.5,1.5,1.5,1.5))
for (i in 1:5){
 plot(flatprobit[,i],type="l",xlab="Iterations",
   ylab=expression(beta[i]))
 hist(flatprobit[1001:10000,i],nclass=50,prob=T,main="",
   xlab=expression(beta[i]))
 acf(flatprobit[1001:10000,i],lag=1000,main="",
   ylab="Autocorrelation",ci=F)
}
\end{verbatim}
which produces the analysis of {\sf bank} with an intercept factor. Figure \ref{fig:bankincpt}
gives the equivalent to Figure 4.4 [in the book]. The intercept $\beta_0$ has a posterior variance equal to
$7558.3$, but this must be put in perspective in that the covariates of {\sf bank} are taking their
values in the magnitude of $100$ for the three first covariates and of $10$ for the last covariate.
The covariance of $x_{i1}\beta_1$ is therefore of order $7000$ as well. A noticeable difference with
Figure 4.4 [in the book] is that, with the inclusion of the intercept, the range of $\beta_1$'s supported
by the posterior is now negative.

\begin{exoset}
Using the latent variable representation of Example 4.2,
introduce $z_i|\beta\sim\mathscr{N}\left( \bx^{i\tee}\beta,1\right)$
$(1\le i\le n)$ such that $y_i=\mathbb{B}_{z_i\le 0}$.
Deduce that
\begin{equation}\label{eq:condiz}
z_i|y_i,\beta\sim\left\{\begin{array}{ll}
\mathscr{N}_+\left( \bx^{i\tee}\beta,1,0\right) & \text{ if}\quad y_i=1 \\
\mathscr{N}_-\left( \bx^{i\tee}\beta,1,0\right) & \text{ if}\quad
y_i=0
\end{array}\right.
\end{equation}
where $\mathscr{N}_+\left(\mu,1,0\right)$ and $\mathscr{N}_-\left(\mu,1,0\right)$ are the normal
distributions with mean $\mu$ and variance $1$ that are left-truncated and right-truncated at $0$, respectively.
Check that those distributions can be simulated using the {\sf R} commands
\begin{verbatim}
   > xp=qnorm(runif(1)*pnorm(mu)+pnorm(-mu))+mu
   > xm=qnorm(runif(1)*pnorm(-mu))+mu
\end{verbatim}
Under the flat prior $\pi(\beta)\propto 1$, show that
\begin{equation*}
\beta|\by,\bz\sim\mathscr{N}_k\left((X^\tee X)^{-1}X^\tee
\bz,(X^\tee X)^{-1}\right)\,,
\end{equation*}
where $\bz=(z_1,\ldots,z_n)$ and derive the corresponding Gibbs sampler, sometimes called the
{\em Albert--Chib} sampler.  ({\em Hint:} A good starting point is the maximum likelihood
estimate of $\beta$.) Compare the application to {\sf bank} with the output in Figure 4.4.
({\em Note:} Account for computing time differences.)
\end{exoset}

If $z_i|\beta\sim\mathscr{N}\left( \bx^{i\tee}\beta,1\right)$ is a latent [unobserved] variable,
it can be related to $y_i$ via the function
$$
y_i=\mathbb{I}_{z_i\le 0}\,,
$$
since $P(y_i=1)=P(z_i\ge 0)=1-\Phi\left(-\bx^{i\tee}\beta\right)=\Phi\left(\bx^{i\tee}\beta\right)$. 
The conditional distribution of $z_i$
given $y_i$ is then a constrained normal distribution: if $y_i=1$, $z_i\le 0$ and therefore
$$
z_i|y_i=1,\beta\sim\mathscr{N}_+\left( \bx^{i\tee}\beta,1,0\right)\,.
$$
(The symmetric case is obvious.)

The command \verb&qnorm(runif(1)*pnorm(mu)+pnorm(-mu))+mu& is a simple application of the inverse
cdf transform principle given in Exercise \ref{exo:invcdf}: the cdf of the $\mathscr{N}_+\left(\mu,1,0\right)$
distribution is
$$
F(x) = \frac{\Phi(x-\mu) - \Phi(-\mu)}{\Phi(\mu)}\,.
$$
If we condition on both $\bz$ and $\by$ [the conjunction of which is defined as the ``completed model"], the
$y_i$'s get irrelevant and we are back to a linear regression model, for which the posterior distribution under
a flat prior is given in Section 3.3.1 and is indeed $\mathscr{N}_k\left((X^\tee X)^{-1}X^\tee
\bz,(X^\tee X)^{-1}\right)$.

\begin{figure}
\begin{center}
\includegraphics[width=\textwidth,height=12cm]{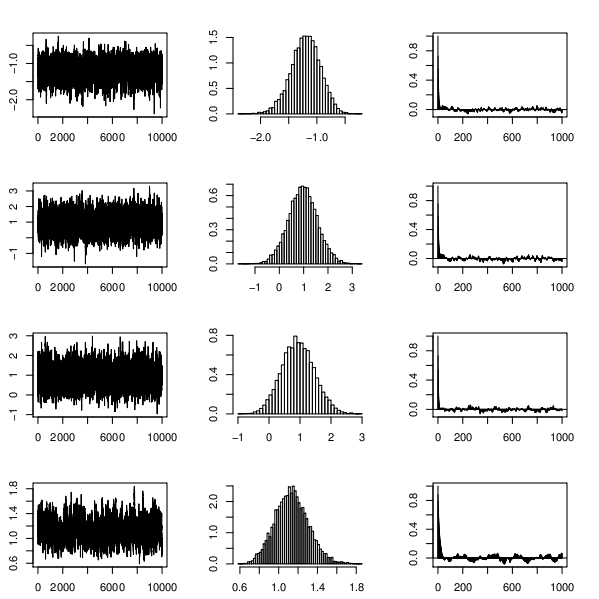}
\caption{\label{fig:gibbsprob}
{\sf bank}: estimation of the probit coefficients [including one intercept $\beta_0$]
by a Gibbs sampler 4.2 under a flat prior.
{\em Left:} $\beta_i$'s ($i=0,\ldots,4$); {\em center:} histogram
over the last $9,000$ iterations; {\em right:} auto-correlation over the last $9,000$ iterations.
}
\end{center}
\end{figure}

This closed-form representation justifies the introduction of the latent variable $\bz$ in the simulation process
and leads to the Gibbs sampler that simulates $\beta$ given $\bz$ and $\bz$ given $\beta$ and $\by$ as in \eqref{eq:condiz}.
The {\sf R} code of this sampler is available in the file \verb+#4.R+ as the function \verb+gibbsprobit+. The output
of this function is represented on Figure \ref{fig:gibbsprob}. Note that the output is somehow smoother than on
Figure \ref{fig:bankincpt}. (This does not mean that the Gibbs sampler is converging faster but rather than its
component-wise modification of the Markov chain induces slow moves and smooth transitions.)

When comparing the computing times, the increase due to the simulation of the $z_i$'s is not noticeable: for the
{\sf bank} dataset, using the codes provided in \verb+#4.txt+ require $27s$ and $26s$ over $10,000$ iterations for
\verb+hmflatprobit+ and \verb+gibbsprobit+. respectively. 

\begin{exoset}\label{exo:proprobo}
Find conditions on $\sum_i y_i$ and on $\sum_i (1-y_i)$ 
for the posterior distribution defined by (4.3) to be proper.
\end{exoset}

There is little difference with Exercise \ref{exo:proprobi} because the additional term 
$\left(\beta^\tee(X^\tee X)\beta\right)^{-(2k-1)/4}$ is creating a problem only when $\beta$ goes to
$0$. This difficulty is however superficial since the power in $||X\beta||^{(2k-1)/2}$ is small enough to
be controlled by the power in $||X\beta||^{k-1}$ in an appropriate polar change of variables. Nonetheless,
this is the main reason why we need a $\pi(\sigma^2)\propto\sigma^{-3/2}$ prior rather than the traditional
$\pi(\sigma^2)\propto\sigma^{-2}$ which is not controlled in $\beta=0$. (This is the limiting case, in the
sense that the posterior is well-defined for $\pi(\sigma^2)\propto\sigma^{-2+\epsilon}$ for all $\epsilon>0$.)
% Warning:
% This seems to imply that the condition does not bear on $\sum_i y_i$ and on $\sum_i (1-y_i)$
% directly....

\begin{exoset}\label{exo:bankx1x4}
For {\sf bank}, compute the Bayes factor associated with the
null hypothesis $H_0:\beta_2=\beta_3=0$.
\end{exoset}

The Bayes factor is given by
\begin{eqnarray*}
B^\pi_{01} &=& \frac{\pi^{-k/2}\Gamma((2k-1)/4)}
{\pi^{-(k-2)/2}\Gamma\{(2k-5)/4\}
}\\
&\times&\frac{ \int \left(\beta^\tee(X^\tee X)\beta\right)^{-(2k-1)/4}
\prod_{i=1}^n\,\Phi(\bx^{i\tee}\beta)^{y_i}\left[1-\Phi(
\bx^{i\tee}\beta)\right]^{1-y_i}\,\text{d}\beta
}{
\int \left\{ (\beta^0)^\tee(X_0^\tee X_0)\beta^0\right\}^{-(2k-5)/4}
\prod_{i=1}^n\,\Phi(x_0^{i\tee}\beta^0)^{y_i} \left[1-\Phi(x_0^{i\tee}\beta^0)\right]^{1-y_i} \text{d}\beta^0
}\,.
\end{eqnarray*}
For its approximation, we can use simulation from a multivariate normal as suggested in the book or
even better from a multivariate $\mathscr{T}$: a direct adaptation from the code in \verb+#4.txt+ is
\begin{verbatim}
noinfprobit=hmnoinfprobit(10000,y,X,1)

library(mnormt)

mkprob=apply(noinfprobit,2,mean)
vkprob=var(noinfprobit)
simk=rmvnorm(100000,mkprob,2*vkprob)
usk=probitnoinflpost(simk,y,X)-
  dmnorm(simk,mkprob,2*vkprob,log=TRUE)

noinfprobit0=hmnoinfprobit(10000,y,X[,c(1,4)],1)
mk0=apply(noinfprobit0,2,mean)
vk0=var(noinfprobit0)
simk0=rmvnorm(100000,mk0,2*vk0)
usk0=probitnoinflpost(simk0,y,X[,c(1,4)])-
  dmnorm(simk0,mk0,2*vk0,log=TRUE)
bf0probit=mean(exp(usk))/mean(exp(usk0))
\end{verbatim}
(If a multivariate $\mathscr{T}$ is used, the \verb+dmnorm+ function must be replaced with the
density of the multivariate $\mathscr{T}$.)
The value contained in \verb+bf0probit+ is $67.74$, which is thus an approximation to $B_{10}^\pi$ [since
we divide the approximate marginal under the full model with the approximate marginal under the restricted
model]. Therefore, $H_0$ is quite unlikely to hold, even though, independently, the Bayes factors associated
with the componentwise hypotheses $H_0^2: \beta_2=0$ and $H_0^3: \beta_3=0$ support those hypotheses.

\begin{exoset}\label{exo:121prob}
Compute the Jacobian $|\partial p_1\cdots\partial p_k \big/ \partial \beta_1\cdots\partial \beta_k|$ and
deduce that the transform of the prior density $\pi(p_1,\ldots,p_k)$ in the prior density above is correct.
\end{exoset}

Since $p_i=\Phi(\tilde\bx^{i\tee}\beta)$, we have
$$
\frac{\partial}{\partial\beta_j} p_i =
\frac{\partial}{\partial\beta_j} \Phi(\tilde\bx^{i\tee}\beta) =
x_{ij} \varphi(\tilde\bx^{i\tee}\beta)\,,
$$
which means that the Jacobian $\mathfrak{J}$ is the determinant of the matrix made of $\tilde X$ multiplied by 
$\varphi(\tilde\bx^{i\tee}\beta)$ on each row. Therefore,
$$
\mathfrak{J} = \prod_{i=1}^k \varphi(\tilde\bx^{i\tee}\beta) |X|\,.
$$
(Note that this is a very special case where $\tilde X$ is a square matrix, hence $|\tilde X|$ is well-defined.)
Since $|\tilde X|$ does not depend on $\beta$, it does need to appear in $\pi(\beta)$, i.e.
$$
\pi(\beta)\propto \prod_{i=1}^k\,\Phi(\tilde \bx^{i\tee}
\beta)^{K_ig_i-1}\left[ 1-\Phi(\tilde \bx^{i\tee}
\beta)\right]^{K_i(1-g_i)-1}\phi(\tilde \bx^{i\tee}\beta)\,.
$$

\begin{exoset}\label{exo:121logit}
In the case of the logit model, i.e.~when $p_i=\exp \tilde
\bx^{i\tee}\beta\big/ \{1+\exp \tilde \bx^{i\tee}\beta\}$ $(1\le
i\le k)$, derive the prior distribution on $\beta$ associated with the prior (4.5)
on $(p_1,\ldots,p_k)$.
\end{exoset}

The only difference with Exercise \ref{exo:121prob}
is in the use of a logistic density, hence both the Jacobian and the probabilities
are modified:
\begin{align*}
\pi(\beta)&\propto \prod_{i=1}^k\,\frac{\exp(\{K_ig_i-1\}\tilde \bx^{i\tee}\beta)}
{\left\{ 1+\exp(\tilde \bx^{i\tee}\beta) \right\}^{K_i-2}}
\frac{\exp(\tilde \bx^{i\tee}\beta)}{\left\{ 1+\exp(\tilde \bx^{i\tee}\beta) \right\}^2}\\
&= \frac{\ds \exp\left(\sum_{i=1}^n K_ig_i \tilde \bx^{i\tee}\beta\right)}
{\ds \prod_{i=1}^k\,\left\{ 1+\exp(\tilde \bx^{i\tee}\beta) \right\}^{K_i}}\,.
\end{align*}

\begin{exoset}
Examine whether or not the sufficient conditions for propriety of
the posterior distribution found in Exercise \ref{exo:proprobo} for
the probit model are the same for the logit model.
\end{exoset}

There is little difference with Exercises \ref{exo:proprobi} and \ref{exo:proprobo} because
the only change is [again] in the use of a logistic density, which has asymptotics similar to the
normal density. The problem at $\beta=0$ is solved in the same manner.

\begin{exoset}\label{exo:bankx1x4+1}
For {\sf bank} and the logit model, compute the Bayes factor associated with the null hypothesis
$H_0:\beta_2=\beta_3=0$ and compare its value with the value obtained for the probit model in Exercise
\ref{exo:bankx1x4}.
\end{exoset}

This is very similar to Exercise \ref{exo:bankx1x4}, except that the parameters are now estimated for the
logit model. The code is provided in file \verb+#4.txt+ as
\begin{verbatim}
# noninformative prior and random walk HM sample
noinflogit=hmnoinflogit(10000,y,X,1)

# log-marginal under full model
mklog=apply(noinflogit,2,mean)
vklog=var(noinflogit)
simk=rmnorm(100000,mklog,2*vklog)
usk=logitnoinflpost(simk,y,X)-
	dmnorm(simk,mklog,2*vklog,log=TRUE)

# noninformative prior and random walk HM sample
# for restricted model
noinflogit0=hmnoinflogit(10000,y,X[,c(1,4)],1)

# log-marginal under restricted model
mk0=apply(noinflogit0,2,mean)
vk0=var(noinflogit0)
simk0=rmnorm(100000,mk0,2*vk0)
usk0=logitnoinflpost(simk0,y,X[,c(1,4)])-
	dmnorm(simk0,mk0,2*vk0,log=TRUE)

bf0logit=mean(exp(usk))/mean(exp(usk0))
\end{verbatim}
The value of \verb+bf0logit+ is $127.2$, which, as an approximation to $B^\pi_{10}$, argues rather
strongly against
the null hypothesis $H_0$. It thus leads to the same conclusion as in the probit model of Exercise 
\ref{exo:bankx1x4}, except that the numerical value is almost twice as large. Note that, once again,
the Bayes factors associated with the componentwise hypotheses $H_0^2: \beta_2=0$ and $H_0^3: 
\beta_3=0$ support those hypotheses.

\begin{exoset}\label{exo:marcon,marcon} % debut de la marcayaise
In the case of a $2\times 2$ contingency table with fixed total
count $n=n_{11}+n_{12}+n_{21}+n_{22}$, we denote by
$\theta_{11},\theta_{12},\theta_{21},\theta_{22}$ the corresponding
probabilities. If the prior on those probabilities is a Dirichlet
$\mathcal{D}_4(1/2,\ldots,1/2)$, give the corresponding marginal
distributions of $\alpha=\theta_{11}+\theta_{12}$ and of
$\beta=\theta_{11}+\theta_{21}$. Deduce the associated Bayes factor
if $H_0$ is the hypothesis of independence between the factors and
if the priors on the margin probabilities $\alpha$ and $\beta$ are
those derived above.
\end{exoset}

A very handy representation of the Dirichlet $\mathcal{D}_k(\delta_1,\ldots,\delta_k)$
distribution is 
$$
\frac{(\xi_1,\ldots,\xi_k)}{\xi_1+\ldots+\xi_k)} \sim \mathcal{D}_k(\delta_1,\ldots,\delta_k)
\quad\text{when}\quad
\xi_i\sim\mathscr{G}a(\delta_i,1)\,,\ i=1,\ldots,k\,.
$$
Therefore, if
$$
(\theta_{11},\theta_{12},\theta_{21},\theta_{22}) =
\frac{(\xi_{11},\xi_{12},\xi_{21},\xi_{22})}{\xi_{11}+\xi_{12}+\xi_{21}+\xi_{22}}\,,
\xi_{ij}\stackrel{\text{iid}}{\sim}\mathscr{G}a(1/2,1)\,,
$$
then
$$
(\theta_{11}+\theta_{12},\theta_{21}+\theta_{22}) =
\frac{(\xi_{11}+\xi_{12},\xi_{21}+\xi_{22})}{\xi_{11}+\xi_{12}+\xi_{21}+\xi_{22}}\,,
$$
and 
$$
(\xi_{11}+\xi_{12}),(\xi_{21}+\xi_{22})\stackrel{\text{iid}}{\sim}\mathscr{G}a(1,1)
$$
implies that $\alpha$ is a $\mathscr{B}e(1,1)$ random
variable, that is, a uniform $\mathscr{U}(01,)$ variable. The same applies to $\beta$.
(Note that $\alpha$ and $\beta$ are dependent in this representation.)

Since the likelihood under the full model is multinomial,
$$
\ell(\mathbf{\theta}|\mathcal{T}) = {n\choose n_{11}\,n_{12}\,n_{21}}
\theta_{11}^{n_{11}}\,\theta_{12}^{n_{12}}\,\theta_{21}^{n_{21}}\,\theta_{22}^{n_{22}}\,,
$$
where $\mathcal{T}$ denotes the contingency table [or the dataset $\{n_{11},n_{12},n_{21},n_{22}\}$],
the [full model] marginal is
\begin{align*}
m(\mathcal{T}) &= \frac{ \displaystyle{{n\choose n_{11}\,n_{12}\,n_{21}}} }{\pi^2}\,
\int{ \theta_{11}^{n_{11}-1/2}\,\theta_{12}^{n_{12}-1/2}\,\theta_{21}^{n_{21}-1/2}\,
	\theta_{22}^{n_{22}-1/2}\,\text{d}\mathbf{\theta}}\\
&=\frac{ \displaystyle{{n\choose n_{11}\,n_{12}\,n_{21}}} }{\displaystyle \pi^2}\,
\frac{\displaystyle \prod_{i,j}\Gamma(n_{ij}+1/2)}{\displaystyle \Gamma(n+2)}\\
&=\frac{ \displaystyle{{n\choose n_{11}\,n_{12}\,n_{21}}} }{\displaystyle \pi^2}\,
\frac{\displaystyle \prod_{i,j}\Gamma(\displaystyle n_{ij}+1/2)}{(n+1)!}\\
&=\frac{ 1}{\displaystyle (n+1) \pi^2}\, \displaystyle \prod_{i,j} 
\frac{\displaystyle \Gamma(n_{ij}+1/2)}{\displaystyle \Gamma(n_{ij}+1)}\,,
\end{align*}
where the $\pi^2$ term comes from $\Gamma(1/2)=\sqrt{\pi}$.

In the restricted model, $\theta_{11}$ is replaced with $\alpha\beta$, $\theta_{12}$
by $\alpha(1-\beta)$, and so on. Therefore, the likelihood under the restricted model
is the product 
$$
{n\choose n_{1\cdot}}\,\alpha^{n_{1\cdot}}(1-\alpha)^{n-n_{1\cdot}}\,
\times
{n\choose n_{\cdot1}}\,\beta^{n_{\cdot1}}(1-\beta)^{n-n_{\cdot1}}\,,
$$
where $n_{1\cdot}=n_{11}+n_{12}$ and $n_{\cdot1}=n_{11}+n_{21}$,
and the restricted marginal under uniform priors on both $\alpha$ and $\beta$ is
\begin{align*}
m_0(\mathcal{T}) &= {n\choose n_{1\cdot}}\,{n\choose n_{\cdot1}}\,
\int_0^1\,\alpha^{n_{1\cdot}}(1-\alpha)^{n-n_{1\cdot}}\,\text{d}\alpha\,
\int_0^1\,\beta^{n_{\cdot1}}(1-\beta)^{n-n_{\cdot1}}\,\text{d}\beta\\
&={n\choose n_{1\cdot}}\,{n\choose n_{\cdot1}}\,
\frac{\ds (n_{1\cdot}+1)!(n-n_{1\cdot}+1)!}{\ds (n+2)!}\,
\frac{\ds (n_{\cdot1}+1)!(n-n_{\cdot1}+1)!}{\ds (n+2)!}\\
&=\frac{\ds (n_{1\cdot}+1)(n-n_{1\cdot}+1)}{\ds (n+2)(n+1)}\,
\frac{\ds (n_{\cdot1}+1)(n-n_{\cdot1}+1)}{\ds (n+2)(n+1)}\,.
\end{align*}
The Bayes factor $B^\pi_{01}$ is then the ratio  $m_0(\mathcal{T})/m(\mathcal{T})$.

\begin{exoset}
Given a contingency table with four categorical variables, determine
the number of submodels to consider.
\end{exoset}

Note that the numbers of classes for the different variables do not matter since,
when building a non-saturated submodel, a variable is in or out. There are 
\begin{enumerate}
\item $2^4$ single-factor models [including the zero-factor model];
\item $(2^6-1)$ two-factor models [since there are ${4 \choose 2}=6$ ways of 
picking a pair of variables out of $4$ and since the complete single-factor
model is already treated];
\item $(2^4-1)$ three-factor models.
\end{enumerate}
Thus, if we exclude the saturated model, there are $2^6+2^5-2=94$ different submodels.

\begin{exoset}
Find sufficient conditions on $(\by,X)$ for this posterior distribution proportional 
to be proper.
\end{exoset}

First, as in Exercise \ref{exo:proprobo}, the term
$\left(\beta^\tee(X^\tee X)\beta\right)^{-(2k-1)/4}$ is not a major problem when $\beta$ goes to
$0$, since it is controlled by the power in the Jacobian $||X\beta||^{k-1}$ in an adapted polar change of 
variables.  Moreover, if the matrix $X$ of regressors is of full rank $k$, then, when $\beta_j$ $(j=1,\ldots,k)$
goes to $\pm\infty$, there exists at least one $1\le i\le n$ such that $\bx^{i\tee}\beta$ goes
to either $+\infty$ or $-\infty$. In the former case, the whole exponential term goes to $0$,
while, in the later case, it depends on $y_i$. For instance, if all $y_i$'s are equal to $1$,
the above quantity is integrable.

\chapter{Capture--Recapture Experiments}\label{ch:cap}

\begin{exoset}
Show that the posterior distribution $\pi(N|n^+)$ given by (5.1)
%$$
%\pi(N|n^+) \propto \frac{1}{N(N+1)}\,\mathbb{I}_{N\geq n^+\vee 1}\,.
%$$
%where $n^+\vee 1=\max(n^+,1)$, 
while associated with an improper prior, is defined for all values of $n^+$. 
Show that the normalization factor of (5.1) is $n^+\vee 1$ and deduce that the posterior median is equal to $2(n^+\vee 1)$.
Discuss the relevance of this estimator and show that it corresponds to a Bayes estimate of $p$ equal to $1/2$.
\end{exoset}

Since the main term of the series is equivalent to $N^{-2}$, the series converges. The posterior distribution
can thus be normalised. Moreover,
\begin{align*}
\sum_{i=n_0}^{\infty}\,\frac{1}{i(i+1)}
&= \sum_{i=n_0}^\infty \left( \frac{1}{i}-\frac{1}{i+1}\right) \\
&= \frac{1}{n_0} - \frac{1}{n_0+1} + \frac{1}{n_0+1} - \frac{1}{n_0+2}+\ldots\\
&= \frac{1}{n_0}\,.
\end{align*}
Therefore, the normalisation factor is available in closed form and is equal to $n^+\vee 1$.
The posterior median is the value $N^\star$ such that $\pi(N\ge N^\star|n^+)=1/2$, i.e.~
\begin{eqnarray*}
\sum_{i=N^\star}^{\infty}\,\frac{1}{i(i+1)}&=& \frac{1}{2}\, \frac{1}{n^+\vee 1}\\
	&=& \frac{1}{N^\star}\,,
\end{eqnarray*}
which implies that $N^\star=2(n^+\vee 1)$. This estimator is rather intuitive in that $\mathbb{E}[n^+|
N,p]=pN$: since the expectation of $p$ is $1/2$, $\mathbb{E}[n^+|N]=N/2$ and $N^\star=2n^+$ is a 
moment estimator of $N$.

\begin{exoset}\label{exo:pintail}
Under the prior $\pi(N,p)\propto N^{-1}$, derive the marginal posterior density of $N$
in the case where $n_1^+\sim\mathscr{B}(N,p)$ and where $k-1$ iid observations 
$$
n_2^+,\ldots,n_k^+\stackrel{\text{iid}}{\sim} \mathscr{B}(n_1^+,p)
$$
are observed (the later are in fact recaptures). Apply to the sample
$$
(n^+_{1},n_2^+,\ldots,n^+_{11}) = (32,20,8,5,1,2, 0,2,1,1,0)\,,
$$
which describes a series of tag recoveries over $11$ years.
\end{exoset}

In that case, if we denote $n_{\cdot}^+=n_1^++\cdots+n_k^+$ the total number of captures, 
the marginal posterior density of $N$ is
\begin{align*}
\pi(N|n_1^+,\ldots,n_k^+) &\propto  \frac{N!}{(N-n_1^+)!}\,N^{-1}\mathbb{I}_{N\ge n_1^+}\\
&\qquad \int_0^1 p^{n_1^++\cdots+n_k^+}(1-p)^{N-n_1^++(n_1+-n_2^++\cdots+n_1^+-n_k^+}\text{d}p \\
&\propto  \frac{(N-1)!}{(N-n_1^+)!}\,\mathbb{I}_{N\ge n_1^+}
\int_0^1 p^{n_\cdot^+}(1-p)^{N+kn_1^+-n_\cdot^+} \text{d}p \\
& \propto \frac{(N-1)!}{(N-n_1^+)!}\,\frac{(N+kn_1^+-n_\cdot^+)!}{(N+kn_1^++1)!}\,\mathbb{I}_{N\geq n_1^+\vee 1}\,, 
\end{align*}
which does not simplify any further. Note that the binomial coefficients 
$$
{n_1^+\choose n_j^+} \qquad(j\ge 2)
$$
are irrelevant for the posterior of $N$ since they only depend on the data.

The {\sf R} code corresponding to this model is as follows:
\begin{verbatim}
n1=32
ndo=sum(32,20,8,5,1,2,0,2,1,1,0)

# unnormalised posterior
post=function(N){
   exp(lfactorial(N-1)+lfactorial(N+11*n1-ndo)-
    lfactorial(N-n1)-lfactorial(N+11*n1+1))
   }

# normalising constant and 
# posterior mean

posv=post((n1:10000))

cons=sum(posv)
pmean=sum((n1:10000)*posv)/cons
pmedi=sum(cumsum(posv)<.5*cons)
\end{verbatim}
The posterior mean is therefore equal to $282.4$, while the posterior median is $243$. Note that a crude analysis estimating 
$p$ by $\hat p=(n_2^++\ldots+n_{11})/(10 n_1^+)=0.125$ and $N$ by $n_1^+/\hat p$ would produce the value $\hat N=256$.

\begin{exoset}
For the two-stage capture-recapture model, show that the distribution of $m_2$ conditional 
on both samples sizes $n_1$ and $n_2$ is given by (5.2)
%$$
%m_2|n_1,n_2 \sim \mathscr{H}(N,n_2,n_1/N)\,,
%$$
and does not depend on $p$. Deduce the expectation $\mathbb{E}[m_2|n_1,n_2,N]$.
\end{exoset}

Since
$$
n_1     \sim\mathscr{B}(N,p)\,,\quad
m_2|n_1 \sim\mathscr{B}(n_1,p)
$$and$$%\quad\hbox{and}\quad
n_2-m_2|n_1,m_2 \sim\mathscr{B}(N-n_1,p)\,,
$$
the conditional distribution of $m_2$ is given by
\begin{align*}
f(m_2|n_1,n_2) &\propto {n_1\choose m_2} p^{m_2}(1-p)^{n_1-m_2} {N-n_1\choose n_2-m_2} p^{n_2-m_2} (1-p)^{N-n_1-n_2+m_2}\\
&\propto {n_1\choose m_2} {N-n_1\choose n_2-m_2}\,p^{m_2+n_2-m_2}(1-p)^{n_1-m_2+N-n_1-n_2+m_2}\\
&\propto {n_1\choose m_2} {N-n_1\choose n_2-m_2}\\
&\propto \frac{{n_1\choose m_2} {N-n_1\choose n_2-m_2}}{{N\choose n_2}}\,,
\end{align*}
which is the hypergeometric $\mathscr{H}(N,n_2,n_1/N)$ distribution. Obviously, this distribution does not depend on $p$
and its expectation is
$$
\mathbb{E}[m_2|n_1,n_2]=\frac{n_1 n_2}{N}\,.
$$

\begin{exoset}
In order to determine the number $N$ of buses in a town, a capture--recapture strategy goes
as follows. We observe  $n_1=20$ buses during the first day and keep track of their identifying numbers. Then we repeat
the experiment the following day by recording the number of buses that have already
been spotted on the previous day, say $m_2=5$, out of the $n_2=30$ buses observed the
second day. For the Darroch model, give the posterior expectation of $N$ under the prior
$\pi(N)=1/N$. %^{-1}\mathbb{I}_{\mathbb{N}^*}(N)$.
\end{exoset}

Using the derivations of the book, we have that
\begin{align*}
\pi(N|n_1,n_2,m_2) &\propto \frac{1}{N}\,{N \choose n^+}\,B(n^c+1,2N-n^c+1)\mathbb{I}_{N\geq n^+} \\
    &\propto \frac{(N-1)!}{(N-n^+)!}\,\frac{(2N-n^c)!}{(2N+1)!}\,\mathbb{I}_{N\geq n^+}
\end{align*}
with $n^+=45$ and $n^c=50$. For $n^+=45$ and $n^c=50$, the posterior mean [obtained by an {\sf R} code
very similar to the one of Exercise \ref{exo:pintail}] is equal to $130.91$.

\begin{exoset}Show that the maximum
likelihood estimator of $N$ for the Darroch model is $\hat  N = n_1 / \left(m_2 /n_2 \right)$
and deduce that it is not defined when $m_2=0$.
\end{exoset}

The likelihood for the Darroch model is proportional to
$$
\ell(N) = \frac{(N-n_1)!}{(N-n_2)!}\,\frac{(N-n^+)!}{N!}\,\mathbb{I}_{N\geq n^+}\,.
$$
Since 
$$
\frac{\ell(N+1)}{\ell(N)} = \frac{(N+1-n_1)(N+1-n_2)}{(N+1-n^+)(N+1)} \ge 1
$$
for 
\begin{eqnarray*}
(N+1)^2-(N+1)(n_1+n_2) + n_1n_2 &\ge& (N+1)^2 -(N+1)n^+ \\
(N+1)(n_1+n_2-n^+) &\ge& n_1n_2\\
(N+1) &\le& \frac{n_1n_2}{m_2}\,,
\end{eqnarray*}
the likelihood is increasing for $N\le n_1n_2/m2$ and decreasing for $N\ge n_1n_2/m2$.
Thus $\hat N=n_1n_2/m2$ is the maximum likelihood estimator [assuming this quantity is
an integer]. If $m_2=0$, the likelihood is increasing with $N$ and therefore there 
is no maximum likelihood estimator.

\begin{exoset} 
Give the likelihood of the extension of Darroch's model
when the capture--recapture experiments are repeated $K$ times
with capture sizes and recapture observations $n_k$ $(1\le k\le K)$ and $m_k$ $(2\le k\le K)$,
respectively. ({\em Hint}: Exhibit first the two-dimensional sufficient statistic associated with
this model.)
\end{exoset}

When extending the two-stage capture-recapture model to a $K$-stage model, we observe $K$ capture
episodes, with $n_i\sim\mathscr{B}(N,p)$ $(1\le i\le K)$, and $K-1$ recaptures,
$$
m_i|n_1,n_2,m_2,\ldots,n_{i-1},m_{i-1},n_i \sim
\mathscr{H}(N,n_i,n_1+n_2-m_2+\cdots-m_{i-1})\,.
$$
The likelihood is therefore
\begin{align*}
\prod_{i=1}^K {N\choose n_i} & p^{n_i} (1-p)^{N-n_i}\,
\prod_{i=2}^K \frac{{n_1-m_2+\cdots-m_{i-1}\choose m_i}{N-n_1+\cdots+m_{i-1}\choose n_i-m_i}}{{N\choose n_i}}\\
&\propto \frac{N!}{(N!-n^+)!}\,p^{n^c}(1-p)^{KN-n^c}\,,
\end{align*}
where $n^+=n_1-m_2+\cdots-m_{K}$ is the number of captured individuals and where $n^c=n_1+\cdots+n_K$ is the
number of captures. These two statistics are thus sufficient for the $K$-stage capture-recapture model.

\begin{exoset} 
Give both conditional posterior distributions in the case $n^+=0$.
\end{exoset}

When $n^+=0$, there is no capture at all during both capture episodes. The likelihood is thus $(1-p)^{2N}$
and, under the prior $\pi(N,p)=1/N$, the conditional posterior distributions of $p$ and $N$ are
\begin{align*}
p|N,n^+=0 &\sim \mathscr{B}e(1,2N+1)\,,\\
N|p,n^+=0 &\sim \frac{(1-p)^{2N}}{N}\,.
\end{align*}
That the joint distribution $\pi(N,p|n^+=0)$ exists is ensured by the fact that $\pi(N|n^+=0)\propto 1/N(2N+1)$,
associated with a converging series.

\begin{exoset}
Show that, when the prior on $N$ is a $\mathscr{P}(\lambda)$
distribution, the conditional posterior on $N-n_+$ is
$\mathscr{P}(\lambda(1-p)^2)$.
\end{exoset}

The posterior distribution of $(N,p)$ associated with the informative prior $\pi(N,p)=\lambda^N e^{-\lambda}/N!$
is proportional to
$$
\frac{N!}{(N-n^+)!N!}\,\lambda^N\,p^{n^c}(1-p)^{2N-n^c}\,\mathbb{I}_{N\geq n^+}\,.
$$
The corresponding conditional on $N$ is thus proportional to 
$$
\frac{\lambda^N}{(N-n^+)!} \,p^{n^c}(1-p)^{2N-n^c}\,\mathbb{I}_{N\geq n^+}
\propto \frac{\lambda^{N-n^+}}{(N-n^+)!} \,p^{n^c}(1-p)^{2N-n^c}\,\mathbb{I}_{N\geq n^+}
$$
which corresponds to a Poisson $\mathscr{P}(\lambda(1-p)^2)$ distribution
on $N-n_+$.

\begin{exoset}
An extension of the $T$-stage capture-recapture model is to consider that the
capture of an individual modifies its probability of being captured from $p$ to $q$ for
future captures. Give the likelihood $\ell(N,p,q|n_1,n_2,m_2\ldots,n_T,m_T)$.
\end{exoset}

When extending the $T$-stage capture-recapture model with different probabilities of being captured
and recaptured, after the first capture episode, where $n_1\sim\mathscr{B}(N,p)$, we observe $T-1$ 
new captures $(i=2,\ldots,T)$
$$
n_i-m_i|n_1,n_2,m_2,\ldots,n_{i-1},m_{i-1}\sim\mathscr{B}(N-n_1-n_2+m_2+\ldots+m_{i-1},p)\,,
$$ 
and $T-1$ recaptures $(i=2,\ldots,T)$,
$$
m_i|n_1,n_2,m_2,\ldots,n_{i-1},m_{i-1} \sim
\mathscr{B}(n_1+n_2-m_2+\ldots-m_{i-1},q)\,.
$$
The likelihood is therefore
\begin{align*}
{N\choose n_1}\,p^{n_1}(1-p)^{N-n_1} &\prod_{i=2}^T {N-n_1+\ldots-m_{i-1}\choose n_i-m_i} 
p^{n_i-m_i} (1-p)^{N-n_1+\ldots+m_i}\\
&\quad\times\prod_{i=2}^T {n_1+n_2-\ldots-m_{i-1}\choose m_i} q^{m_i} (1-q)^{n_1+\ldots-m_{i}} \\
&\propto \frac{N!}{(N-n^+)!}\,p^{n^+}(1-p)^{TN-n^*}\,q^{m^+}(1-q)^{n^*-n_1},
\end{align*}
where $n^+=n_1-m_2+\cdots-m_{T}$ is the number of captured individuals,
$$
n^* = Tn_1 + \sum_{j=2}^T (T-j+1)(n_j-m_j)
$$
and where $m^+=m_1+\cdots+m_T$ is the number of recaptures. The four statistics $(n_1,n^+,n^*,m^+)$
are thus sufficient for this version of the $T$-stage capture-recapture model.

\begin{exoset}
Another extension of the $2$-stage capture-recapture model is to allow
for mark losses. If we introduce $q$ as the probability of losing the mark, $r$ as the probability of
recovering a lost mark and $k$ as the number of
recovered lost marks, give the associated likelihood $\ell(N,p,q,r|n_1,n_2,m_2,k)$. 
\end{exoset}

There is an extra-difficulty in this extension in that it contains a latent variable: let us denote by $z$
the number of tagged individuals that have lost their mark. Then $z\sim\mathscr{B}(n_1,q)$ is not observed,
while $k\sim\mathscr{B}(z,r)$ is observed. Were we to observe $(n_1,n_2,m_2,k,z)$, the [completed] likelihood would
be
\begin{align*}
\ell^\star(N,p,q,r&|n_1,n_2,m_2,k,z)={N\choose n_1}\,p^{n_1}(1-p)^{N-n_1}\,{n_1\choose z}\,q^z(1-q)^{n_1-z}\\
&\quad \times {z\choose k}\,r^k(1-r)^{z-k}\,{n_1-z\choose m_2}\,p^{m_2}(1-p)^{n_1-z-m_2}\\
&\quad \times {N-n_1+z\choose n_2-m_2}\,p^{n_2-m_2}(1-p)^{N-n_1+z-n_2+m_2}\,,
\end{align*}
since, for the second round, the population gets partitioned into individuals that keep their tag and are/are 
not recaptured, those that loose their tag and are/are not recaptured, and those that are captured for the first time.
Obviously, it is not possible to distinguish between the two last categories. Since $z$ is not known, the [observed]
likelihood is obtained by summation over $z$:
\begin{align*}
\ell(N,p,q,r&|n_1,n_2,m_2,k) \propto \frac{N!}{(N-n_1)!} \,p^{n_1+n_2}(1-p)^{2N-n_1-n_2}\\
&\sum_{z=k\vee N-n_1-n_2+m_2}^{n_1-m_2} {n_1\choose z} \,{n_1-z\choose m_2}\\
&\times {N-n_1+z\choose n_2-m_2}\, q^z(1-q)^{n_1-z}\,r^k(1-r)^{z-k}\,.
\end{align*}
Note that, while a proportionality sign is acceptable for the computation of the likelihood, the terms depending on
$z$ must be kept within the sum to obtain the correct expression for the distribution of the observations. A simplified
version is thus
\begin{align*}
\ell(N,p,q,r&|n_1,n_2,m_2,k) \propto \frac{N!}{(N-n_1)!} \,p^{n_1+n_2}(1-p)^{2N-n_1-n_2}\,
q^{n_1}(r/(1-r))^k\\
&\sum_{z=k\vee N-n_1-n_2+m_2}^{n_1-m_2} \frac{(N-n_1+z)![q(1-r)/(1-q)]^z
}{z!(n_1-z-m_2)!(N-n_1-n_2+m_2+z)!}\,,
\end{align*}
but there is no close-form solution for the summation over $z$.

\begin{exoset}
Reproduce the analysis of {\sf eurodip} when switching the prior from
$\pi(N,p)\propto \lambda^N/N!$ to $\pi(N,p)\propto N^{-1}$.
\end{exoset}

\begin{figure}
\begin{center}
\includegraphics[width=\textwidth,height=6cm]{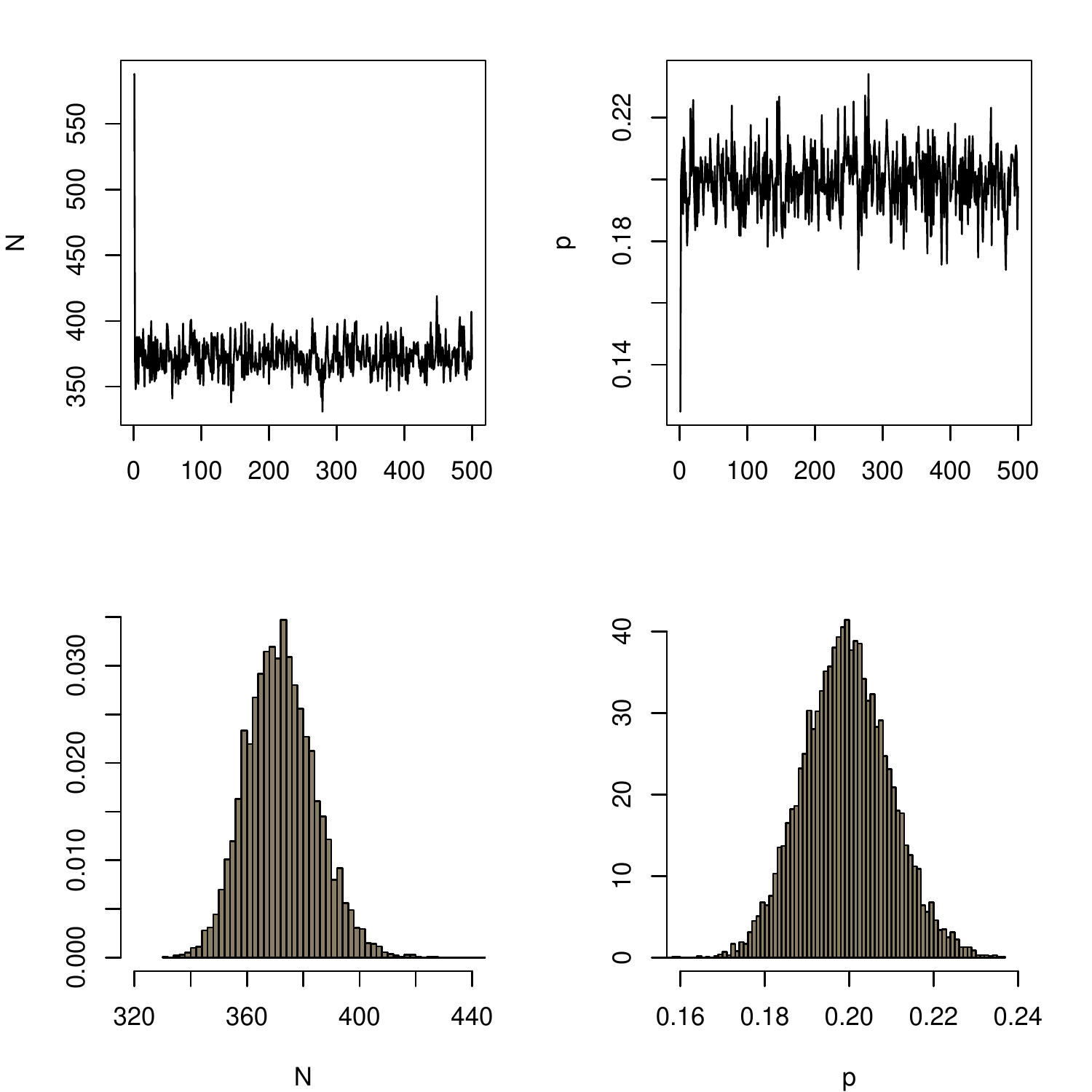}
\caption{\label{fig:euronew}{\sf eurodip}: MCMC simulation under the
prior $\pi(N,p)\propto N^{-1}$.}
\end{center}
\end{figure}

The main purpose of this exercise is to modify the code of the function \verb+gibbs1+ in 
the file \verb+#5.R+ on the webpage, since the marginal posterior distribution of $N$
is given in the book as
$$
\pi(N|n^+,n^c) \propto {(N-1)! \over (N-n^+)!}\, { (TN-n^c)! \over (TN+1)!}\, 
\mathbb{I}_{N\ge n^+\vee 1}\,.
$$
(The conditional posterior distribution of $p$ does not change.) This distribution being
non-standard, it makes direct simulation awkward and we prefer to use a Metropolis-Hastings
step, using a modified version of the previous Poisson conditional as proposal $q(N^\prime|N,p)$. 
We thus simulate 
$$
N^\star-n^+\sim \mathscr{P}\left( N^{(t-1)}(1-p^{(t-1)})^T \right)
$$
and accept this value with probability
$$
\frac{\pi(N^\star|n^+,n^c)}{\pi(N^{(t-1)}|n^+,n^c)}\,
\frac{q(N^{(t-1)}|N^\star,p^{(t-1)})}{q(N^\star|N^{(t-1)},p^{(t-1)})} \wedge 1\,.
$$
The corresponding modified {\sf R} function is
\begin{verbatim}
gibbs11=function(nsimu,T,nplus,nc)
{
# conditional posterior
rati=function(N){
  lfactorial(N-1)+lfactorial(T*N-nc)-
    lfactorial(N-nplus)-lfactorial(T*N+1)
  }

N=rep(0,nsimu)
p=rep(0,nsimu)

N[1]=2*nplus
p[1]=rbeta(1,nc+1,T*N[1]-nc+1)
for (i in 2:nsimu){

  # MH step on N
  N[i]=N[i-1]
  prop=nplus+rpois(1,N[i-1]*(1-p[i-1])^T)
  if (log(runif(1))<rati(prop)-rati(N[i])+
        dpois(N[i-1]-nplus,prop*(1-p[i-1])^T,log=T)-
        dpois(prop-nplus,N[i-1]*(1-p[i-1])^T,log=T)) 
     N[i]=prop
  p[i]=rbeta(1,nc+1,T*N[i]-nc+1)
  }
list(N=N,p=p)
}
\end{verbatim}
The output of this program is given in Figure \ref{fig:euronew}.

\begin{exoset}
Show that the conditional distribution of $r_1$ is indeed proportional
to the product (5.4).
\end{exoset}

The joint distribution of $\mathcal{D}^*=(n_1,c_2,c_3,r_1,r_2)$ is
given in the book as 
\begin{align*}
{N\choose n_1}&p^{n_1}(1-p)^{N-n_1}\,{n_1\choose r_1}\,q^{r_1}(1-q)^{n_1-r_1}{n_1-r_1\choose c_2}\,p^{c_2}(1-p)^{n_1-r_1-c_2}\\
&\times{n_1-r_1\choose r_2} q^{r_2}(1-q)^{n_1-r_1-r_2}{n_1-r_1-r_2\choose c_3}\,p^{c_3}(1-p)^{n_1-r_1-r_2-c_3}\,.
\end{align*}
Therefore, if we only keep the terms depending on $r_1$, we indeed recover
\begin{align*}
&\frac{1}{r_1!(n_1-r_1)!}\,q^{r_1}(1-q)^{n_1-r_1}\,\frac{(n_1-r_1)!}{(n_1-r_1-c_2)!}\,(1-p)^{n_1-r_1-c_2}\\
&\quad\times\frac{(n_1-r_1)!}{(n_1-r_1-r_2)!}\,(1-q)^{n_1-r_1-r_2}\,\frac{(n_1-r_1-r_2)!}{(n_1-r_1-r_2-c_3)!}\,
(1-p)^{n_1-r_1-r_2-c_3}\\
&\propto\frac{(n_1-r_1)!}{r_1!(n_1-r_1-c_2)!(n_1-r_1-r_2-c_3)!}\,\left\{ \frac{q}{(1-q)^2(1-p)^2}\right\}^{r_1}\\
&\propto{n_1-c_2\choose r_1}\,{n_1-r_1\choose r_2+c_3}\,\left\{ \frac{q}{(1-q)^2(1-p)^2}\right\}^{r_1}\,,
\end{align*}
under the constraint that $r_1\le \min(n_1,n_1-r_2,n_1-r_2-c_3,n_1-c_2)=\min(n_1-r_2-c_3,n_1-c_2)$.

\begin{exoset}\label{exo:margot}
Show that $r_2$ can be integrated out in the above joint distribution and leads to
the following distribution on $r_1$:
\begin{align}\label{eq:margir1}
\pi(r_1|p,q,n_1,c_2,c_3) \propto\, &\frac{(n_1-r_1)!(n_1-r_1-c_3)!}{r_1!(n_1-r_1-c_2)!}\\
&\times \left(\frac{q}{(1-p)(1-q)[q+(1-p)(1-q)]}\right)^{r_1}\,.\nonumber
\end{align}
Compare the computational cost of a Gibbs sampler based on this approach with 
a Gibbs sampler using the full conditionals.
\end{exoset}

Following the decomposition of the likelihood in the previous exercise, the terms depending on $r_2$ are
\begin{align*}
\frac{1}{r_2!(n_1-r_1-r_2)!}&\left(\frac{q}{(1-p)(1-q)}\right\}^{r_2} \frac{(n_1-r_1-r_2)!}{(n_1-r_1-r_2-c_3)!}\\
&=\frac{1}{r_2!(n_1-r_1-r_2-c_3)!}\left(\frac{q}{(1-p)(1-q)}\right\}^{r_2}\,.
\end{align*}
If we sum over $0\le r_2\le n_1-r_1-c_3$, we get
\begin{align*}
\frac{1}{(n_1-r_1-c_3)!}\,&\sum_{k=0}^{n_1-r_1-c_3} {n_1-r_1-c_3\choose k} \left(\frac{q}{(1-p)(1-q)}\right\}^k\\
&= \left\{1+\frac{q}{(1-p)(1-q)}\right\}^{n_1-r_1-c_3}
\end{align*}
that we can agregate with the remaining terms in $r_1$
$$
\frac{(n-r_1)!}{r_1!(n_1-r_1-c_2)!}\left\{ \frac{q}{(1-q)^2(1-p)^2}\right\}^{r_1}
$$
to recover \eqref{eq:margir1}.

Given that a Gibbs sampler using the full conditionals is simulating from standard distributions while a Gibbs sampler
based on this approach requires the simulation of this non-standard distribution on $r_1$, it appears that one
iteration of the latter is more time-consuming than for the former.

\begin{exoset}
Show that the likelihood associated with an open population can be written as
\begin{align*}
\ell(N,p|\mathcal{D}^*) &= \sum_{(\epsilon_{it},\delta_{it})_{it}}\prod_{t=1}^T
\prod_{i=1}^N q_{\epsilon_{i(t-1)}}^{\epsilon_{it}}(1-q_{\epsilon_{i(t-1)}})^{1-\epsilon_{it}} \\
&\qquad\qquad \times p^{(1-\epsilon_{it})\delta_{it}} (1-p)^{(1-\epsilon_{it})(1-\delta_{it})} \,,
\end{align*}
where $q_0=q$, $q_1=1$, and $\delta_{it}$ and $\epsilon_{it}$
are the capture and exit indicators, respectively. Derive the order of complexity
of this likelihood, that is, the number of elementary operations necessary to
compute it.
\end{exoset}

This is an alternative representation of the model where each individual capture and life history
is considered explicitely. This is also the approach adopted for the Arnason-Schwarz model of Section 
5.5. We can thus define the history of individual $1\le i\le N$ as a pair of sequences $(\epsilon_{it})$
and $(\delta_{it})$, where $\epsilon_{it}=1$ at the exit time $t$ and forever after. For the model given 
at the beginning of Section 5.3, there are $n_1$ $\delta_{i1}$'s equal to $1$, $r_1$ $\epsilon_{i1}$'s equal 
to $1$, $c_2$ $\delta_{i2}$'s equal to $1$ among the $i$'s for which $\delta_{i1}=1$ and so on. If we do not
account for these constraints, the likelihood is of order $\text{O}(3^{NT})$ [there are three possible
cases for the pair $(\epsilon_{it},\delta_{it})$ since $\delta_{it}=0$ if $\epsilon_{it}=1$]. 
Accounting for the constraints on the total number of $\delta_{it}$'s equal to $1$ increases 
the complexity of the computation.

\begin{exoset}\label{exo:noneed4con}
Show that, for $M>0$, if $g$ is replaced with $Mg$ in $\mathscr{S}$
and if $(X,U)$ is uniformly distributed on $\mathscr{S}$, the marginal distribution
of $X$ is still $g$. Deduce that the density $g$ only needs to be known up to a
normalizing constant.
\end{exoset}

The set
$$ 
\mathscr{S}=\{(x,u):0<u<Mg(x)\}
$$
has a surface equal to $M$. Therefore, the uniform distribution on $\mathscr{S}$ has density $1/M$
and the marginal of $X$ is given by
$$
\int \mathbb{I}_{(0,Mg(x))}\,\frac{1}{M}\,\text{d}u = \frac{Mg(x)}{M}=g(x)\,.
$$
This implies that uniform simulation in $\mathscr{S}$  provides an output from $g$ no matter what the
constant $M$ is. In other words, $g$ does not need to be normalised.

\begin{exoset}\label{ex:2Hard}
For the function $g(x)=(1+\sin^2(x))
(2+\cos^4(4x))\exp[-x^4\{1+\sin^6(x)\}]$ on $[0,2\pi]$, examine the feasibility of running a uniform
sampler on the associated set $\mathscr{S}$.\end{exoset}

The function $g$ is non-standard but it is bounded [from above] by the function
$\overline{g}(x) = 6\exp[-x^4]$ since both $\cos$ and $\sin$ are bounded by $1$ or
even $\overline{g}(x) = 6$. Simulating uniformly over the set $\mathscr{S}$ associated
with $g$ can thus be achieved by simulating uniformly
over the set $\mathscr{S}$ associated with $\overline{g}$ until the output falls within
the set $\mathscr{S}$ associated with $g$.  This is the basis of accept-reject algorithms.

\begin{exoset}
Show that the probability of acceptance in Step 2 of Algorithm 5.2 is $1/M$,
and that the number of trials until a variable is accepted has a geometric distribution
with parameter $1/M$. Conclude that the expected number of trials per simulation is $M$.
\end{exoset}

The probability that $U\le g(X)/(Mf(X))$ is the probability that a uniform draw in the set
$$
\mathscr{S}=\{(x,u):0<u<Mg(x)\}
$$
falls into the subset
$$
\mathscr{S}_0=\{(x,u):0<u<f(x)\}.
$$
The surfaces of $\mathscr{S}$ and $\mathscr{S}_0$ being $M$ and $1$, respectively, the probability
to fall into $\mathscr{S}_0$ is $1/M$.

Since steps {\sf 1.} and {\sf 2.} of Algorithm 5.2 are repeated independently, each round has a
probability $1/M$ of success and the rounds are repeated till the first success. The number of
rounds is therefore a geometric random variable with parameter $1/M$ and expectation $M$.

\begin{exoset}
For the conditional distribution of $\alpha_t$ derived from (5.3),
%$$
%\pi(\alpha_t|N,n_t) \; \propto  (1 + e^{\alpha_{t}})^{-N} \\
%\;\exp \left\{\alpha_t n_t - \frac{1}{2\sigma^2} (\alpha_{t} - \mu_{t})^{2} \right\} \,.
%$$
construct an Accept-Reject
algorithm based on a normal bounding density $f$ and study its performances for $N=53$,
$n_t=38$, $\mu_t=-0.5$, and $\sigma^2=3$.
\end{exoset}

That the target is only known up to a constant is not a problem, as demonstrated in Exercise 
\ref{exo:noneed4con}. To find a bound on $\pi(\alpha_t|N,n_t)$ [up to a constant], we just have
to notice that 
$$
(1 + e^{\alpha_{t}})^{-N} < e^{-N\alpha_{t}}
$$
and therefore
\begin{align*}
(1 + e^{\alpha_{t}})^{-N} &
\;\exp \left\{\alpha_t n_t - \frac{1}{2\sigma^2} (\alpha_{t} - \mu_{t})^{2} \right\} \\
&\le \exp \left\{\alpha_t (n_t-N) - \frac{1}{2\sigma^2} (\alpha_{t} - \mu_{t})^{2} \right\} \\
&= \exp \left\{ - \frac{\alpha_t^2}{2\sigma^2} + 2\frac{\alpha_t}{2\sigma^2}(\mu_t
-\sigma^2(N-n_t)) - \frac{\mu_t^2}{2\sigma^2} \right\} \\
&= \frac{1}{\sqrt{2\pi}\sigma} \exp \left\{ - \frac{1}{2\sigma^2} (\alpha_{t} - \mu_t + \sigma^2(N-n_t))^2 \right\}\\
&\quad\times
\sqrt{2\pi}\sigma \exp \left\{ - \frac{1}{2\sigma^2} (\mu_t^2 - [\mu_t - \sigma^2(N-n_t)]^2 )\right\}\,.
\end{align*}
The upper bound thus involves a normal $\mathscr{N}(\mu_t - \sigma^2(N-n_t),\sigma^2)$ distribution and
the corresponding constant. The {\sf R} code associated with this decomposition is 
\begin{verbatim}
# constants
N=53
nt=38
mut=-.5
sig2=3
sig=sqrt(sig2)

# log target
ta=function(x){
  -N*log(1+exp(x))+x*nt-(x-mut)^2/(2*sig2)
  }

#bounding constant
bmean=mut-sig2*(N-nt)
uc=0.5*log(2*pi*sig2)+(bmean^2-mut^2)/(2*sig2)

prop=rnorm(1,sd=sig)+bmean
ratio=ta(prop)-uc-dnorm(prop,mean=bmean,sd=sig,log=T)

while (log(runif(1))>ratio){

  prop=rnorm(1,sd=sig)+bmean
  ratio=ta(prop)-uc-dnorm(prop,mean=bmean,sd=sig,log=T)
  }
\end{verbatim}
The performances of this algorithm degenerate very rapidly when $N-n_t$ is [even moderately] large.

\begin{exoset}
When uniform simulation on $\mathscr{S}$ is impossible, construct a Gibbs sampler based on
the conditional distributions of $u$ and $x$. ({\em Hint}: Show that both conditionals are uniform
distributions.) This special case of the Gibbs sampler is called the {\em slice sampler} (see
Robert and Casella, 2004, Chapter 8). Apply to the distribution of Exercise \ref{ex:2Hard}.
\end{exoset}

Since the joint distribution of $(X,U)$ has the constant density
$$
t(x,u)=\mathbb{I}_{0\le u\le g(x)}\,,
$$
the conditional distribution of $U$ given $X=x$ is $\mathscr{U}(0,g(x))$ and the conditional
distribution of $X$ given $U=u$ is $\mathscr{U}(\{x;g(x)\ge u\})$, which is uniform over the
set of highest values of $g$. Both conditionals are therefore uniform and this special Gibbs
sampler is called the {\em slice sampler}. In some settings, inverting the condition $g(x)\ge u$
may prove formidable! 

If we take the case of Exercise \ref{ex:2Hard} and of $\overline{g}(x)=
\exp(-x^4)$, the set $\{x;\overline{g}(x)\ge u\}$ is equal to 
$$
\left\{x;\overline{g}(x)\ge u\right\}=
\left\{x;x\le (-\log(x))^{1/4}\right\},
$$
which thus produces a closed-form solution.

\begin{exoset}Reproduce the above analysis for the marginal distribution of $r_1$ computed
in Exercise \ref{exo:margot}.\end{exoset}

The only change in the codes provided in \verb+#5.R+ deals with \verb+seuil+, called by \verb+ardipper+,
and with \verb+gibbs2+ where the simulation of $r_2$ is no longer required.

\begin{exoset}
Show that, given a mean and a $95\%$ confidence interval in $[0,1]$, there exists at
most one beta distribution $\mathscr{B}(a,b)$ with such a mean and confidence interval.
\end{exoset}

\begin{figure}
\begin{center}
\includegraphics[width=\textwidth,height=5cm]{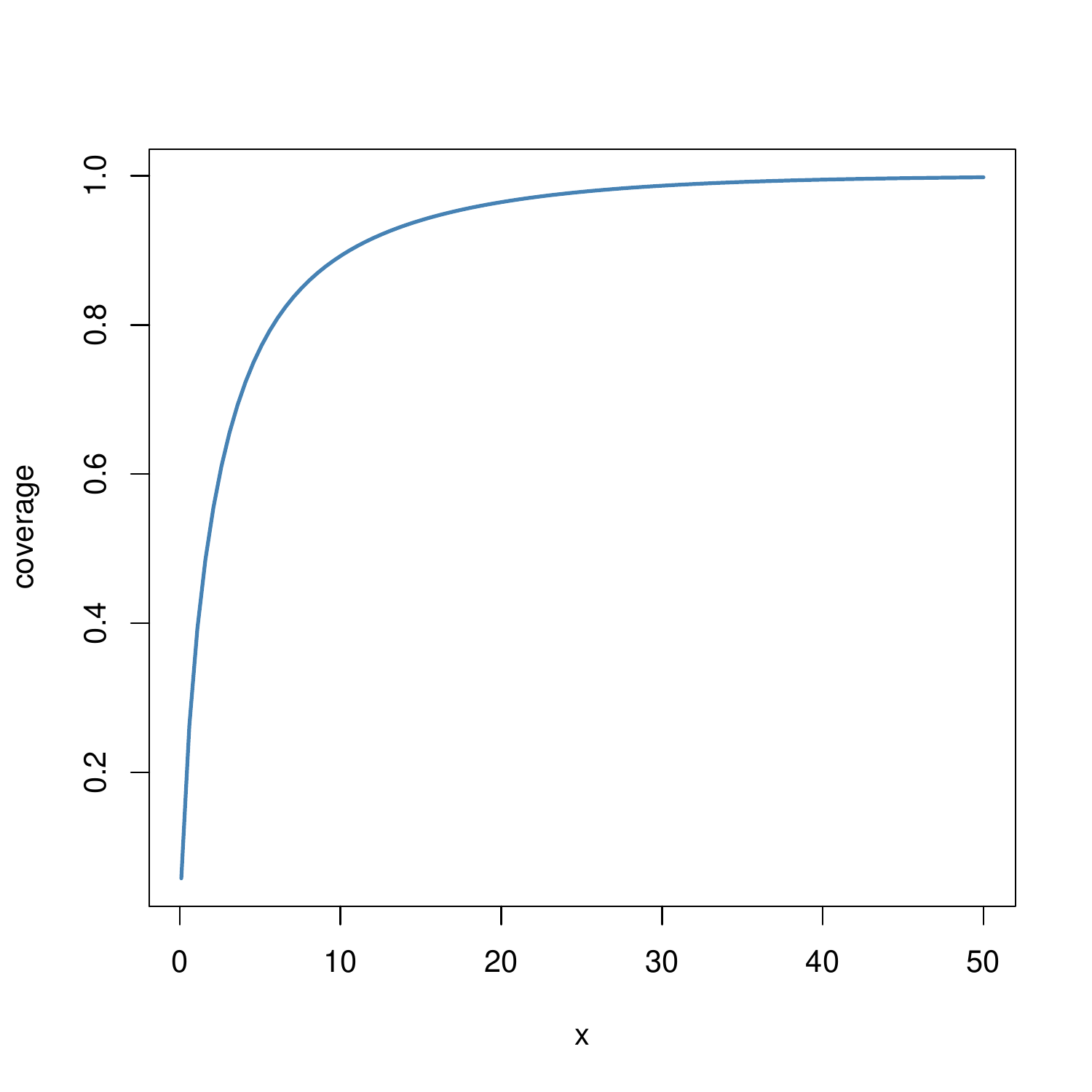}
\caption{\label{fig:betacova}Coverage of the interval $(\ell,u)=(0.1,0.6)$ by a 
$\mathscr{B}e(0.4\alpha ,0.6\alpha)$ distribution when $\alpha$ varies.}
\end{center}
\end{figure}

If $0<m<1$ is the mean $m=a/(a+b)$ of a beta $\mathscr{B}e(a,b)$ distribution, then this 
distribution is necessarily a beta $\mathscr{B}e(\alpha m,\alpha(1-m))$ distribution, with 
$\alpha>0$. For a given confidence interval $[\ell,u]$, with $0<\ell<m<u<1$, we have that
$$
\lim_{\alpha\to 0} \int_\ell^u \frac{\Gamma(\alpha)}{\Gamma(\alpha m)\Gamma(\alpha(1-m)}\,
x^{\alpha m-1}(1-x)^{\alpha(1-m)-1}\,\text{d}x = 0
$$
[since, when $\alpha$ goes to zero, the mass of the beta $\mathscr{B}e(\alpha m,\alpha(1-m))$
distribution gets more and more concentrated around $0$ and $1$, with masses $(1-m)$ and $m$,
respectively] and
$$
\lim_{\alpha\to \infty} \int_\ell^u \frac{\Gamma(\alpha)}{\Gamma(\alpha m)\Gamma(\alpha(1-m))}\,
x^{\alpha m-1}(1-x)^{\alpha(1-m)-1}\,\text{d}x = 1
$$
[this is easily established using the gamma representation introduced 
in Exercise \ref{exo:marcon,marcon} and the law of large numbers].
Therefore, due to the continuity [in $\alpha$] of the coverage probability, there must exist one
value of $\alpha$ such that
$$
B(\ell,u|\alpha,m) = \int_\ell^u \frac{\Gamma(\alpha)}{\Gamma(\alpha m)\Gamma(\alpha(1-m)}\,
x^{\alpha m-1}(1-x)^{\alpha(1-m)-1}\,\text{d}x = 0.9\,.
$$
Figure \ref{fig:betacova} illustrates this property by plotting $B(\ell,u|\alpha,m)$ for $\ell=0.1$,
$u=0.6$, $m=0.4$ and $\alpha$ varying from $0.1$ to $50$.

\begin{exoset}
Show that groups of consecutive unknown locations are independent of one another,
conditional on the observations. Devise a way to simulate these groups by blocks
rather than one at a time, that is, using the joint posterior distributions of the groups
rather than the full conditional distributions of the states.
\end{exoset}

As will become clearer in Chapter 7, the Arnason-Schwarz model is a very special case of
[partly] hidden Markov chain: the locations $z_{(i,t)}$ of an individual $i$ along time
constitute a Markov chain that is only observed at times $t$ when the individual is captured.
Whether or not $z_{(i,t)}$ is observed has no relevance on the fact that, given $z_{(i,t)}$,
$(z_{(i,t-1)},z_{(i,t-2)},\ldots)$ is independent from $(z_{(i,t+1)},z_{(i,t+2)},\ldots)$. 
Therefore, conditioning on any time $t$ and on the corresponding value of $z_{(i,t)}$ makes the
past and the future locations independent. In particular, conditioning on the observed locations
makes the blocks of unobserved locations in-between independent.

Those blocks could therefore be generated independently and parallely, an alternative 
which would then speed up the Gibbs sampler compared with the implementation in Algorithm 5.3.
In addition, this would bring additional freedom in the choice of the proposals for the simulation
of the different blocks and thus could further increase efficiency.

\chapter{Mixture Models}\label{ch:msg}

\begin{exoset}
Show that a mixture of Bernoulli distributions is again a Bernoulli distribution.
Extend this to the case of multinomial distributions.
\end{exoset}

By definition, if 
$$
x\sim\sum_{i=1}^k p_i \mathscr{B}(q_i)\,,
$$
then $x$ only takes the values $0$ and $1$ with probabilities
$$
\sum_{i=1}^k p_i (1-q_i) = 1  - \sum_{i=1}^k p_iq_i
\quad\text{and}\quad
\sum_{i=1}^k p_iq_i\,,
$$
respectively. This mixture is thus a Bernoulli distribution
$$
\mathscr{B}\left( \sum_{i=1}^k p_iq_i \right)\,.
$$

When considering a mixture of multinomial distributions,
$$
x\sim\sum_{i=1}^k p_i \mathscr{M}_k(\mathbf{q}_i)\,,
$$
with $\mathbf{q}_i=(q_{i1},\ldots,q_{ik})$,
$x$ takes the values $1\le j\le k$ with probabilities
$$
\sum_{i=1}^k p_i q_{ij}
$$
and therefore this defines a multinomial distribution.
This means that a mixture of multinomial distributions cannot be identifiable unless some
restrictions are set upon its parameters.

\begin{exoset}
Show that the number of nonnegative integer solutions of the decomposition of $n$ into $k$ parts
such that $n_1+\ldots+n_k=n$ is equal to
$$
\mathfrak{r}={n+k-1 \choose n} \,.
$$
Deduce that the number of partition sets is of order $\hbox{O}(n^{k-1})$.
\end{exoset}

This is a usual combinatoric result, detailed for instance in Feller (1970). 
A way to show that $\mathfrak{r}$ is the solution is to use the ``bottomless box" trick:
consider a box with $k$ cases and $n$ identical balls to put into those cases. If we remove
the bottom of the box, one allocation of the $n$ balls is represented by a sequence of balls (O)
and of case separations ($|$) or, equivalently, of $0$'s and $1$'s, of which there are $n$ and $k-1$
respectively [since the box itself does not count, we have to remove the extreme separations]. 
Picking $n$ positions out of $n+(k-1)$ is
exactly $\mathfrak{r}$.

This value is thus the number of ``partitions" of an $n$ sample into $k$ groups [we write
``partitions" and not partitions because, strictly speaking, all sets of a partition are non-empty]. 
Since
$$
{n+k-1 \choose n} = \frac{(n+k-1)!}{n!(k-1)!} \approx \frac{n^{k-1}}{(k-1)!}\,,
$$
when $n\gg k$, there is indeed an order $\hbox{O}(n^{k-1})$ of partitions.

\begin{exoset}\label{exo:2N}
For a mixture of two normal distributions with all parameters unknown,
$$
p\mathcal{N}(\mu_1,\sigma_1^2) +(1-p)\mathcal{N}(\mu_2,\sigma_2^2)\,,
$$
and for the prior distribution $(j=1,2)$
$$
\mu_j|\sigma_j\sim\mathscr{N}(\xi_j,\sigma_i^2/n_j)\,,\quad
\sigma_j^2\sim\mathscr{IG}(\nu_j/2,s_j^2/2)\,,\quad
p\sim\mathscr{B}e(\alpha,\beta)\,,
$$
show that
$$
p|\bx,\bz\sim\mathscr{B}e(\alpha+\ell_1,\beta+\ell_2),
$$
$$
\mu_j|\sigma_j,\bx,\bz \sim\mathscr{N}\left(\xi_1(\bz),{\sigma_j^2\over n_j+\ell_j}\right)\,,\quad \sigma_j^2|\bx,\bz \sim\mathscr{IG}((\nu_j+\ell_j)/2,s_j(\bz)/2)
$$
where $\ell_j$ is the number of $z_i$ equal to $j$, $\bar x_j(\bz)$ and $\hat s_j(\bz)$
are the empirical mean and variance for the subsample with $z_i$ equal to $j$, and
$$
\xi_j(\bz) = {n_j\xi_j+\ell_j \bar x_j(\bz) \over n_j+\ell_j}\,,\quad
s_j(\bz) = s^2_j+\hat s_j^2(\bz) + {n_j\ell_j\over n_j+\ell_j}
  (\xi_j-\bar x_j(\bz))^2\,.
$$
Compute the corresponding weight $\omega(\bz)$. % introduced in eqn.~(6.5).
\end{exoset}

If the latent (or missing) variable $\bz$ is introduced, the joint distribution of $(\bx,\bz)$
[equal to the completed likelihood] decomposes into
\begin{align}\label{eq:compmimi}
\prod_{i=1}^n p_{z_i}\,f(x_i|\theta_{z_i}) &= \prod_{j=1}^2 \prod_{i;z_i=j} p_j \,f(x_i|\theta_j)\nonumber\\
	&\propto \prod_{j=1}^k p_j^{\ell_j}\,\prod_{i;z_i=j} \frac{e^{-(x_i-\mu_j)^2/2\sigma_j^2}}{\sigma_j}\,,
\end{align}
where $p_1=p$ and $p_2=(1-p)$.
Therefore, using the conjugate priors proposed in the question, we have a decomposition of the posterior
distribution of the parameters given $(\bx,\bz)$ in
$$
p^{\ell_1+\alpha-1}(1-p)^{\ell2+\beta-1}\,\prod_{j=1}^2 \frac{e^{-(x_i-\mu_j)^2/2\sigma_j^2}}{\sigma_j}
\pi(\mu_j,\sigma_j^2)\,.
$$
This implies that $p|\bx,\bz\sim\mathscr{B}e(\alpha+\ell_1,\beta+\ell_2)$ and that the posterior distributions
of the pairs $(\mu_j,\sigma_j^2)$ are the posterior distributions associated with the normal observations allocated
(via the $z_i$'s) to the corresponding component. The values of the hyperparameters are therefore those already 
found in Chapter 2 (see, e.g., eqn.~(4.6) and Exercise 2.13).

The weight $\omega(\bz)$ is the marginal [posterior] distribution of $\bz$, since
$$
\pi(\btheta,p|\bx) = \sum_{\bz} \omega(\bz) \pi(\btheta,p|\bx,\bz)\,.
$$
Therefore, if $p_1=p$ and $p_2=1-p$,
\begin{align*}
\omega(\bz) &\propto \int \prod_{j=1}^2 p_j^{\ell_j}\,\prod_{i;z_i=j} 
\frac{e^{-(x_i-\mu_j)^2/2\sigma_j^2}}{\sigma_j} 
\pi(\btheta,p)\,\text{d}\btheta\text{d}p \\
&\propto %\frac{\Gamma(\alpha+\beta)}{\Gamma(\alpha)\Gamma(\beta)}\,
\frac{\Gamma(\alpha+\ell_1)\Gamma(\beta+\ell_2)}{\Gamma(\alpha+\beta+n)}\\
&\quad\int\,\prod_{j=1}^2 %\frac{\sqrt{n_j}}{(s_j^2/2)^{\nu_j/2}\Gamma(\nu_j/2)}\\
\exp\left[\frac{-1}{2\sigma_j^2}\left\{(n_j+\ell_j)(\mu_j-\xi_j(\bz))^2+s_j(\bz)
\right\}\right] \sigma_j^{-\ell_j-\nu_j-3}\,\text{d}\theta \\ 
&\propto %\frac{\Gamma(\alpha+\beta)}{\Gamma(\alpha)\Gamma(\beta)}\,
\frac{\Gamma(\alpha+\ell_1)\Gamma(\beta+\ell_2)}{\Gamma(\alpha+\beta+n)}\,
\prod_{j=1}^2 %\frac{\sqrt{n_j}}{(s_j^2/2)^{\nu_j/2}\Gamma(\nu_j/2)}\,
\frac{\Gamma((\ell_j+\nu_j)/2)(s_j(\bz)/2)^{(\nu_j+\ell_j)/2}}{\sqrt{n_j+\ell_j}}
\end{align*}
and the proportionality factor can be derived by summing up the rhs over all $\bz$'s.
(There are $2^n$ terms in this sum.) 

\begin{exoset}
For the normal mixture model of Exercise 6.3, compute the function
$Q(\theta_0,\theta)$ and derive both steps of the EM algorithm. Apply
this algorithm to a simulated dataset and test the influence of the starting point $\theta_0$.
\end{exoset}

Starting from the representation \eqref{eq:compmimi} above,
$$
\log \ell(\btheta,p|\bx,\bz) = \sum_{i=1}^n \left\{
\mathbb{I}_1(z_i)\log(p\,f(x_i|\theta_1)
+ \mathbb{I}_2(z_i)\log((1-p)\,f(x_i|\theta_2)\right\}\,,
$$
which implies that
\begin{align*}
Q\{(\btheta^{(t)},&p^{(t)}),(\btheta,p)\} 
= \mathbb{E}_{(\theta^{(t)},p^{(t)})}\left[\log \ell(\btheta,p|\bx,\bz)|\bx\right] \\
&= \sum_{i=1}^n \left\{
\text{P}_{(\theta^{(t)},p^{(t)})}\left( z_i=1|\bx \right)\log(p\,f(x_i|\btheta_1)\right.\\
&\qquad\left.+\text{P}_{(\btheta^{(t)},\bp^{(t)})}\left( z_i=2|\bx \right)\log((1-p)\,f(x_i|\btheta_2)\right\}\\
&= \log(p/\sigma_1) \sum_{i=1}^n \text{P}_{(\btheta^{(t)},p^{(t)})}\left( z_i=1|\bx \right)\\
&\quad +\log((1-p)/\sigma_2) \sum_{i=1}^n \text{P}_{(\btheta^{(t)},p^{(t)})}\left( z_i=2|\bx \right)\\
&\quad -\sum_{i=1}^n \text{P}_{(\btheta^{(t)},p^{(t)})}\left( z_i=1|\bx \right) \frac{(x_i-\mu_1)^2}{2\sigma^2_1}\\
&\quad -\sum_{i=1}^n \text{P}_{(\btheta^{(t)},p^{(t)})}\left( z_i=2|\bx \right) \frac{(x_i-\mu_2)^2}{2\sigma^2_2}\,.
\end{align*}
If we maximise this function in $p$, we get that
\begin{align*}
p^{(t+1)} &= \frac{1}{n}\,\sum_{i=1}^n \text{P}_{(\btheta^{(t)},p^{(t)})}\left( z_i=1|\bx \right)\\
&= \frac{1}{n}\,\sum_{i=1}^n \frac{p^{(t)} f(x_i|\btheta_1^{(t)})}{p^{(t)} f(x_i|\btheta_1^{(t)})+
	(1-p^{(t)}) f(x_i|\btheta_2^{(t)})}
\end{align*}
while maximising in $(\mu_j,\sigma_j)$ $(j=1,2)$ leads to
\begin{align*}
\mu_j^{(t+1)} &= \sum_{i=1}^n \text{P}_{(\btheta^{(t)},p^{(t)})}\left( z_i=j|\bx \right) x_i \bigg/ 
	\sum_{i=1}^n \text{P}_{(\btheta^{(t)},p^{(t)})}\left( z_i=j|\bx \right)\\
	      &= \frac{1}{np_j^{(t+1)}}\,\sum_{i=1}^n \frac{x_i p_j^{(t)} f(x_i|\btheta_j^{(t)})}{p^{(t)} 
	f(x_i|\btheta_1^{(t)})+ (1-p^{(t)}) f(x_i|\btheta_2^{(t)})} \,,\\
\sigma_j^{2(t+1)} &= \sum_{i=1}^n \text{P}_{(\btheta^{(t)},p^{(t)})}\left( z_i=j|\bx \right) 
	(x_i-\mu_j^{(t+1)})^2\bigg/ \sum_{i=1}^n \text{P}_{(\btheta^{(t)},p^{(t)})}\left( z_i=j|\bx \right)\\
   &= \frac{1}{np_j^{(t+1)}}\,\sum_{i=1}^n \frac{\left[x_i-\mu_j^{(t+1)}\right]^2 p_j^{(t)} f(x_i|\btheta_j^{(t)})}{p^{(t)}
        f(x_i|\btheta_1^{(t)})+ (1-p^{(t)}) f(x_i|\btheta_2^{(t)})}\,,
\end{align*}
where $p_1^{(t)}=p^{(t)}$ and $p_2^{(t)}=(1-p^{(t)})$.

A possible implementation of this algorithm in {\sf R} is given below:
\begin{verbatim}
# simulation of the dataset
n=324
tz=sample(1:2,n,prob=c(.4,.6),rep=T)
tt=c(0,3.5)
ts=sqrt(c(1.1,0.8))
x=rnorm(n,mean=tt[tz],sd=ts[tz])

para=matrix(0,ncol=50,nrow=5)
likem=rep(0,50)

# initial values chosen at random
para[,1]=c(runif(1),mean(x)+2*rnorm(2)*sd(x),rexp(2)*var(x))
likem[1]=sum(log( para[1,1]*dnorm(x,mean=para[2,1],
  sd=sqrt(para[4,1]))+(1-para[1,1])*dnorm(x,mean=para[3,1],
  sd=sqrt(para[5,1])) ))

# 50 EM steps
for (em in 2:50){

   # E step
   postprob=1/( 1+(1-para[1,em-1])*dnorm(x,mean=para[3,em-1],
     sd=sqrt(para[5,em-1]))/( para[1,em-1]*dnorm(x,
     mean=para[2,em-1],sd=sqrt(para[4,em-1]))) )

   # M step
   para[1,em]=mean(postprob)
   para[2,em]=mean(x*postprob)/para[1,em]
   para[3,em]=mean(x*(1-postprob))/(1-para[1,em])
   para[4,em]=mean((x-para[2,em])^2*postprob)/para[1,em]
   para[5,em]=mean((x-para[3,em])^2*(1-postprob))/(1-para[1,em])

   # value of the likelihood
   likem[em]=sum(log(para[1,em]*dnorm(x,mean=para[2,em],
     sd=sqrt(para[4,em]))+(1-para[1,em])*dnorm(x,mean=para[3,em],
     sd=sqrt(para[5,em])) ))
}
\end{verbatim}

\begin{figure}
\begin{center}
\includegraphics[width=.8\textwidth,height=6cm]{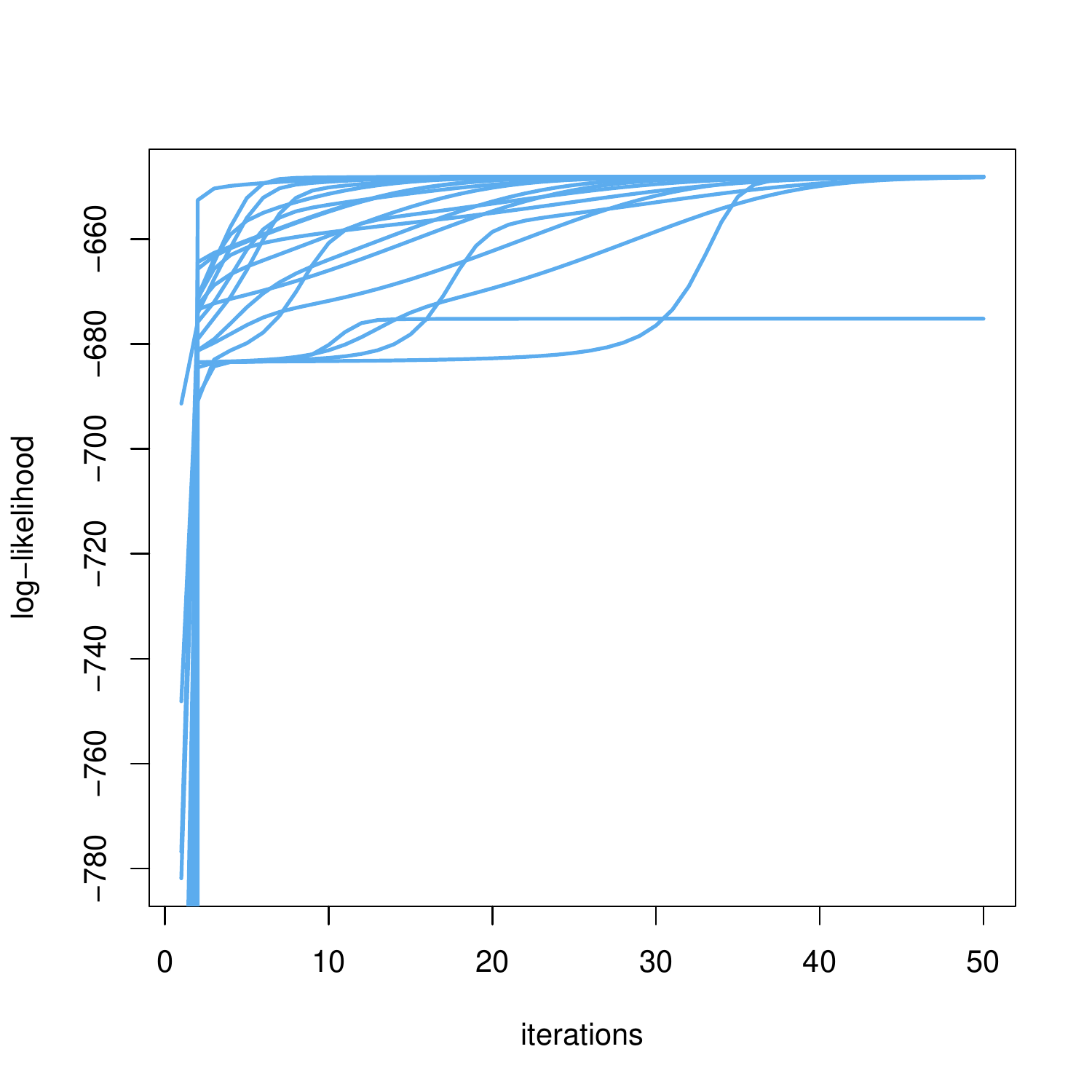}
\caption{\label{fig:let'em}Increase of the log-likelihood along EM iterations for
$20$ different starting points.}
\end{center}
\end{figure}

Figure \ref{fig:let'em} represents the increase in the log-likelihoods along EM iterations for
$20$ different starting points [and the same dataset $x$]. While most starting points lead to
the same value of the log-likelihood after $50$ iterations, one starting point induces a different
convergence behaviour.

\begin{exoset}
Show that the $\theta_j$'s in model (6.2) are dependent on each other given (only) $\bx$.
\end{exoset}

The likelihood associated with model (6.2) being
$$
\ell(\btheta,p|\bx)=\prod_{i=1}^n \left[ \sum_{j=1}^k p_j\,f(x_i|\btheta_j) \right]\,,
$$
it is clear that the posterior distribution will not factorise as a product of functions
of the different parameters. It is only given $(\bx,\bz)$ that the $\btheta_j$'s are independent.

\begin{exoset}\label{exo:newP}
Construct and test the Gibbs sampler associated with the $(\xi,\mu_0)$ parameterization of (6.3)
%\begin{equation}\label{eq:mix.mean}
%p\,\mathscr{N}(\mu_1,1)+(1-p)\,\mathscr{N}(\mu_2,1)\,,
%\end{equation}
when $\mu_1=\mu_0-\xi$ and $\mu_2=\mu_0+\xi$.
\end{exoset}

The simulation of the $z_i$'s is unchanged [since it does not depend on the
parameterisation of the components. The conditional distribution of $(\xi,\mu_0)$
given $(\bx,\bz)$ is
$$
\pi(\xi,\mu_0|\bx,\bz)\propto \exp\frac{-1}{2}\left\{ \sum_{z_i=1} (x_i-\mu_0+\xi)^2
+ \sum_{z_i=2} (x_i-\mu_0-\xi)^2 \right\}\,.
$$
Therefore, $\xi$ and $\mu_0$ are not independent given $(\bx,\bz)$, with
\begin{eqnarray*}
\mu_0|\xi,\bx,\bz &\sim& \mathscr{N}\left( \frac{n\overline x +(\ell_1-\ell_2)\xi}{n},\frac{1}{n}\right)\,,\\
\xi|\mu_0,\bx,\bz &\sim& \mathscr{N}\left( \frac{\sum_{z_i=2} (x_i-\mu_0) - \sum_{z_i=1} (x_i-\mu_0) }{n},\frac{1}{n}\right)
\end{eqnarray*}

The implementation of this Gibbs sampler is therefore a simple modification of the code given in \verb+#6.R+ on the
webpage: the MCMC loop is now
\begin{verbatim}
for (t in 2:Nsim){

  # allocation
  fact=.3*sqrt(exp(gu1^2-gu2^2))/.7
  probs=1/(1+fact*exp(sampl*(gu2-gu1)))
  zeds=(runif(N)<probs)

  # Gibbs sampling
  mu0=rnorm(1)/sqrt(N)+(sum(sampl)+xi*(sum(zeds==1)
    -sum(zeds==0)))/N
  xi=rnorm(1)/sqrt(N)+(sum(sampl[zeds==0]-mu0)
    -sum(sampl[zeds==1]-mu0))/N

  # reparameterisation
  gu1=mu0-xi
  gu2=mu0+xi
  muz[t,]=(c(gu1,gu2))

}
\end{verbatim}

\begin{figure}
\begin{center}
\includegraphics[width=.8\textwidth,height=7cm]{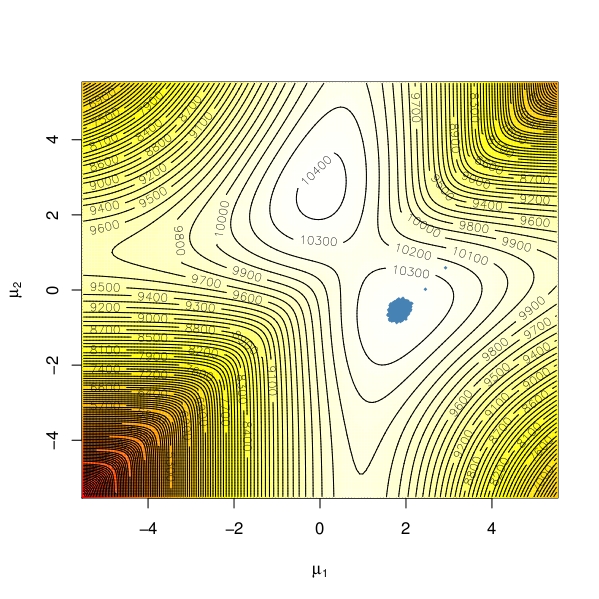}\\
\includegraphics[width=.8\textwidth,height=7cm]{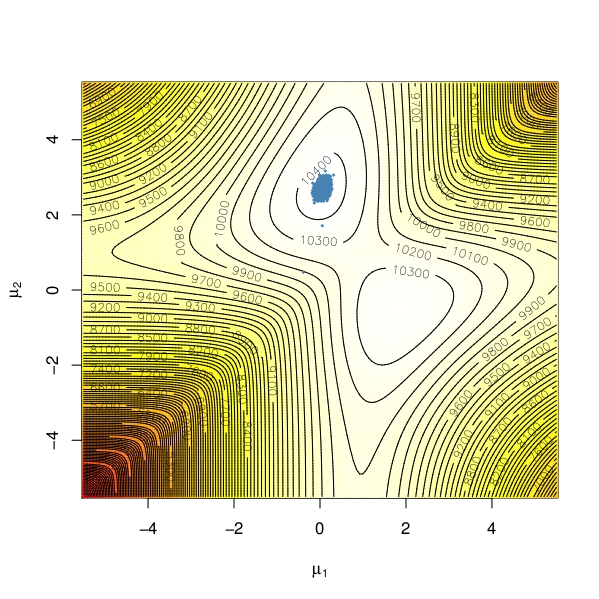}
\caption{\label{fig:captimode}Influence of the starting value on the convergence
of the Gibbs sampler associated with the location parameterisation of the mean
mixture ($10,000$ iterations).}
\end{center}
\end{figure}

If we run repeatedly this algorithm, the Markov chain produced is highly dependent on
the starting value and remains captive of local modes, as illustrated on Figure \ref{fig:captimode}.
This reparameterisation thus seems less robust than the original parameterisation.

\begin{exoset}
Give the ratio corresponding to (6.7)
%$$
%\frac{\pi(\widetilde{w_j})}{\pi\left(w_j^{(t-1)}\right)}\,\frac{\widetilde{w_j}}{w_j^{(t-1)}}\wedge 1
%$$
when the parameter of interest is in $[0,1]$ and the random walk 
proposal is on the logit transform $\log\theta/(1-\theta)$.
\end{exoset}

Since
$$
\frac{\partial }{\partial\theta} \log\left[\theta/(1-\theta)\right] = 
\frac{1}{\theta}+\frac{1}{1-\theta)} =
\frac{1}{\theta(1-\theta)}\,,
$$
the Metropolis--Hastings acceptance ratio for the logit transformed random walk is
$$
\frac{\pi(\widetilde{\theta_j})}{\pi(\theta_j^{(t-1)})}\,
\frac{\widetilde{\theta_j}(1-\widetilde{\theta_j})}{\theta_j^{(t-1)}(1-\theta_j^{(t-1)})}\wedge 1\,.
$$

\begin{exoset}
Show that, if an exchangeable prior $\pi$ is used on the vector of weights $(p_1,\ldots,p_k)$, 
then, necessarily, $\mathbb{E}^\pi[p_j]=1/k$ and, if the prior on the other parameters
$(\theta_1,\ldots,\theta_k)$ is also exchangeable, then $\mathbb{E}^\pi[p_j|x_1,\ldots,x_n]=1/k$ 
for all $j$'s.
\end{exoset}

If 
$$
\pi(p_1,\ldots,p_k) = \pi(p_{\sigma(1)},\ldots,p_{\sigma(k)})
$$
for any permutation $\sigma\in\mathfrak{S}_k$, then
$$
\mathbb{E}^\pi[p_j]=\int p_j \pi(p_1,\ldots,p_j,\ldots,p_k) \,\text{d}\mathbf{p}
=\int p_j \pi(p_j,\ldots,p_1,\ldots,p_k) \,\text{d}\mathbf{p} = \mathbb{E}^\pi[p_1]\,.
$$
Given that $\sum_{j=1}^k p_j=1$, this implies $\mathbb{E}^\pi[p_j]=1/k$.

When both the likelihood and the prior are exchangeable in $(p_j,\theta_j)$, the same
result applies to the posterior distribution.

\begin{exoset}
Show that running an MCMC algorithm with target $\pi(\theta|\bx)^\gamma$
will increase the proximity to the MAP estimate when $\gamma>1$ is large.
Discuss the modifications required in Algorithm 6.2.%\ref{al:mix.mean}
to achieve simulation from $\pi(\theta|\bx)^\gamma$ when $\gamma\in\mathbb{N}^*$.
\end{exoset}

The power distribution $\pi_\gamma(\theta)\propto\pi(\theta)^\gamma$ shares the same modes as $\pi$, 
but the global mode gets more and more mass as $\gamma$ increases. If $\theta^\star$ is the global
mode of $\pi$ [and of $\pi_\gamma$], then $\{\pi(\theta)/\pi(\theta^\star)\}^\gamma$ goes to $0$ as $\gamma$ goes to $\infty$
for all $\theta$'s different from $\theta^\star$. Moreover, for any $0<\alpha<1$, if we define the $\alpha$
neighbourhood $\mathfrak{N}_\alpha$ of $\theta^\star$ as the set of $\theta$'s such that $\pi(\theta)\ge \alpha 
\pi(\theta^\star)$, then $\pi_\gamma(\mathfrak{N}_\alpha)$ converges to $1$ as $\gamma$ goes to $\infty$.

The idea behind {\em simulated annealing} is that, first, the distribution $\pi_\gamma(\theta)\propto\pi(\theta)^\gamma$
is more concentrated around its main mode than $\pi(\theta)$ if $\gamma$ is large and, second, that it is not
necessary to simulate a whole sample from $\pi(\theta)$, then a whole sample from $\pi(\theta)^2$ and so on to achieve
a convergent approximation of the MAP estimate. Increasing $\gamma$ slowly enough along iterations leads to the same
result with a much smaller computing requirement.

When considering the application of this idea to a mean mixture as (6.3) [in the book], the modification of
Algorithm 6.2 is rather immediate: since we need to simulate from $\pi(\btheta,p|\bx)^\gamma$ [up to a 
normalising constant], this is equivalent to simulate from $\ell(\btheta,p|\bx)^\gamma\times\pi(\btheta,p)^\gamma$.
This means that, since the prior is [normal] conjugate, the prior hyperparameter $\lambda$ is modified into $\gamma\lambda$
and that the likelihood is to be completed $\gamma$ times rather than once, i.e.
$$
\ell(\btheta,p|\bx)^\gamma = \left( \int f(\bx,\bz|\btheta,p)\,\text{d}\bz \right)^\gamma
= \prod_{j=1}^\gamma \int f(\bx,\bz_j|\btheta,p)\,\text{d}\bz_j\,.
$$
Using this duplication trick, the annealed version of Algorithm 6.2 writes as
\begin{algo} {\bf Annealed Mean Mixture Gibbs Sampler}
\begin{itemize}
\item[]  {\sffamily Initialization.} Choose $\mu_1^{(0)}$ and $\mu_2^{(0)}$,
\item[]  {\sffamily Iteration $t$ $(t\ge 1)$.}
\begin{enumerate}
\item[{\sf 1.}] For $i=1,\ldots,n$, $j=1,\ldots,\gamma$, generate $z_{ij}^{(t)}$ from
\begin{eqnarray*}
\mathbb{P}\left(z_{ij}=1\right)&\propto& p\,\exp\left\{-\frac{1}{2}\left(x_i-\mu_1^{(t-1)}\right)^2\right\}\\
\mathbb{P}\left(z_{ij}=2\right)&\propto& (1-p)\,\exp\left\{-\frac{1}{2}\left(x_i-\mu_2^{(t-1)}\right)^2\right\}
\end{eqnarray*}
\item[{\sf 2.}] Compute 
$$
\ell=\sum_{j=1}^\gamma\sum_{i=1}^n\mathbb{I}_{z_{ij}^{(t)}=1}
\quad\text{and}\quad
\bar x_u\left(\bz\right)=\sum_{j=1}^\gamma\sum_{i=1}^n\mathbb{I}_{z_{ij}^{(t)}=u}x_i
$$
\item[{\sf 3.}] Generate $\mu_1^{(t)}$ from
$\displaystyle \mathscr{N}\left(\frac{\gamma\lambda\delta+bar x_1\left(\bz\right)}
{\gamma\lambda+\ell},\frac{1}{\gamma\lambda+\ell}\right)$
\item[{\sf 4.}] Generate $\mu_2^{(t)}$ from $\displaystyle \mathscr{N}
\left(\frac{\gamma\lambda\delta+\bar x_2\left(\bz\right)}{\gamma\lambda+\gamma n-\ell},
\frac{1}{\gamma \lambda+\gamma n-\ell}\right)$.
\end{enumerate}
\end{itemize}
\end{algo}
This additional level of completion means that the Markov chain will have difficulties to move around, compared
with the original Gibbs sampling algorithm. While closer visits to the global mode are guaranteed in theory, they
may require many more simulations in practice. 

\begin{exoset}
In the setting of the mean mixture (6.3), run an MCMC simulation experiment to
compare the influence of a $\mathscr{N}(0,100)$ and of a $\mathscr{N}(0,10000)$ prior on
$(\mu_1,\mu_2)$ on a sample of $500$ observations.
\end{exoset}

This is straightforward in that the code in \verb+#6.R+ simply needs to be modified from 
\begin{verbatim}
# Gibbs samplin
  gu1=rnorm(1)/sqrt(.1+length(zeds[zeds==1]))+
   (sum(sampl[zeds==1]))/(.1+length(zeds[zeds==1]))
  gu2=rnorm(1)/sqrt(.1+length(zeds[zeds==0]))+
   (sum(sampl[zeds==0]))/(.1+length(zeds[zeds==0]))
\end{verbatim}
to
\begin{verbatim}
# Gibbs samplin
  gu1=rnorm(1)/sqrt(.01+length(zeds[zeds==1]))+
    (sum(sampl[zeds==1]))/(.01+length(zeds[zeds==1]))
  gu2=rnorm(1)/sqrt(.01+length(zeds[zeds==0]))+
    (sum(sampl[zeds==0]))/(.01+length(zeds[zeds==0]))
\end{verbatim}
for the $\mathscr{N}(0,100)$ prior and to
\begin{verbatim}
# Gibbs samplin
  gu1=rnorm(1)/sqrt(.0001+length(zeds[zeds==1]))+
    (sum(sampl[zeds==1]))/(.0001+length(zeds[zeds==1]))
  gu2=rnorm(1)/sqrt(.0001+length(zeds[zeds==0]))+
    (sum(sampl[zeds==0]))/(.0001+length(zeds[zeds==0]))
\end{verbatim}
for the $\mathscr{N}(0,10^4)$ prior. While we do not reproduce the results here, it appears that
the sampler associated with the $\mathscr{N}(0,10^4)$ prior has a higher probability to escape the
dubious mode.

\begin{exoset}
Show that, for a normal mixture $0.5\,\mathscr{N}(0,1)+0.5\,\mathscr{N}(\mu,\sigma^2)$,
the likelihood is unbounded. Exhibit this feature by plotting the likelihood of a simulated
sample, using the {\sf R image} procedure.
\end{exoset}

This follows from the decomposition of the likelihood
$$
\ell(\btheta|\bx)=\prod_{i=1}^n \left[ \sum_{j=1}^2 0.5\,f(x_i|\btheta_j) \right]\,,
$$
into a sum [over all partitions] of the terms 
$$
\prod_{i=1}^n f(x_i|\btheta_{z_i})
=\prod_{i;z_i=1} \varphi(x_i) \prod_{i;z_i=2} \frac{\varphi\{(x_i-\mu)/\sigma\}}{\sigma}\,.
$$
In exactly $n$ of those $2^n$ partitions, a single observation is allocated to the second component,
i.e.~there is a single $i$ such that $z_i=2$. For those particular partitions, if we choose $\mu=x_i$,
the second product reduces to $1/\sigma$ which is not bounded when $\sigma$ goes to $0$. Since the
observed likelihood is the sume of all those terms, it is bounded from below by terms that are unbounded
and therefore it is unbounded.

\begin{figure}
\begin{center}
\includegraphics[width=.8\textwidth,height=7cm]{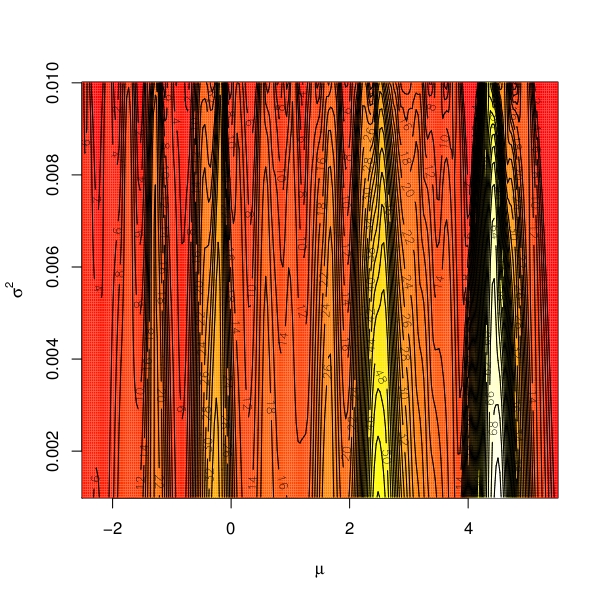}
\caption{\label{fig:infimix}Illustration of an unbounded mixture likelihood.}
\end{center}
\end{figure}

An {\sf R} code illustrating this behaviour is
\begin{verbatim}
# Sample construction
N=100
sampl=rnorm(N)+(runif(N)<.3)*2.7

# Grid
mu=seq(-2.5,5.5,length=250) 
sig=rev(1/seq(.001,.01,length=250))  # inverse variance
mo1=mu%*%t(rep(1,length=length(sig)))
mo2=(rep(1,length=length(mu)))%*%t(sig)
ca1=-0.5*mo1^2*mo2
ca2=mo1*mo2
ca3=sqrt(mo2)
ca4=0.5*(1-mo2)

# Likelihood surface
like=0*mo1
for (i in 1:N)
  like=like+log(1+exp(ca1+sampl[i]*ca2+sampl[i]^2*ca4)*ca3)
like=like-min(like)

sig=rev(1/sig)
image(mu,sig,like,xlab=expression(mu),
  ylab=expression(sigma^2),col=heat.colors(250))
contour(mu,sig,like,add=T,nlevels=50)

\end{verbatim}
and Figure \ref{fig:infimix} exhibits the characteristic stripes of an explosive likelihood as
$\sigma$ approaches $0$ for values of $\mu$ close to the values of the sample.

\begin{exoset}
Show that the ratio (6.8)
%$$
%\left(\frac{\pi(\btheta^\prime,\bp^\prime|\bx)}
     %{\pi(\btheta,\bp|\bx)}\right)^\alpha\,
%\frac{q(\btheta,\bp|\btheta^\prime,\bp^\prime)}
     %{q(\btheta^\prime,\bp^\prime|\btheta,\bp)}\wedge 1
%$$
goes to $1$ when $\alpha$ goes to $0$ when the proposal $q$ is a random walk. 
Describe the average behavior of this ratio in the case of an independent proposal.
\end{exoset}

This is obvious since, when the proposal is a random walk [without reparameterisation], 
the ratio $q(\btheta,\bp|\btheta^\prime,\bp^\prime)/q(\btheta^\prime,\bp^\prime|\btheta,\bp)$
is equal to $1$. The Metropolis--Hastings ratio thus reduces to the ratio of the targets to
the power $\alpha$, which [a.e.] converges to $1$ as $\alpha$ goes to $0$.

In the case of an independent proposal,
$$
\left(\frac{\pi(\btheta^\prime,\bp^\prime|\bx)}
     {\pi(\btheta,\bp|\bx)}\right)^\alpha\,
\frac{q(\btheta,\bp)}{q(\btheta^\prime,\bp^\prime)}\wedge 1
$$
is equivalent to $q(\btheta,\bp)/q(\btheta^\prime,\bp^\prime)$ and therefore does not
converge to $1$. This situation can however be avoided by picking $q^\alpha$ rather than $q$,
in which case the ratio once more converges to $1$ as $\alpha$ goes to $0$.

\begin{exoset}
If one needs to use importance sampling weights, show that the
simultaneous choice of several powers $\alpha$ requires
the computation of the normalizing constant of $\pi_\alpha$.
\end{exoset}

If samples $(\theta_{i\alpha})_i$ from several tempered versions $\pi_\alpha$ of $\pi$ are
to be used simultaneously, the importance weights associated with those samples $\pi(\theta_{i\alpha})/
\pi_\alpha(\theta_{i\alpha})$ require the computation of the normalizing constants, which is
most often impossible. This difficulty explains the appeal of the ``pumping mechanism" of Algorithm
6.5, which cancels the need for normalizing constants by using the same $\pi_\alpha$ twice, once in
the numerator and once in the denominator (see Exercice \ref{ex:dessous}).

\begin{exoset}\label{ex:dessous}
Check that Algorithm 6.5 does not require the normalizing constants of the $\pi_{\alpha_i}$'s
and show that $\pi$ is the corresponding stationary distribution. 
\end{exoset}

Since the acceptance probability
$$
\min \left\{ 1 , \frac{\pi_{\alpha_1}(x^{(t)})}{\pi(x^{(t)})} \cdots
\frac{\pi_{\alpha_p}(x_{p-1}^{(t)})}{\pi_{\alpha_{p-1}}(x_{p-1}^{(t)})}
\frac{\pi_{\alpha_{p-1}}(x_{p}^{(t)})}{\pi_{\alpha_p}(x_{p}^{(t)})}
        \cdots
\frac{\pi(x_{2p-1}^{(t)})}{\pi_{\alpha_1}(x_{2p-1}^{(t)})} \right\}
$$
uses twice each power $\alpha_j$ of $\pi$, the unknown normalizing constants of the $\pi_{\alpha_i}$'s
vanish, which is one of the main reasons for using this algorithm. 

The fact that $\pi$ is
stationary can be derived from the ``detailed balance" condition: let us assume that each of the
MCMC kernels satisfy the corresponding detailed balance equation
$$
\pi_{\alpha_j}(x_0)\hbox{MCMC}(x_1|x_0,\pi_{\alpha_j}) 
= \pi_{\alpha_j}(x_1)\hbox{MCMC}(x_0|x_1,\pi_{\alpha_j})\,.
$$
Then
\begin{align*}
\pi(x^{(t)})&\hbox{MCMC}(x_1^{(t)}|x^{(t)},\pi_{\alpha_1})\cdots
\hbox{MCMC}(x_{2p}^{(t)}|x_{2p-1}^{(t)},\pi_{\alpha_1})\\
&\times \min \left\{ 1 , \frac{\pi_{\alpha_1}(x^{(t)})}{\pi(x^{(t)})} \cdots
\frac{\pi_{\alpha_p}(x_{p-1}^{(t)})}{\pi_{\alpha_{p-1}}(x_{p-1}^{(t)})}
\frac{\pi_{\alpha_{p-1}}(x_{p}^{(t)})}{\pi_{\alpha_p}(x_{p}^{(t)})}
        \cdots
\frac{\pi(x_{2p-1}^{(t)})}{\pi_{\alpha_1}(x_{2p-1}^{(t)})} \right\}\\
&=\pi_{\alpha_1}(x^{(t)}) \hbox{MCMC}(x_1^{(t)}|x^{(t)},\pi_{\alpha_1})\cdots
\hbox{MCMC}(x_{2p}^{(t)}|x_{2p-1}^{(t)},\pi_{\alpha_1})\\
&\times \min \left\{\frac{\pi(x^{(t)})}{\pi_{\alpha_1}(x^{(t)})},
\frac{\pi_{\alpha_2}(x_1^{(t)})}{\pi_{\alpha_1}(x_1^{(t)})} \cdots 
\frac{\pi(x_{2p-1}^{(t)})}{\pi_{\alpha_1}(x_{2p-1}^{(t)})} \right\}\\
&=\hbox{MCMC}(x^{(t)}|x_1^{(t)},\pi_{\alpha_1}) \pi_{\alpha_1}(x_1^{(t)}) 
\hbox{MCMC}(x_2^{(t)}|x_1^{(t)},\pi_{\alpha_2})\cdots \\
&\times \hbox{MCMC}(x_{2p}^{(t)}|x_{2p-1}^{(t)},\pi_{\alpha_1})
\min \left\{\frac{\pi(x^{(t)})}{\pi_{\alpha_1}(x^{(t)})},
\frac{\pi_{\alpha_2}(x_1^{(t)})}{\pi_{\alpha_1}(x_1^{(t)})} \cdots
\frac{\pi(x_{2p-1}^{(t)})}{\pi_{\alpha_1}(x_{2p-1}^{(t)})} \right\}\\
&=\hbox{MCMC}(x^{(t)}|x_1^{(t)},\pi_{\alpha_1})
\hbox{MCMC}(x_1^{(t)}|x_2^{(t)},\pi_{\alpha_2})\pi_{\alpha_2}(x_2^{(t)}) \cdots \\
&\times \hbox{MCMC}(x_{2p}^{(t)}|x_{2p-1}^{(t)},\pi_{\alpha_1})
\min \left\{\frac{\pi(x^{(t)})\pi_{\alpha_1}(x_1^{(t)})}{\pi_{\alpha_1}(x^{(t)})\pi_{\alpha_2}(x_1^{(t)})},
\frac{\pi_{\alpha_3}(x_2^{(t)})}{\pi_{\alpha_2}(x_2^{(t)})} \cdots \right\}\\
%\frac{\pi(x_{2p-1}^{(t)})}{\pi^{\alpha_1}(x_{2p-1}^{(t)})} \right\}\\
&= \cdots \\
&=\hbox{MCMC}(x^{(t)}|x_1^{(t)},\pi_{\alpha_1})\cdots
\hbox{MCMC}(x_{2p-1}^{(t)}|x_{2p}^{(t)},\pi_{\alpha_1}) \pi(x_{2p}^{(t)})\\
&\times \min \left\{\frac{\pi(x^{(t)})}{\pi_{\alpha_1}(x^{(t)})}\cdots
\frac{\pi_{\alpha_1}(x_{2p}^{(t)})}{\pi(x_{2p}^{(t)})},1 \right\}
\end{align*}
by a ``domino effect" resulting from the individual detailed balance conditions.
This generalised detailed balance condition then ensures that $\pi$ is the stationary
distribution of the chain $(x^{(t)})_t$.

\begin{exoset}
Show that the decomposition (6.9)
%\begin{align*}
%\mathbb{E}^\pi&[\mathbb{E}[x]|x_1,\ldots,x_n] =\nonumber\\
%&\sum_k \hbox{P}({\mathfrak M}_k|x_1,\ldots,x_n)
%\iint x\,f_k(x|\theta_k) \hbox{d}x\,
%\pi_k(\theta_k|x_1,\ldots,x_n) \hbox{d}\theta
%\end{align*}
is correct by representing the generic parameter
$\theta$ as $(k,\theta_k)$ and by introducing the submodel marginals,
$m_k(x_1,\ldots,x_n) = \int f_k(x_1,\ldots,x_n|\theta_k) \pi_k(\theta_k)\, \hbox{d}\theta_k$.
\end{exoset}

In a variable dimension model, the sampling distribution is
$$
f(x_1,\ldots,x_n|\theta) = f(x_1,\ldots,x_n|(k,\theta_k)) = f_k(x_1,\ldots,x_n|\theta_k).
$$
Decomposing the prior distribution as 
$$
\pi(\theta) = \pi((k,\theta_k)) = \text{P}({\mathfrak M}_k) \pi_k(\theta_k)\,,
$$
the joint distribution of $(x_1,\ldots,x_n)$ and of $\theta$ is
$$
f(x_1,\ldots,x_n|\theta)\pi(\theta) = \text{P}({\mathfrak M}_k) \pi_k(\theta_k) f_k(x_1,\ldots,x_n|\theta_k)
$$
and the marginal distribution of $(x_1,\ldots,x_n)$ is therefore derived by
\begin{align*}
m(x_1,\ldots,x_n) &= \sum_k \int \text{P}({\mathfrak M}_k) \pi_k(\theta_k) f_k(x_1,\ldots,x_n|\theta_k) \text{d}\theta_k\\
&= \sum_k \text{P}({\mathfrak M}_k) m_k(x_1,\ldots,x_n)\,.
\end{align*}
Similarly, the predictive distribution $f(x|x_1,\ldots,x_n)$ can be expressed as
\begin{align*}
f(x|x_1,\ldots,x_n) &= \int f(x|\theta) \pi(\theta|x_1,\ldots,x_n)\,\text{d}\theta\\
&= \sum_k \int f_k(x|\theta_k) \frac{\text{P}({\mathfrak M}_k) \pi_k(\theta_k)}{m(x_1,\ldots,x_n)} \,\text{d}\theta_k\\
&= \sum_k \frac{\text{P}({\mathfrak M}_k) m_k(x_1,\ldots,x_n)}{m(x_1,\ldots,x_n)}\,
\int f_k(x|\theta_k) \frac{\pi_k(\theta_k)}{m_k(x_1,\ldots,x_n)} \,\text{d}\theta_k\\
&= \sum_k \hbox{P}({\mathfrak M}_k|x_1,\ldots,x_n)
\int f_k(x|\theta_k) \pi_k(\theta_k|x_1,\ldots,x_n) \,\text{d}\theta_k\,.
\end{align*}
Therefore,
$$
\mathbb{E}[x|x_1,\ldots,x_n] = \sum_k \hbox{P}({\mathfrak M}_k|x_1,\ldots,x_n)
\iint f_k(x|\theta_k) \pi_k(\theta_k|x_1,\ldots,x_n) \,\text{d}\theta_k\,.
$$

\begin{exoset}
For a finite collection of submodels  ${\mathfrak M}_k$ $(k=1,\ldots,K)$, with
respective priors $\pi_k(\theta_k)$ and weights $\varrho_k$, write a generic
importance sampling algorithm that approximates the posterior distribution.
\end{exoset}

The formal definition of an importance sampler in this setting is straightforward: all
that is needed is a probability distribution $(\omega_1,\ldots,\omega_K)$ and a collection
of importance distributions $\eta_k$ with supports at least as large as $\text{supp}(\pi_k)$.
The corresponding importance algorithm is then made of the three following steps:
\begin{algo}
{\bf Importance Sampled Model Choice}
\begin{enumerate}
\item Generate $k\sim (\omega_1,\ldots,\omega_K)$;
\item Generate $\theta_k \sim \eta_k(\theta_k)$;
\item Compute the importance weight $\varrho_k \pi_k(\theta_k)/\omega_k\eta_k(\theta_k)$\,.
\end{enumerate}
\end{algo}
Obviously, the difficulty in practice is to come up with weights $\omega_k$ that are not too
different from the $\varrho_k$'s [for efficiency reasons] while selecting pertinent importance
distributions $\eta_k$. This is most often impossible, hence the call to reversible jump techniques
that are more local and thus require less information about the target.

\begin{exoset}\label{ex:2pi}
Show that, if we define the acceptance probability
$$
\varrho = \frac{\pi_2(x^\prime)}{\pi_1(x)}\,\frac{q(x|x^\prime)}{q(x^\prime|x)}\wedge 1
$$
for moving from $x$ to $x^\prime$ and
$$
\varrho^\prime = \frac{\pi_1(x)}{\pi_2(x^\prime)}\,\frac{q(x^\prime|x)}{q(x|x^\prime)}\wedge 1
$$
for the reverse move, the detailed balance condition is modified in such a way that,
if $X_t\sim\pi_1(x)$ and if a proposal is made based on $q(x|x_t)$, $X_{t+1}$ is distributed
from $\pi_2(x)$.  Relate this property to Algorithm 6.5 and its acceptance probability.
\end{exoset}

If $K$ denotes the associated Markov kernel, we have that
\begin{align*}
\pi_1(x) K(x,x^\prime) &= \pi_1(x)\left\{ q(x^\prime|x) \varrho(x,x^\prime) 
+ \delta_x(x^\prime) \int q(z|x) [1-\varrho(x,z)] \text{d}z \right\} \\
&=  \min\left\{ \pi_1(x) q(x^\prime|x), \pi_1(x^\prime) q(x|x^\prime) \right\}\\
&\ + \delta_x(x^\prime) \int \max\left\{0,q(z|x)\pi_1(x)-q(x|z)\pi_2(x)\right\} \text{d}z \\
&= \pi_2(x) \widetilde{K}(x^\prime,x)
\end{align*}
under the assumption that the reverse acceptance probability for the reverse move 
is as proposed [a rather delicate assumption that makes the whole exercise 
definitely less than rigorous].

In Algorithm 6.5, the derivation is perfectly sound because the kernels are used twice, once
forward and once backward.

\begin{exoset}
Show that the marginal distribution of $p_1$ when $(p_1,\ldots,p_k)\sim
\mathscr{D}_k(a_1,\ldots,a_k)$ is a $\mathscr{B}e(a_1,a_2+\ldots+a_k)$ distribution.
\end{exoset}

This result relies on the same representation as Exercise \ref{exo:marcon,marcon}: if
$(p_1,\ldots,p_k)\sim \mathscr{D}_k(a_1,\ldots,a_k)$, $p_1$ is distributed identically to
$$
\frac{\xi_1}{\xi_1+\ldots+\xi_k}\,,\quad \xi_j\sim\mathscr{G}a(a_j)\,.
$$
Since $\xi_2+\ldots+\xi_k\sim\mathscr{G}a(a_2+\cdots+a_k)$, this truly corresponds
to a $\mathscr{B}e(a_1,a_2+\ldots+a_k)$ distribution.

\chapter{Dynamic Models}\label{ch:dyn}

\begin{exoset}
Consider the process $(x_t)_{t\in\mathbb{Z}}$ defined by
$$
x_t=a+bt+y_t\,,
$$
where $(y_t)_{t\in\mathbb{Z}}$ is an iid sequence of random variables with mean
$0$ and variance $\sigma^2$, and where $a$ and $b$ are constants. Define
$$
w_t=(2q+1)^{-1}\sum_{j=-q}^q x_{t+j}\,.
$$
Compute the mean and the autocovariance function of $(w_t)_{t\in\mathbb{Z}}$. Show that
$(w_t)_{t\in\mathbb{Z}}$ is not stationary but that its autocovariance function $\gamma_w(t+h,t)$
does not depend on $t$.
\end{exoset}

We have
\begin{eqnarray*}
\mathbb{E}[w_t] & = & \mathbb{E}\left[(2q+1)^{-1}\sum_{j=-q}^q x_{t+j}\right] \\
                & = & (2q+1)^{-1} \sum_{j=-q}^q \mathbb{E}\left[a+b(t+j)+y_t\right] \\
                & = &  a+bt\,.
\end{eqnarray*}
The process $(w_t)_{t\in\mathbb{Z}}$ is therefore not stationary. Moreover
\begin{eqnarray*}
\mathbb{E}[w_tw_{t+h}] & = & \mathbb{E}\left[\left(a+bt+\frac{1}{2q+1}\sum_{j=-q}^q y_{t+j}\right)
                             \left(a+bt+bh+\sum_{j=-q}^q y_{t+h+j}\right)\right] \\
                       & = & (a+bt)(a+bt+bh)+\mathbb{E}\left[ \sum_{j=-q}^q y_{t+j}\sum_{j=-q}^q y_{t+h+j}\right] \\
                       & = & (a+bt)(a+bt+bh)+\mathbb{I}_{|h|\leq q}(q+1-|h|)\sigma^2\,.
\end{eqnarray*}
Then,
\begin{equation*}
\text{cov}(w_t,w_{t+h})=\mathbb{I}_{|h|\leq q}(q+1-|h|)\sigma^2
\end{equation*}
and,
\begin{equation*}
\gamma_w(t+h,t)=\mathbb{I}_{|h|\leq q}(q+1-|h|)\sigma^2\,.
\end{equation*}

\begin{exoset}
Suppose that the process $(x_t)_{t\in\mathbb{N}}$ is such that $x_0\sim\mathscr{N}(0,\tau^2)$ 
and, for all $t\in\mathbb{N}$,
$$
x_{t+1}|\bx_{0:t} \sim \mathscr{N}(x_t/2,\sigma^2)\,,\qquad \sigma>0\,.
$$
Give a necessary condition on $\tau^2$ for $(x_t)_{t\in\mathbb{N}}$ to be a (strictly) stationary process.
\end{exoset}

We have
\begin{equation*}
\mathbb{E}[x_1]=\mathbb{E}[\mathbb{E}[x_1|x_0]]=\mathbb{E}[x_0/2]=0\,.
\end{equation*}
Moreover,
\begin{equation*}
\mathbb{V}(x_1)=\mathbb{V}(\mathbb{E}[x_1|x_0])+\mathbb{E}[\mathbb{V}(x_1|x_0)]=\tau^2/4+\sigma^2\,.
\end{equation*}
Marginaly, $x_1$ is then distributed as a $\mathscr{N}(0,\tau^2/4+\sigma^2)$ variable, with the same
distribution as $x_0$ only if $\tau^2/4+\sigma^2=\tau^2$, i.e.~if $\tau^2=4\sigma^2/3$.

\begin{exoset}
Suppose that $(x_t)_{t\in\mathbb{N}}$ is a {\em Gaussian random walk} on $\mathbb{R}$:
$x_0\sim\mathscr{N}(0,\tau^2)$ and, for all $t\in\mathbb{N}$,
$$
x_{t+1}|\bx_{0:t} \sim \mathscr{N}(x_t,\sigma^2)\,,\qquad \sigma>0\,.
$$
Show that, whatever the value of $\tau^2$ is,
$(x_t)_{t\in\mathbb{N}}$ is not a (strictly) stationary process.
\end{exoset}

We have
\begin{equation*}
\mathbb{E}[x_1]=\mathbb{E}[\mathbb{E}[x_1|x_0]]=\mathbb{E}[x_0]=0\,.
\end{equation*}
Moreover,
\begin{equation*}
\mathbb{V}(x_1)=\mathbb{V}(\mathbb{E}[x_1|x_0])+\mathbb{E}[\mathbb{V}(x_1|x_0)]=\tau^2+\sigma^2\,.
\end{equation*}
The marginal distribution of $x_1$ is then a $\mathscr{N}(0,\tau^2+\sigma^2)$ distribution which cannot
be equal to a $\mathscr{N}(0,\tau^2)$ distribution.

\begin{exoset}\label{exo:studbaker} Consider the process $(x_t)_{t\in\mathbb{N}}$ such that $x_0=0$ and, for all $t\in\mathbb{N}$,
$$
x_{t+1}|\bx_{0:t}\sim\mathscr{N}(\varrho\,x_t,\sigma^2)\,.
$$
Suppose that $\pi(\varrho,\sigma)=1/\sigma$ and that there is no constraint on $\varrho$.
Show that the conditional posterior distribution of $\varrho$, 
conditional on the observations $\bx_{0:T}$ and on $\sigma^2$
is a $\mathscr{N}(\mu_T,\omega_T^2)$ distribution, with
$$
\mu_T = \sum_{t=1}^T x_{t-1}x_t\bigg/ \sum_{t=1}^T x_{t-1}^2
\quad\text{ and }\quad
\omega_T^2 = \sigma^2\bigg/ \sum_{t=1}^T x_{t-1}^2\,.
$$
Show that the marginal posterior distribution of $\varrho$ is a Student $\mathscr{T}(T-1,\mu_T,
\nu_T^2)$ distribution,
with
$$
% ???New typo???
%\nu_T^2 = \frac{1}{T}\,\left(\sum_{t=1}^T x_t^2 - \sum_{t=1}^T x_{t-1}x_t\right)\bigg/\sum_{t=1}^T x_{t-1}^2\,.
\nu_T^2 = \frac{1}{T-1}\,\left(\sum_{t=1}^T x_t^2 \bigg/ \sum_{t=0}^{T-1} x_t^2 - \mu_T^2 \right)\,.
$$
Apply this modeling to the {\sf AEGON} series in {\sf Eurostoxx 50}
and evaluate its predictive abilities.
\end{exoset}

\noindent{\bf Warning!} The text above replaces the text of Exercise \ref{exo:studbaker} in the first printing
of the book, with $T-1$ degrees of freedom instead of $T$ and a new expression for $\nu_T^2$.

% Merci Jean-Mi'!
The posterior conditional density of $\varrho$ is proportional to 
\begin{align*}
\prod_{t=1}^T &\exp\left\{ -(x_t-\varrho\,x_{t-1})^2 / 2\sigma^2 \right\} \\
	&\propto \exp\left\{ \left[- \varrho^2 \sum_{t=0}^{T-1} x_t^2 
	+ 2 \varrho \sum_{t=0}^{T-1} x_t x_{t+1} \right] \big/ 2\sigma^2 \right\}\,,
\end{align*}
which indeed leads to a $\mathscr{N}(\mu_T,\omega_T^2)$ conditional distribution as
indicated above.

Given that the joint posterior density of $(\varrho,\sigma)$ is proportional to
$$
\sigma^{-T-1} \prod_{t=1}^T \exp\left\{ -(x_t-\varrho\,x_{t-1})^2 / 2\sigma^2 \right\} \,
$$
integrating out $\sigma$ leads to a density proportional to
\begin{align*}
\int &\left(\sigma^2\right)^{-T/2-1/2}
\exp\left(\sum_{t=1}^T(x_t-\rho x_{t-1})^2/(2\sigma^2)\right) \text{d}\sigma\\
&=\int \left(\sigma^2\right)^{-T/2-1}
\exp\left(\sum_{t=1}^T(x_t-\rho x_{t-1})^2/(2\sigma^2)\right) \text{d}\sigma^2\\
&=\left\{ \sum_{t=1}^T (x_t-\varrho\,x_{t-1})^2  \right\}^{-T/2}
\end{align*}
when taking into account the Jacobian. We thus get a 
Student $\mathscr{T}(T-1,\mu_T,\nu_T^2)$ distribution
and the parameters can be derived from expanding the sum of squares:
$$
\sum_{t=1}^T (x_t-\varrho\,x_{t-1})^2 = \sum_{t=0}^{T-1} x_t^2 \left(
\varrho^2  - 2 \varrho \mu_T \right) + \sum_{t=1}^{T} x_t^2 
$$
into 
\begin{align*}
\sum_{t=0}^{T-1} x_t^2 &(\varrho - \mu_T) ^2 + \sum_{t=1}^{T} x_t^2 - \sum_{t=0}^{T-1} x_t^2 \mu_T^2\\
&\propto \frac{(\varrho - \mu_T) ^2}{T-1} 
+  \frac{1}{T-1} \left( \frac{\sum_{t=1}^{T} x_t^2}{\sum_{t=0}^{T-1} x_t^2} - \mu_T^2 \right) \\
&= \frac{(\varrho - \mu_T) ^2}{T-1} + \nu_T^2 \,.
\end{align*}

%Et on applique gentillement a AEGON...
The main point with this example is that, when $\varrho$ is unconstrained, the joint posterior 
distribution of $(\varrho,\sigma)$ is completely closed-form. Therefore, the predictive distribution
of $x_{T+1}$ is given by
$$
\int \frac{1}{\sqrt{2\pi}\sigma} \exp\{ -(x_{T+1}-\varrho x_T)^2/2\sigma^2 \}\,
	\pi(\sigma,\varrho|\bx_{0:T}) \text{d}\sigma \text{d}\varrho
$$
which has again a closed-form expression:
\begin{align*}
\int \frac{1}{\sqrt{2\pi}\sigma} \exp\{ &-(x_{T+1}-\varrho x_T)^2/2\sigma^2 \}\,
        \pi(\sigma,\varrho|\bx_{0:T}) \text{d}\sigma \text{d}\varrho\\
&\propto \int \sigma^{-T-2} \exp\{ -\sum_{t=0}^{T} (x_{t+1}-\varrho x_t)^2/2\sigma^2 \}
				       \text{d}\sigma \text{d}\varrho\\
&\propto \int \left\{ \sum_{t=0}^T (x_{t+1}-\varrho\,x_{t})^2  \right\}^{-(T+1)/2} \text{d}\varrho\\
&\propto \left( \sum_{t=0}^T x_{t}^2 \right)^{-(T+1)/2} \int \left\{ \frac{(\varrho 
	- \mu_{T+1}) ^2}{T} + \nu_{T+1}^2 \right\}^{-(T+2)/2} \text{d}\varrho\\
&\propto \left( \sum_{t=0}^T x_{t}^2 \right)^{-(T+1)/2}\, \nu_T^{-T-1} \\
&\propto \left( \sum_{t=0}^T x_{t}^2 \sum_{t=0}^T x_{t+1}^2 - \left\{ \sum_{t=0}^T x_{t}x_{t+1}\right\}^2
	 \right)^{(T+1)/2}\,.
\end{align*}
This is a Student $\mathcal{T}(T,\delta_T,\omega_T)$ distribution, with
$$
\delta_T = x_T \sum_{t=0}^{T-1} x_{t}x_{t+1} / \sum_{t=0}^{T-1} x_{t}^2 = \hat \rho_T x_T
$$
and
$$
\omega_T = \left\{ \sum_{t=0}^T x_t^2 \sum_{t=0}^T x_t^2 - \left( \sum_{t=0}^T x_{t}x_{t+1}
\right)^2 \right\} \bigg/ T \sum_{t=0}^{T-1} x_{t}^2\,.
$$
The predictive abilities of the model are thus in providing a point estimate for the next
observation $\hat x_{T+1} = \hat \rho_T x_T$, and a confidence band around this value.

\begin{exoset}
Give the necessary and sufficient condition under which an AR$(2)$ process with autoregressive polynomial
$\mathcal{P}(u)=1-\varrho_1 u-\varrho_2 u^2$ (with $\varrho_2\neq 0$) is causal.
\end{exoset}

The AR$(2)$ process with autoregressive polynomial $\mathcal{P}$ is causal if and only if the roots of $\mathcal{P}$
are outside the unit circle in the complex plane. The roots of $\mathcal{P}$ are given by 
$$
u^-=\frac{-\varrho_1-\sqrt{\varrho_1^2+4\varrho_2}}{-2\varrho_2} \quad\text{and}\quad
u^+=\frac{-\varrho_1+\sqrt{\varrho_1^2+4\varrho_2}}{-2\varrho_2}
$$
with the convention that $\sqrt{x}=\iota\sqrt{-x}\in\mathbb{C}$ if $x<0$. 
(Because of the symmetry of the roots wrt $\rho_1$, the causality region will be symmetric in $\rho_1$.)

\begin{figure}[h]
\centerline{\includegraphics[width=.7\textwidth]{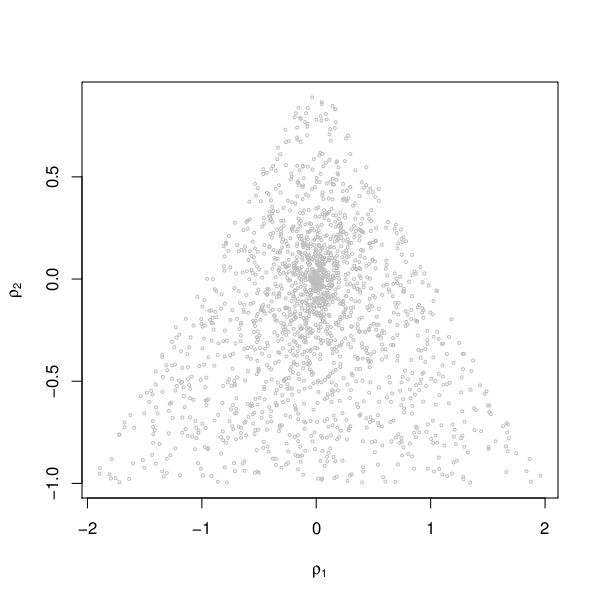}}
\caption{\label{fig:pods}
Acceptable values of $(\rho_1,\rho_2)$ for the $AR(2)$ model obtained by simulation.}
\end{figure}

A first empirical approach based on simulation is to produce a sample of $(\varrho_1,\varrho_2)$'s over the 
sphere of radius $6$ ($6$ is chosen arbitrarily and could be changed if this is too small a bound)
and to plot only those $(\varrho_1,\varrho_2)$'s for which the roots $u^-$ and $u^+$ are outside
the unit circle.
\begin{verbatim}
# Number of points
N=10^4 	

# Sample of rho's
rho1=rnorm(N)
rho2=rnorm(N)
rad=6*runif(N)/sqrt(rho1^2+rho2^2)
rho1=rad*rho1
rho2=rad*rho2
R=matrix(1,ncol=3,nrow=N)
R[,2]=-rho1
R[,3]=-rho2

roots=apply(R,1,polyroot)
indx=(1:N)[(Mod(roots[1,])>1)]
indx=indx[(Mod(roots[2,indx])>1)]
plot(rho1[indx],rho2[indx],col="grey",cex=.4,
  xlab=expression(rho[1]),ylab=expression(rho[2]))

\end{verbatim}
The output of this program is given on Figure \ref{fig:pods} but, while it looks like a triangular shape,
this does not define an analytical version of the restricted parameter space.

If we now look at the analytical solution,
there are two cases to consider: either $\varrho_1^2+4\varrho_2<0$ and the roots are then complex numbers,
or $\varrho_1^2+4\varrho_2>0$ and the roots are then real numbers.

If $\varrho_1^2+4\varrho_2<0$, then 
$$
\left|\frac{-\varrho_1\pm i\sqrt{-(\varrho_1^2+4\varrho_2)}}{2\varrho_2}\right|^2>1
$$
implies that $-1<\varrho_2$, which, together with the constraint $\varrho_1^2+4\varrho_2<0$, provides a first
region of values for $(\varrho_1,\varrho_2)$:
$$
\mathcal{C}_1 = \left\{ (\varrho_1,\varrho_2);\,|\varrho_1|\le 4,\varrho_2<-\varrho_1^2/4 \right\}\,.
$$

If $\varrho_1^2+4\varrho_2>0$, then the condition $|u^\pm|>1$ turns into
$$
\sqrt{\varrho_1^2+4\varrho_2} - |\varrho_1| > 2 \varrho_2\quad\mbox{if}\quad \varrho_2>0 \quad\quad\mbox{and}
|\varrho_1| - \sqrt{\varrho_1^2+4\varrho_2} > -2 \varrho_2\quad\mbox{if}\quad \varrho_2<0\,.
$$
Thus, this amounts to compare $\sqrt{\varrho_1^2+4\varrho_2}$ with $2 | \varrho_2+ |\varrho_1|$ or, 
equivalently, $4\varrho_2$ with $4\varrho_2^2 + 4 | \varrho_1 \varrho_2|$, ending up with the global
condition $\varrho_2<1-|\varrho_1$. The second part of the causality set is thus
$$
\mathcal{C}_2 = \left\{ (\varrho_1,\varrho_2);\,|\varrho_1|\le 4,\varrho_2<-\varrho_1^2/4 \right\}\,,
$$
and cumulating both regions leads to the triangle 
$$
\mathcal{T} = \left\{ (\varrho_1,\varrho_2);\,\varrho_2>-1,\varrho_2<1-|\varrho_1| \right\}\,.
$$

%Figure \ref{fig:arkady} provides the causality region for the AR(2) process.
%\begin{figure}[h]
%\centerline{\includegraphics[width=.7\textwidth]{../figures/arcau.pdf}}
%\caption{\label{fig:arkady}
%Causality region for the $AR(2)$ model.} 
%\end{figure}

\begin{exoset}
Show that the stationary distribution of $\bx_{-p:-1}$ is a
$\mathscr{N}_p(\mu{\mathbf 1}_p,\mathbf{A})$ distribution, and
give a fixed point equation satisfied by the covariance matrix $\mathbf{A}$.
\end{exoset}

%J2M
If we denote
$$
\bz_t=\left(x_t,x_{t-1},\ldots,x_{t+1-p}\right)\,,
$$
then
$$
\bz_{t+1}=\mu{\mathbf 1}_p+B\left(\bz_t-\mu{\mathbf 1}_p\right)+\epsilon_{t+1}\,.
$$
Therefore,
$$
\mathbb{E}\left[\bz_{t+1}|\bz_t\right]=\mu{\mathbf 1}_p+B\left(\bz_t-\mu{\mathbf 1}_p\right)
$$
and
$$
\mathbb{V}\left(\bz_{t+1}|\bz_t\right)=\mathbb{V}\left(\epsilon_{t+1}\right)=\left[
\begin{array}{llll}
 \sigma^2 & 0      & \ldots & 0 \\
 0        & 0      & \ldots & 0 \\
 \vdots   & \vdots & \vdots & \vdots \\
 0        & 0      & \ldots & 0
\end{array}
\right]=V\,.
$$
Then,
$$
\bz_{t+1}|\bz_{t}\sim\mathcal{N}_p\left(\mu{\mathbf 1}_p+B\left(\bz_t-\mu{\mathbf 1}_p\right),V\right)\,.
$$
Therefore, if $\bz_{-1}=\bx_{-p:-1}\sim\mathcal{N}_p\left(\mu{\mathbf 1}_p,A\right)$ is Gaussian, then 
$\bz_t$ is Gaussian. Suppose that $\bz_t\sim\mathcal{N}_p(M,A)$, we get 
$$
\mathbb{E}\left[\bz_{t+1}\right)=\mu{\mathbf 1}_p+B\left(M-\mu{\mathbf 1}_p\right]
$$
and $\mathbb{E}\left[\bz_{t+1}\right]=\mathbb{E}\left[\bz_t\right]$ if
$$
\mu{\mathbf 1}_p +B\left(M-\mu{\mathbf 1}_p\right)=M\,,
$$
which means that $M=\mu{\mathbf 1}_p$. Similarly,
$\mathbb{V}\left(\bz_{t+1}\right)=\mathbb{V}\left(\bz_t\right)$ if and only if 
$$
BAB'+V=A\,,
$$
which is the "fixed point" equation satisfied by $A$.

\begin{exoset}
Show that the posterior distribution on $\boldsymbol{\theta}$ associated
with the prior $\pi(\boldsymbol{\theta})=1/\sigma^2$ is well-defined for $T > p$ observations.
\end{exoset}

\noindent{\bf Warning:} The prior $\pi(\boldsymbol{\theta})=1/\sigma$ was wrongly used
in the first printing of the book.

The likelihood conditional on the initial values $\bx_{0:(p-1)}$ is proportional to
$$
\sigma^{-T+p-1} \prod_{t=p}^T \exp\left\{-\left(x_t-\mu - \sum_{i=1}^p \varrho_i 
(x_{t-i}-\mu) \right)^2 \big/ 2\sigma^2 \right\}\,.
$$
A traditional noninformative prior is $\pi(\mu,\varrho_1,\ldots,\varrho_p,\sigma^2)=1/\sigma^2$. 
In that case, the probability density of the posterior distribution is proportional to
$$
\sigma^{-T+p-3} \prod_{t=p}^T \exp\left\{-\left(x_t-\mu - \sum_{i=1}^p \varrho_i 
	(x_{t-i}-\mu) \right)^2 \big/ 2\sigma^2 \right\}\,.
$$
And
$$
\int (\sigma^2)^{-(T-p+3)/2} \prod_{t=p}^T \exp\left\{-\left(x_t-\mu - \sum_{i=1}^p 
  \varrho_i (x_{t-i}-\mu) \right)^2 \big/ 2\sigma^2 \right\}\text{d}\sigma^2<\infty
$$
holds for $T-p+1>0$, i.e., $T>p-1$. This integral is equal to
$$
\left\{-\left(x_t-\mu - \sum_{i=1}^p
  \varrho_i (x_{t-i}-\mu) \right)^2 \big/ 2\sigma^2 \right\}^{(p-T-1)/2}\,,
$$
which is integrable in $\mu$ for $T-p>0$, i.e.~$T > p$.
The other parameters $\varrho_j$ $(j=1,\ldots,p0$ being bounded, the remaining integrand
is clearly integrable in $\boldsymbol{\varrho}$.

\begin{exoset}\label{ex:root2pol}
Show that the coefficients of the polynomial $\mathcal{P}$ can be derived in $\hbox{O}(p^2)$
time from the inverse roots $\lambda_i$ using the recurrence relations $(i=1,\ldots,p,j=0,\ldots,p)$
$$
\psi^i_0=1\,,\qquad \psi^i_j = \psi^{i-1}_j-\lambda_i\psi_{j-1}^{i-1}\,,
$$
where $\psi^0_0=1$ and $\psi^i_j=0$ for $j>i$, and setting $\varrho_j= -\psi^p_j$ $(j=1,\ldots,p)$.
\end{exoset}

\noindent{\bf Warning:} The useless sentence {\em ``Deduce that the likelihood is computable in $\hbox{O}(Tp^2)$ time"} 
found in the first printing of the book has been removed.

Since
$$
\prod_{i=1}^p (1-\lambda_i x) = 1 - \sum_{j=1}^j \varrho_j x^j\,,
$$
we can expand the lhs one root at a time. If we set
$$
\prod_{j=1}^i (1-\lambda_j x) = \sum_{j=0}^i \psi^i_j x^j\,,
$$`
then
\begin{eqnarray*}
\prod_{j=1}^{i+1} (1-\lambda_j x) &=&  (1-\lambda_{i+1} x) \prod_{j=1}^i (1-\lambda_j x)\\
	&=& (1-\lambda_{i+1} x) \sum_{j=0}^i \psi^i_j x^j \\
        &=& 1 + \sum_{j=1}^i (\psi^i_j - \lambda_{i+1} \psi^i_{j-1}) x^j -\lambda_{i+1} \psi^i_i x^{i+1} \,,
\end{eqnarray*}
which establishes the $\psi^{i+1}_j = \psi^{i}_j-\lambda_{i+1}\psi_{j-1}^{i}$ recurrence relation.

This recursive process requires the allocation of $i$ variables at the $i$th stage; the coefficients of
$\mathcal{P}$ can thus be derived with a complexity of $\hbox{O}(p^2)$.

\begin{exoset}
Show that, if the proposal on $\sigma^2$ is a log-normal distribution $\mathcal{LN}(\log(\sigma^2_{t-1}),\tau^2)$
and if the prior distribution on $\sigma^2$ is the noninformative prior $\pi(\sigma^2)=1/\sigma^2$, the acceptance ratio
also reduces to the likelihood ratio because of the Jacobian.
\end{exoset}

\noindent{\bf Warning:} In the first printing of the book, there is a $\log$ missing in the 
mean of the log-normal distribution.

If we write the Metropolis--Hastings ratio for a current value $\sigma_0^2$ and a proposed value
$\sigma_1^2$, we get
$$
\frac{\pi(\sigma_1^2)\ell(\sigma_1^2)}{\pi(\sigma_0^2)\ell(\sigma_0^2)}\,
\frac{\exp\left( -(\log(\sigma^2_0-\log(\sigma^2_1))^2/2\tau^2\right)/\sigma_0^2}{
\exp\left( -(\log(\sigma^2_0-\log(\sigma^2_1))^2/2\tau^2\right)/\sigma_1^2}
= \frac{\ell(\sigma_1^2)}{\ell(\sigma_0^2)}\,,
$$
as indicated.

\begin{exoset}\label{exo:vanilla}
Write an {\sf R} program that extends the reversible jump
algorithm 7.1 to the case when the order $p$ is unknown and apply it to
the same {\sf Ahold Kon.} series of {\sf Eurostoxx 50}.
\end{exoset}

The modification is rather straightforward if one only considers birth and death moves, adding and
removing real or complex roots in the polynomial. When the new values are generated from the prior,
as in the program provided by {\sf \#7.txt} on the Webpage, the acceptance probability remains equal
to the likelihood ratio (with the usual modifications at the boundaries).

\begin{exoset}
For an MA$(q)$ process, show that $(s\le q)$
$$
\gamma_x(s) = \sigma^2 \sum_{i=0}^{q-|s|} \vartheta_i \vartheta_{i+|s|}\,.
$$
\end{exoset}

We have
\begin{eqnarray*}
\gamma_x(s) & = & \mathbb{E}\left[x_t x_{t-s}\right] \\
            & = & \mathbb{E}\left[\left[\epsilon_t+\vartheta_1\epsilon_{t-1}+\ldots+\vartheta_q\epsilon_{t-q}\right]
                  \left[\epsilon_{t-s}+\vartheta_1\epsilon_{t-s-1}+\ldots+\vartheta_q\epsilon_{t-s-q}\right]\right]\,.
\end{eqnarray*}
Then, if $1\le s\le q$,
$$
\gamma_x(s)=\left[\vartheta_s+\vartheta_{s+1}\vartheta_1+\ldots+\vartheta_q\vartheta_{q-s}\right]\sigma^2
$$
and
$$
\gamma_x(0)=\left[1+\vartheta_1^2+\ldots+\vartheta_q^2\right]\sigma^2\,.
$$
Therefore, if $(0\le s\le q)$ with the convention that $\vartheta_0=1$
$$
\gamma_x(s) = \sigma^2 \sum_{i=0}^{q-s} \vartheta_i \vartheta_{i+s}\,.
$$
The fact that $\gamma_x(s)=\gamma_x(-s)$ concludes the proof.

\begin{exoset}\label{exo:norminov}
Show that the conditional distribution of $(\epsilon_0,\ldots,\epsilon_{-q+1})$
given both $\bx_{1:T}$ and the parameters is a normal distribution.
Evaluate the complexity of computing the mean and covariance matrix of this distribution.
\end{exoset}

%J2M
The distribution of $\bx_{1:T}$ conditional on $(\epsilon_0,\ldots,\epsilon_{-q+1})$ is
proportional to
$$
\sigma^{-T} \prod_{t=1}^T\exp\left\{-\left(x_t-\mu+ \sum_{j=1}^q 
\vartheta_j\widehat \epsilon_{t-j} \right)^2\bigg/ 2\sigma^2 \right\} \,,
$$
Take 
$$
(\epsilon_0,\ldots,\epsilon_{-q+1})\sim\mathcal{N}_q\left(0_q,\sigma^2I_q\right)\,.
$$
In that case, the  conditional distribution of $(\epsilon_0,\ldots,\epsilon_{-q+1})$
given $\bx_{1:T}$ is proportional to
$$
\prod_{i=-q+1}^0 \exp\left\{-\epsilon_i^2/2\sigma^2\right\}\,
\prod_{t=1}^T \exp\left\{-\widehat\epsilon_t^2/2\sigma^2\right\}\,.
$$
% Feignant...  
Due to the recursive definition of $\hat \epsilon_t$,
the computation of the mean and the covariance matrix of this distribution is too costly to be available
for realistic values of $T$. For instance, getting the conditional mean of $\epsilon_i$ requires deriving the
coefficients of $\epsilon_i$ from all terms
$$
\left(x_t-\mu+ \sum_{j=1}^q\vartheta_j\widehat \epsilon_{t-j}\right)^2
$$
by exploiting the recursive relation
$$
\widehat \epsilon_t = x_t -\mu + \sum_{j=1}^q \vartheta_j \widehat\epsilon_{t-j}\,.
$$
If we write $\widehat \epsilon_1 = \delta_1 + \beta_1 \epsilon_i$ and
$\widehat \epsilon_t = \delta_t + \beta_t \epsilon_i$, then we need to
use the recursive formula
$$
\delta_t = x_t - \mu + \sum_{j=1}^q \vartheta_j \delta_{t-j}\,,
\qquad
\beta_t = \sum_{j=1}^q \beta_{t-j}\,,
$$
before constructing the conditional mean of $\epsilon_i$. The corresponding cost for this
single step is therefore $\mbox{O}(Tq)$ and therefore $\mbox{O}(qT^2)$ for the whole series 
of $\epsilon_i$'s. Similar arguments can be used for computing the conditional variances.

\begin{exoset}\label{ex:conMa}
Give the conditional distribution of $\epsilon_{-t}$ given the other $\epsilon_{-i}$'s,
$\bx_{1:T}$, and the $\widehat\epsilon_i$'s. Show that it only depends on the other $\epsilon_{-i}$'s,
$\bx_{1:q-t+1}$, and $\widehat\epsilon_{1:q-t+1}$.
\end{exoset}

The formulation of the exercise is slightly too vague in that the $\widehat\epsilon_i$'s are deterministic
quantities based on the $\epsilon_{-i}$'s and $\bx_{1:T}$. Thus, from a probabilistic point of view, 
the conditional distribution of $\epsilon_{-t}$ only depends on the other $\epsilon_{-i}$'s and $\bx_{1:T}$.
However, from an algorithmic point of view, if we take the $\widehat\epsilon_i$'s as additional observables in
(7.11), spotting $\epsilon_{-\ell}$ in
$$
\sum_{t=1}^T\left(x_t-\mu+ \sum_{j=1}^q \vartheta_j\widehat \epsilon_{t-j} \right)^2
$$
leads to keep only
$$
\sum_{t=1}^{q-\ell} \left(x_t-\mu + \sum_{j=1}^{t-1} \vartheta_j\widehat \epsilon_{t-j} 
	+ \sum_{j=t}^{q} \vartheta_j \epsilon_{t-j} \right)^2 
$$
in the sum since, for $t-q>-\ell$, i.e.~for $t>q-\ell$, $\epsilon_{-\ell}$ does not appear in the distribution. 
(Again, this is a formal construct that does not account for the deterministic derivation of the 
$\widehat\epsilon_i$'s.) The conditional distribution of $\epsilon_{-\ell}$ is then obviously a normal 
distribution.

\begin{exoset}\label{exo:miliT}
Show that the predictive horizon for the MA$(q)$ model is restricted to the first $q$ future observations $x_{t+i}$.
\end{exoset}

%J2M
Obviously, due to the lack of correlation between $x_{T+q+j}$ $(j>0)$ and $\bx_{1:T}$ we have
$$
\mathbb{E}\left[x_{T+q+1}|\bx_{1:T}\right]
=\mathbb{E}\left[x_{T+q+1}\right]=0
$$
and therefore the $MA(q)$ model has no predictive ability further than horizon $q$.

\begin{exoset}\label{exo:hmm=mix}
Show that, when the support $\mathcal{Y}$ is finite and when $(y_t)_{t\in\mathbb{N}}$ is stationary,
the marginal distribution of $x_t$ is the same mixture distribution for all $t$'s.
Deduce that the same identifiability problem as in mixture models occurs in this setting.
\end{exoset}

Since the marginal distribution of $x_t$ is given by
$$
\int f(x_t|y_t) \pi(y_t)\,\text{d}y_t = \sum_{y\in\mathcal{Y}} \pi(y) f(x_t|y)\,,
$$
where $\pi$ is the stationary distribution of $(y_t)$, this is indeed a mixture
distribution. Although this is not the fundamental reason for the unidentifiability
of hidden Markov models, there exists an issue of label switching similar to the
case of standard mixtures.

\begin{exoset}\label{exo:nolike}
Write down the joint distribution of $(y_t,x_t)_{t\in\mathbb{N}}$ in (7.19) 
%\ref{eq:stovo}) 
and deduce that the (observed) likelihood is not available in closed form.
\end{exoset}

% J2M solution ????
Recall that
$y_0\sim\mathcal{N}(0,\sigma^{2})$ and, for $t=1,\ldots,T$,
$$
\begin{cases}
y_t = \varphi y_{t-1} + \sigma \epsilon^*_{t-1}\,, &\cr
x_t = \beta e^{y_t/2} \epsilon_t\,, &\cr
\end{cases}
$$
where both $\epsilon_t$ and $\epsilon^*_t$ are iid $\mathcal{N}(0,1)$ random variables.
The joint distribution of $\left(\bx_{1:T},\by_{0:T}\right)$ is therefore
\begin{align*}
f\left(\bx_{1:T},\by_{0:T}\right)&=f\left(\bx_{1:T}|\by_{0:T}\right)f\left(\by_{0:T}\right)\\
&=\left(\prod_{i=1}^T f(x_i|y_i)\right)f(y_0)f(y_1|y_0)\ldots f(y_T|y_{T-1})\\
&=\frac{1}{\left(2\pi\beta^2\right)^{T/2}}\exp\left\{ -\sum_{t=1}^Ty_t/2\right)
\exp\left(-\frac{1}{2\beta^2}\sum_{t=1}^T x_t^2\exp(-y_t)\right)\\
&\quad\times\frac{1}{\left(2\pi\sigma^2\right)^{(T+1)/2}}\exp\left(-\frac{1}{2\sigma^2}
	\left(y_0^2+\sum_{t=1}^T\left(y_t-\phi y_{t-1}\right)^2\right)\right\}\,.
\end{align*}
Due to the double exponential term $\exp\left(-\frac{1}{2\beta^2}\sum_{t=1}^T x_t^2\exp(-y_t)\right)$, it is 
impossible to find a closed-form of the integral in $\by_{0:T}$.

\begin{exoset}
Show that the counterpart of the prediction filter in the
Markov-switching case is given by
$$
\log p(\bx_{1:t}) = \sum_{r=1}^{t} \log \left[\sum_{i=1}^{\kappa} f(x_r|x_{r-1},y_r=i) \varphi_r(i)\right]\,,
$$
where $\varphi_r(i)=\mathbb{P}(y_r=i|\bx_{1:r-1})$ is given by the recursive formula
$$
\varphi_r(i) \propto \sum_{j=1}^\kappa p_{ji} f(x_{r-1}|x_{r-2},y_{r-1}=j) \varphi_{r-1}(j)\,.
$$
\end{exoset}

\noindent{\bf Warning!} There is a typo in the first printing of the book where $\varphi_r$ is defined
conditional on $\bx_{1:t-1}$ instead of $\bx_{1:r-1}$.

This exercise is more or less obvious given the developments provided in the book. The distribution of
$y_r$ given the past values $\bx_{1:r-1}$ is the marginal of $(y_r,y_{r-1})$ given the past values $\bx_{1:r-1}$:
\begin{eqnarray*}
\mathbb{P}(y_r=i|\bx_{1:t-1}) &=& \sum_{j=1}^\kappa \mathbb{P}(y_r=i,y_{r-1}=j|\bx_{1:r-1})\\
	&=& \sum_{j=1}^\kappa \mathbb{P}(y_{r-1}=j|\bx_{1:r-1})\,\mathbb{P}(y_r=i|y_{r-1}=j)\\
        &\propto& \sum_{j=1}^\kappa  p_{ji} \mathbb{P}(y_{r-1}=j,x_{r-1}|\bx_{1:r-2})\\
        &=& \sum_{j=1}^\kappa  p_{ji} \mathbb{P}(y_{r-1}=j,|\bx_{1:r-2}) f(x_{r-1}|x_{r-2},y_{r-1}=j)\,,
\end{eqnarray*}
which leads to the update formula for the $\varphi_r(i)$'. The marginal distribution $\bx_{1:t}$ is then
derived by
\begin{eqnarray*}
p(\bx_{1:t}) &=& \prod_{r=1}^t p(x_r|\bx_{1:(r-1)}) \\
             &=& \prod_{r=1}^t \sum_{j=1}^\kappa \mathbb{P}(y_{r-1}=j,x_r|\bx_{1:r-1}) \\
             &=& \prod_{r=1}^t \sum_{j=1}^\kappa f(x_r|x_{r-1},y_r=i) \varphi_r(i)\,,
\end{eqnarray*}
with the obvious convention $\varphi_1(i)=\pi_i$, if $(\pi_1,\ldots,\pi_\kappa)$ is the
stationary distribution associated with $\mathbb{P}=(p_{ij})$.
 
\chapter{Image Analysis}\label{ch:spt}
\
\begin{exoset}
Draw an example with $n=5$ points in $\mathbb{R}^2$ such 
that the $k$-nearest-neighbor relation is not symmetric.
\end{exoset}

\begin{figure}[h]
\centerline{\includegraphics[width=.7\textwidth]{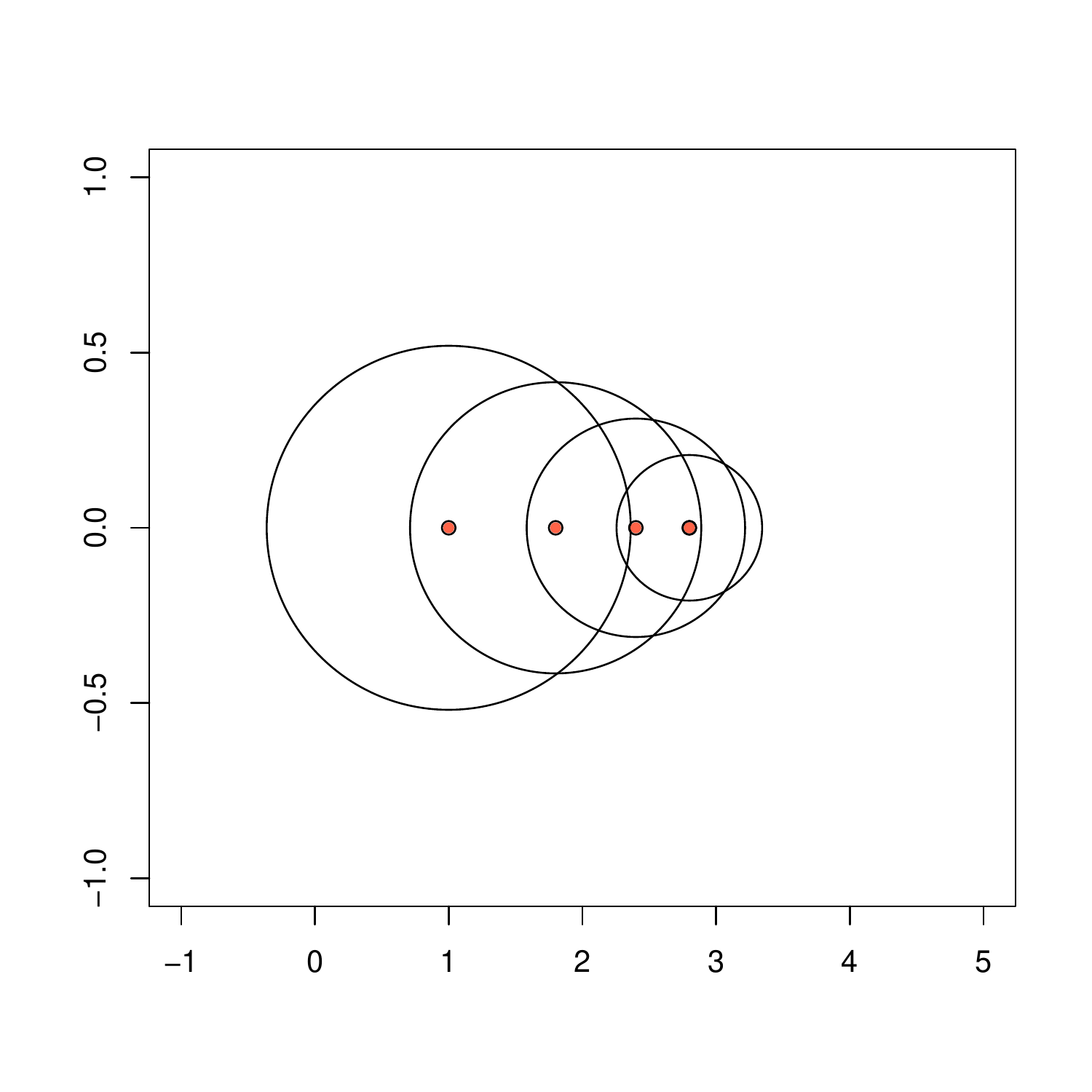}}
\caption{\label{fig:nonsymN}
Sequence of five points with assymetry in the nearest-neighbor relations.}
\end{figure}

The first quadrant of Figure 8.2 in the book is already an illustration
of an assymmetric neighborhood relation.  Once two points are drawn, it is 
sufficient to find sequentialy
new points that are closer to the latest than the earlier points.
Figure \ref{fig:nonsymN} illustrates this case in dimension one,
with decreasing-radius circles to emphasize the assymetry: each point
on the line is such that its right neighbor is its nearest neighbor and
that it is the nearest neighbor of its left neighbor.

\begin{exoset}
For a given pair $(k,n)$ and a uniform distribution of $\bx$ in $[0,1]^3$, 
design a Monte Carlo experiment that evaluates the distribution
of the size of the symmetrized $k$-nearest-neighborhood. 
\end{exoset}

\begin{figure}[h]
\centerline{\includegraphics[height=4truecm,width=.5\textwidth]{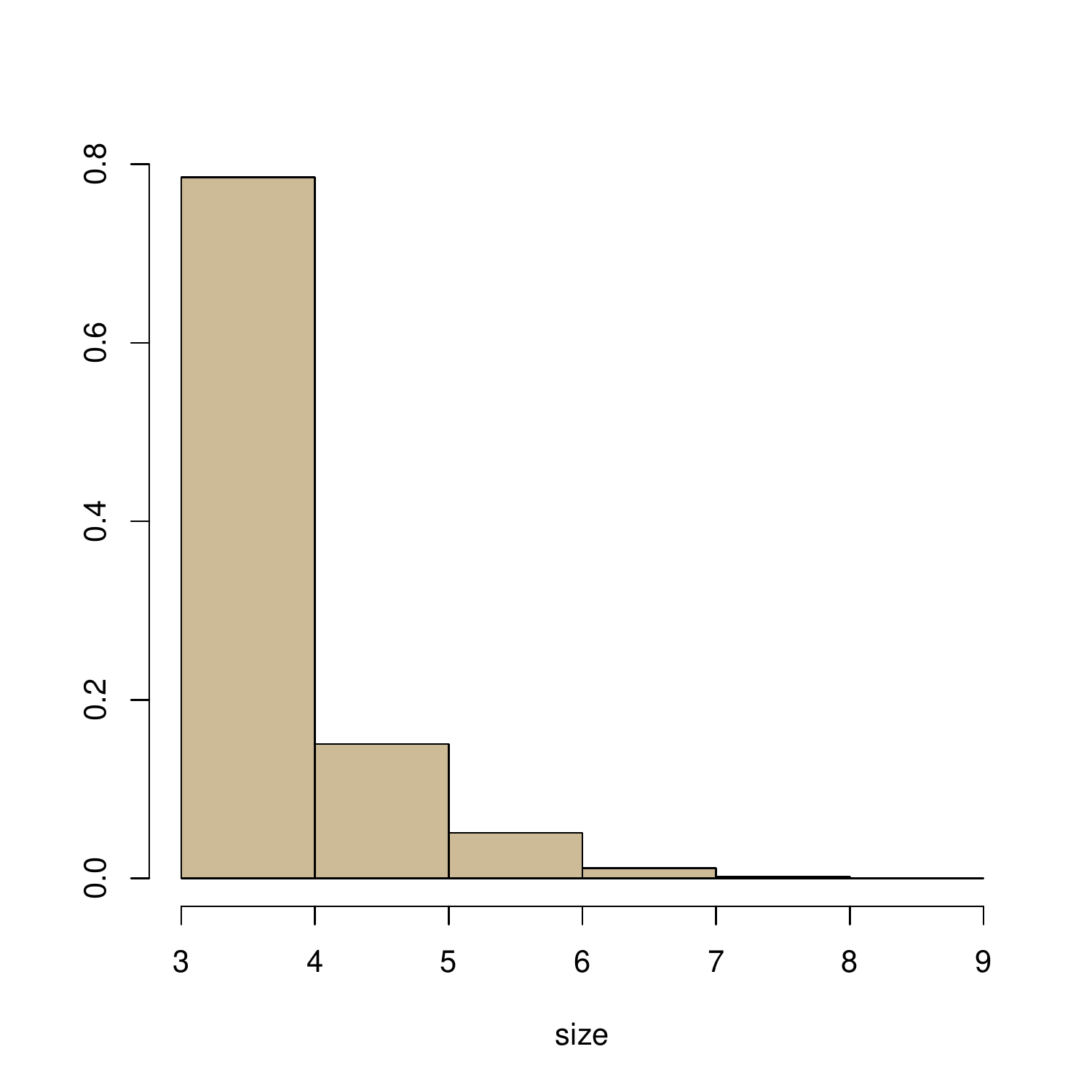}}
\caption{\label{fig:histosize}
Histogram of the size of the symmetrized $k$-nearest-neighborhoods when $k=3$ and $n=200$
based on $N=1000$ simulations.}
\end{figure}

For a given pair $(k,n)$, the Monte Carlo experiment produces $N$ random samples on $[0,1]^3$,
for instance as
\begin{verbatim}
  samp=matrix(runif(3*n),n,3)
\end{verbatim}
compute the $k$-nearest-neighborhood matrix for this sample, by
\begin{verbatim}
  disamp=as.matrix(dist(samp,up=T,diag=T))   #distance matrix

  neibr=t(apply(disamp,1,order))[,-1]        #k nearest neighbours
  knnbr=neibr[,1:k]

  newnbr=matrix(0,n,n)                       #k nearest neighbours
  for (i in 1:n)                             #indicator matrix
    newnbr[i,knnbr[i,]]=1
\end{verbatim}
and compute the sizes of the symmetrized $k$-nearest-neighborhoods for those samples, by
\begin{verbatim}
  size[t,]=apply(newnbr+t(newnbr)>0,1,sum)
\end{verbatim}
ending up with an approximation to the distribution of the size over the samples.
It is then possible to summarize the matrix \verb+size+ on an histogram as in Figure
\ref{fig:histosize}.

\begin{exoset}\label{exo:kliford}
When  $\mathbf{y}=(y_1,y_2)$, show that the joint pdf of $\mathbf{y}$ is given by
$$
f(\mathbf{y}|\mathbf{X},\beta,k) = f(y_1|y_2,\mathbf{X},\beta,k) \bigg/
\sum_{g=1}^G \frac{f(C_g|y_2,\mathbf{X},\beta,k)}{f(y_2|C_g,\mathbf{X},\beta,k)}\,.
$$
Discuss the extension to the general case. ({\em Indication}: The extension is solved via the
Hammersley--Clifford theorem, given in Section 8.3.1.)
\end{exoset}

This exercice is a consequence of Exercice \ref{exo:babyHaC}: we have that, for all $y_2$'s,
\begin{eqnarray*}
f(y_1|\mathbf{X},\beta,k)  &=& \frac{f(y_1|y_2,\mathbf{X},\beta,k)/f(y_2|y_1,\mathbf{X},\beta,k) 
	}{\int f(y_1|y_2,\mathbf{X},\beta,k)/f(y_2|y_1,\mathbf{X},\beta,k) \text{d}y_1}\\
	&=& \frac{f(y_1|y_2,\mathbf{X},\beta,k)/f(y_2|y_1,\mathbf{X},\beta,k)
        }{\sum_{g=1}^G f(C_g|y_2,\mathbf{X},\beta,k)/f(y_2|C_g,\mathbf{X},\beta,k) }\,,
\end{eqnarray*}
since the support of $y_1$ is finite. We can therefore conclude that
$$
f(\mathbf{y}|\mathbf{X},\beta,k) = \frac{f(y_1|y_2,\mathbf{X},\beta,k)}{
	\sum_{g=1}^G f(C_g|y_2,\mathbf{X},\beta,k)/f(y_2|C_g,\mathbf{X},\beta,k) }\,.
$$

As suggested, the extension is solved via the Hammersley--Clifford theorem, given in (8.4). See
Section 8.3.1 for details.

\begin{exoset}\label{exo:pascom}
Find two conditional distributions $f(x|y)$ and $g(y|x)$ such that there is no joint distribution corresponding to
both $f$ and $g$. Find a necessary condition for $f$ and $g$ to be compatible in that respect, i.e.~to correspond
to a joint distribution on $(x,y)$.
\end{exoset}

As stated, this is a rather obvious question: if $f(x|y)=4y\exp(-4yx)$ and if $g(y|x)=6x\exp(-6xy)$, there cannot
be a joint distribution inducing these two conditionals. What is more interesting is that, if
$f(x|y)=4y\exp(-4yx)$ and $g(y|x)=4x\exp(-4yx)$, there still is no joint distribution, despite the formal
agreement between both conditionals: the only joint that would work has the major drawback that it has an
infinite mass! 

\begin{exoset}\label{exo:klimt}
Using the Hammersley-Clifford theorem, show that the full conditional distributions given by (8.1)
are compatible with a joint distribution.
\end{exoset}

\noindent{\bf Note:} In order to expose the error made in the first printing
in using the size of the symmetrized
neighborhood, $N_k(i)$, we will compute the potential joint distribution based on the
pseudo-conditional
$$
\mathbb{P}(y_i=C_j|\mathbf{y}_{-i},\mathbf{X},\beta,k) \propto \exp\left(\beta \sum_{\ell\sim_k i}
\mathbb{I}_{C_j}(y_\ell)\bigg/N_k(i) \right)\,,
$$
even though it is defined for a fixed $N_k(i)=N_k$ in the book.

It follows from (8.4) that, if there exists a joint distribution, it satisfies
$$
\mathbb{P}(\by|\mathbf{X},\beta,k) \propto \prod_{i=0}^{n-1}
\frac{\mathbb{P}(y_{i+1}  |y_1^*,\ldots,y_i^*,y_{i+2},\ldots,y_n,\mathbf{X},\beta,k)}{
      \mathbb{P}(y_{i+1}^*|y_1^*,\ldots,y_i^*,y_{i+2},\ldots,y_n,\mathbf{X},\beta,k)}\,.
$$
Therefore, 
\begin{align*}
\mathbb{P}(\by|\mathbf{X},\beta,k) \propto
&\exp\left\{ \beta \sum_{i=1}^{n} \frac{1}{N_k(i)} \left(
\sum_{\ell<i,\ell\sim_k i} \left[ \mathbb{I}_{y_\ell^*}(y_i)-\mathbb{I}_{y_\ell^*}(y_i^*)\right] + \right.\right.\\
&\qquad\left.\left. 
\sum_{\ell>i,\ell\sim_k i} \left[\mathbb{I}_{y_\ell}(y_i)-\mathbb{I}_{y_\ell}(y_i^*)\right] \right) \right\}
\end{align*}
is the candidate joint distribution. Unfortunately, if we now try to derive the conditional distribution of
$y_j$ from this joint, we get
\begin{align*}
\mathbb{P}(y_i=C_j|\mathbf{y}_{-i},\mathbf{X},\beta,k) \propto \exp\beta
\left\{ \frac{1}{N_k(j)} \sum_{\ell>j,\ell\sim_k j} \mathbb{I}_{y_\ell}(y_j) +
\sum_{\ell<j,\ell\sim_k j} \frac{\mathbb{I}_{y_\ell}(y_j)}{N_k(\ell)} \right.\\
+\left. \frac{1}{N_k(j)} \sum_{\ell<j,\ell\sim_k j} \mathbb{I}_{y_\ell^*}(y_j) -
\sum_{\ell<j,\ell\sim_k j} \frac{\mathbb{I}_{y_\ell^*}(y_j)}{N_k(\ell)} \right\}
\end{align*}
which differs from the orginal conditional if the $N_k(j)$'s differ. In conclusion,
there is no joint distribution if (8.1) is defined as in the first printing. 
Taking all the $N_k(j)$'s equal leads to a coherent joint distribution since the
last line in the above equation cancels.

\begin{exoset}\label{exo:B>0}
If a joint density $\pi(y_1,...,y_n)$ is such that the conditionals 
$\pi(y_{-i}|y_i)$ never cancel on the supports of the marginals 
$m_{-i}(y_{-i})$, show that the support of $\pi$ is equal to the cartesian product of the supports of the marginals.
\end{exoset}

\noindent {\bf Warning!} This exercise replaces the former version of Exercise \ref{exo:B>0}:
{\em ``If $\pi(x_1,\ldots,\allowbreak x_n)$ is a density such that its full conditionals never cancel on its support,
characterize the support of $\pi$ in terms of the supports of the marginal distributions."}

Let us suppose that the support of $\pi$ is not equal to the product of the supports of the marginals.
(This means that the support of $\pi$ is smaller than this product.) Then the conditionals $\pi(\by_{-i}|y_i)$
cannot be positive everywhere on the support of $m(\by_{-i})$. 

\begin{exoset}\label{exo:kliklak}
Describe the collection of cliques $\mathcal{C}$ for an $8$ neighbor
neighborhood structure such as in Figure 8.7 on a regular
$n\times m$ array. Compute the number of cliques.
\end{exoset}

\begin{figure}[h]
\centerline{\includegraphics[width=.7\textwidth]{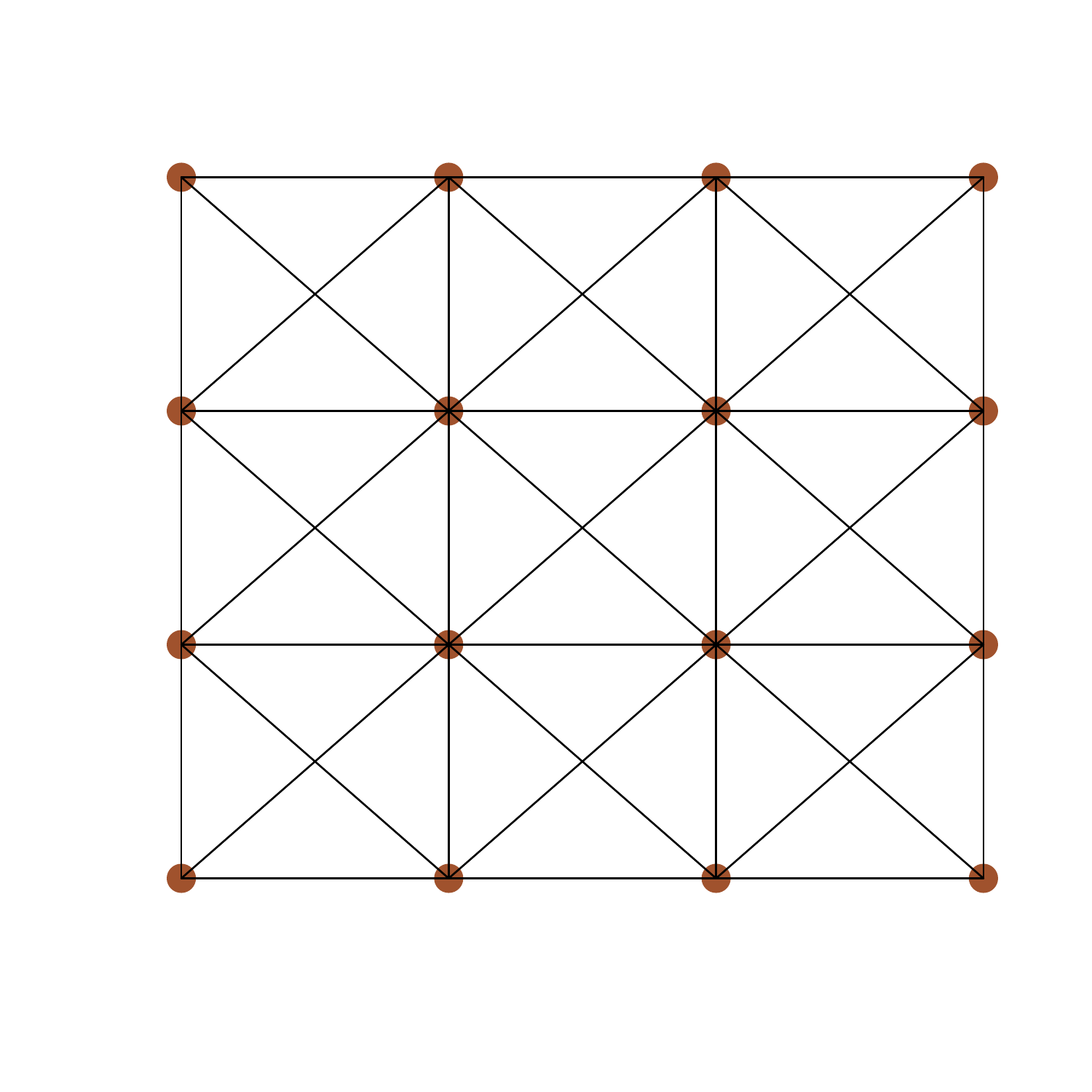}}
\caption{\label{fig:8night}
Neighborhood relations between the points of a $4\times 4$ regular grid for a $8$ neighbor
neighborhood structure.}
\end{figure}

If we draw a detailed graph of the connections on a regular grid as in Figure \ref{fig:8night},
then the maximal structure such that all members are neighbors is made of $4$ points. Cliques are 
thus made of squares of $4$ points and there are $(n-1)\times(m-1)$ cliques on a $n\times m$ array.

\begin{exoset}
Use the Hammersley-Clifford theorem to establish that (8.6)
is the joint distribution associated with the above conditionals. 
Deduce that the Ising model is a MRF.
\end{exoset}

Following the developments in Exercise \ref{exo:klimt}, this is exactly the same
problem as for the distribution (8.1) with a fixed neighborhood structure and the
use of $\beta$ instead of $\beta/N_k$.

\begin{exoset}\label{exo:thecostofpotts}
Draw the function $Z(\beta)$ for a $3\times 5$ array.
Determine the computational cost of the derivation of
the normalizing constant $Z(\beta)$ of (8.6) for a $m\times n$ array.
\end{exoset}

The function $Z(\beta)$ is defined by
$$
Z(\beta) = 1 \bigg/ \sum_{\bx\in\mathcal{X}} \exp\left(\beta \sum_{j\sim i} \mathbb{I}_{x_j=x_i}\right)\,,
$$
which involves a summation over the set $\mathcal{X}$ of size $2^{15}$. The {\sf R} code corresponding
to this summation is
\begin{verbatim}
neigh=function(i,j){	#Neighbourhood indicator function

   (i==j+1)||(i==j-1)||(i==j+5)||(i==j-5)
}

zee=function(beta){

  val=0
  array=rep(0,15)

  for (i in 1:(2^15-1)){

    expterm=0
    for (j in 1:15)
      expterm=expterm+sum((array==array[j])*neigh(i=1:15,j=j))

    val=val+exp(beta*expterm)
    
    j=1
    while (array[j]==1){
        
        array[j]=0
        j=j+1
      }
      array[j]=1
       
    }

  expterm=0
  for (j in 1:15)
      expterm=expterm+sum((array==array[j])*neigh(i=1:15,j=j))

  val=val+exp(beta*expterm)

  1/val
}
\end{verbatim}

\begin{figure}
\centerline{\includegraphics[width=.7\textwidth]{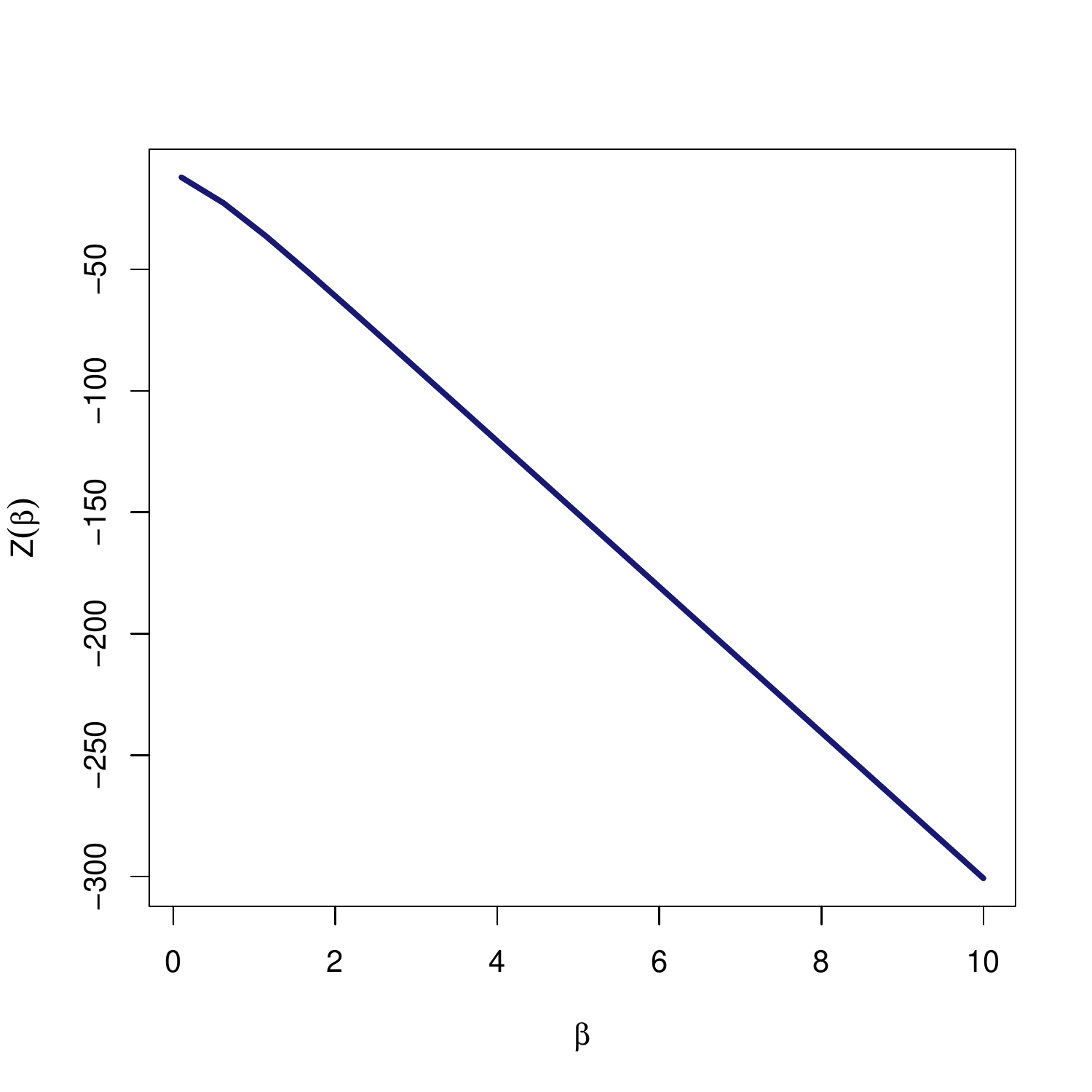}}
\caption{\label{fig:zee}
Plot of the function $Z(\beta)$ for a $3\times 5$ array with a four neighbor structure.}
\end{figure}

It produces the (exact) curve given in Figure \ref{fig:zee}.

In the case of a $m\times n$ array, the summation involves $2^{m\times n}$ and each exponential term
in the summation requires $(m\times n)^2$ evaluations, which leads to a $\text{O}((m\times n)^2\,
2^{m\times n})$ overall cost.

\begin{exoset}\label{exo:sweswa}
For an $n\times m$ array $\mathcal{I}$, if the neighborhood relation
is based on the four nearest neighbors as in Figure 8.7, show that the $x_{i,j}$'s
for which $(i+j)\equiv 0(2)$ are independent conditional on the $x_{i,j}$'s for which $(i+j)
\equiv 1(2)$ $(1\le i\le n,\,1\le j\le m)$. Deduce that the update of the whole image can be
done in two steps by simulating the pixels with even sums of indices and
then the pixels with odd sums of indices. (This modification of Algorithm 8.2
is a version of {\em the Swendsen--Wang} algorithm.)
\end{exoset}

\noindent {\bf Warning!} This exercise replaces the former version of Exercise \ref{exo:sweswa}
{\em ``For an $n\times m$ array $\mathcal{I}$, if the neighborhood relation
is based on the four nearest neighbors, show that the $x_{2i,2j}$'s
are independent conditional on the $x_{2i-1,2j-1}$'s $(1\le i\le
n,\,1\le j\le m)$. Deduce that the update of the whole image can be
done in two steps by simulating the pixels with even indices and
then the pixels with odd indices"}

This exercise is simply illustrating in the simplest case
the improvement brought by the Swendsen-Wang algorithm upon 
the Gibbs sampler for image processing.

As should be obvious from Figure 8.7 in the book, 
the dependence graph between the nodes of the array is such that a given $x_{i,j}$ is independent from
all the other nodes, conditional on its four neighbours. When $(i+j)\equiv 0(2)$, the neighbours have
indices $(i,j)$ such that $(i+j) \equiv 1(2)$, which establishes the first result.

Therefore, a radical alternative to the node-by-node update is to run a Gibbs sampler with two steps: a first
step that updates the nodes $x_{i,j}$ with even $(i+j)$'s and a step that updates the nodes $x_{i,j}$ with 
odd $(i+j)$'s. This is quite a powerful solution in that it achieves the properties of two-stage Gibbs sampling,
as for instance the Markovianity of the subchains generated at each step (see Robert and Casella, 2004, Chapter
9, for details).

\begin{exoset}
Determine the computational cost of the derivation of
the normalizing constant of the distribution (8.7)
for a $m\times n$ array and $G$ different colors.
\end{exoset}

Just as in Exercise \ref{exo:thecostofpotts}, finding the exact normalizing requires summing over all
possible values of $\bx$, which involves $G^{m\times n}$ terms. And each exponential term involves a sum over
$(m\times n)^2$ terms, even though clever programing of the neighborhood system may reduce the computational
cost down to $m\times n$. Overall, the normalizing constant faces a computing cost of at least
$\text{O}(m\times n\times G^{m\times n})$.

\begin{exoset}
Use the Hammersley-Clifford theorem to establish that (8.7) is the joint 
distribution associated with the above conditionals. Deduce that the Potts model is a MRF.
\end{exoset}

Similar to the resolution of Exercise \ref{exo:klimt}, using the Hammersley-Clifford
representation (8.4) and defining an arbitrary order on the set $\mathcal{I}$
leads to the joint distribution
\begin{align*}
\pi(\bx) &\propto \frac{\exp\left\{ \beta \sum_{i\in\mathcal{I}} 
\sum_{j<i,j\sim i} \mathbb{I}_{x_i=x_j} +\sum_{j>i,j\sim i} \mathbb{I}_{x_i=x_j^\star} \right\}}{
\exp\left\{ \beta \sum_{i\in\mathcal{I}}
\sum_{j<i,j\sim i} \mathbb{I}_{x_i^\star=x_j} +\sum_{j>i,j\sim i} \mathbb{I}_{x_i^\star=x_j^\star} \right\}} \\
&\propto \exp\left\{ \beta \left(\sum_{j\sim i,j<i} \mathbb{I}_{x_i=x_j} 
+ \sum_{j\sim i,j>i} \mathbb{I}_{x_i=x_j^\star}
-\sum_{j\sim i, j>i} \mathbb{I}_{x_j^\star=x_i} \right) \right\} \\
&= \exp\left\{ \beta \sum_{j\sim i} \mathbb{I}_{x_i=x_j} \right\}\,. %Another mistake!?!?
\end{align*}
So we indeed recover a joint distribution that is compatible with the initial full conditionals of the
Potts model. The fact that the Potts is a MRF is obvious when considering its conditional distributions.

\begin{exoset}
Derive an alternative to Algorithm 8.3 where the
probabilities in the multinomial proposal are proportional to the
numbers of neighbors $n_{u_\ell,g}$ and compare its performance with
those of Algorithm 8.3.
\end{exoset}

In Step 2 of Algorithm 8.3, another possibility is to select the proposed value of $x_{u_\ell}$ from
a multinomial distribution 
$$
\mathcal{M}_G \left( 1;n_1^{(t)}(u_\ell),\ldots,n_G^{(t)}(u_\ell) \right)
$$
where $n_g^{(t)}(u_\ell)$ denotes the number of neighbors of $u_l$ that take the value $g$. This is likely to
be more efficient than a purely random proposal, especially when the value of $\beta$ is high.

\begin{exoset} Show that the Swendsen-Wang improvement given in Exercise \ref{exo:sweswa} 
also applies to the simulation of $\pi(\bx|\by,\beta,\sigma^2,\bmu)$.
\end{exoset}

This is kind of obvious when considering that taking into account the values of the $y_i$'s does not
modify the dependence structure of the Potts model. Therefore, if there is a decomposition of the grid 
$\mathcal{I}$ into a small number of sub-grids $\mathcal{I}_1,\ldots,\mathcal{I}_k$ such that all the
points in $\mathcal{I}_j$ are independent from one another given the other $\mathcal{I}_\ell$'s, a $k$ 
step Gibbs sampler can be proposed for the simulation of $\bx$.

\begin{exoset}
Using a piecewise-linear interpolation of $f(\beta)$ based on the values $f(\beta^1),\ldots,f(\beta^M)$, with $0<\beta_1<\ldots
<\beta_M=2$, give the explicit value of the integral
$$
\int_{\alpha_0}^{\alpha_1} \hat{f}(\beta)\,\text{d}\beta
$$
for any pair $0\le\alpha_0<\alpha_1\le 2$.
\end{exoset}

This follows directly from the {\sf R} program provided in \verb+#8.txt+, with
$$
\int_{\alpha_0}^{\alpha_1} \hat{f}(\beta)\,\text{d}\beta \approx
\sum_{i,\alpha_0\le \beta_i\le \alpha_1} f(\beta_i) (\beta_{i+1}-\beta_i)\,,
$$
with the appropriate corrections at the boundaries.

\begin{exoset}
Show that the estimators $\widehat\bx$ that minimize the posterior expected losses
$\mathbb{E}[L_1(\bx,\widehat\bx)|\by)]$  and $\mathbb{E}[L_2(\bx,\widehat\bx)|\by)]$
are $\widehat\bx^{MPM}$ and $\widehat\bx^{MAP}$, respectively.
\end{exoset}

Since
$$
L_1(\bx,\widehat\bx) = \sum_{i\in\mathcal{I}} \mathbb{I}_{x_i\ne\hat x_i}\,,
$$
the estimator $\widehat\bx$ associated with $L_1$ is minimising
$$
\mathbb{E}\left[\sum_{i\in\mathcal{I}} \mathbb{I}_{x_i\ne\hat x_i} \big|\by \right]
$$
and therefore, for every $i\in\mathcal{I}$, $\hat x_i$ minimizes $\mathbb{P}(x_i\ne\hat x_i)$,
which indeed gives the MPM as the solution.
Similarly,
$$
L_2(\bx,\widehat\bx) =  \mathbb{I}_{\bx\ne\widehat\bx}
$$
leads to $\widehat\bx$ as the solution to 
$$
\min_{\widehat{\bx}} \mathbb{E}\left[ \mathbb{I}_{\bx\ne\widehat{\bx}} \big| \by \right] = 
\min_{\widehat{\bx}} \mathbb{P} \left( \bx\ne\widehat{\bx} \big| \by \right)\,,
$$
which means that $\widehat{\bx}$ is the posterior mode.

\begin{exoset}\label{exo:classX}
Determine the estimators $\widehat\bx$ associated with two loss functions that penalize
differently the classification errors,
$$
L_3(\bx,\widehat\bx) =  \sum_{i,j\in\mathcal{I}}\mathbb{I}_{x_i=x_j}\,\mathbb{I}_{\hat x_i\ne\hat x_j}
\quad
\text{and}
\quad
L_4(\bx,\widehat\bx) =  \sum_{i,j\in\mathcal{I}}\mathbb{I}_{x_i\ne x_j}\,\mathbb{I}_{\hat x_i=\hat x_j}
$$
\end{exoset}

Even though $L_3$ and $L_4$ are very similar, they enjoy completely different properties. In fact, $L_3$ is
basically useless because $\widehat\bx=(1,\cdots,1)$ is always an optimal solution! 

If we now look at $L_4$, we first notice that this loss function is invariant by permutation of the classes in $\bx$: 
all that matters are the groups of components of $\bx$ taking the same value. Minimizing this loss function then
amounts to finding a clustering algorithm. To achieve this goal, we first look at the difference in the risks when
allocating an arbitrary $\hat x_i$ to the value $a$ and when allocating $\hat x_i$ to the value $b$. 
This difference is equal to
$$
\sum_{j, \hat x_j=a} \mathbb{P}(x_i=x_j) - \sum_{j, \hat x_j=b} \mathbb{P}(x_i=x_j)\,.
$$
It is therefore obvious that, for a given configuration of the other $x_j$'s, we should pick the value $a$ that minimizes
the sum $\sum_{j, \hat x_j=a} \mathbb{P}(x_i=x_j)$. Once $x_i$ is allocated to this value, a new index $\ell$ is to be 
chosen for possible reallocation until the scheme has reached a fixed configuration, that is, no $\hat x_i$ need
reallocation.

This scheme produces a smaller risk at each of its steps so it does necessarily converge to a fixed point. What is
less clear is that this produces the global minimum of the risk. An experimental way of checking this is to run the
scheme with different starting points and to compare the final values of the risk.

\begin{exoset}\label{exo:cheapneal}
Since the maximum of $\pi(\bx|\by)$ is the same as that of
$\pi(\bx|\by)^\kappa$ for every $\kappa\in\mathbb{N}$, show that
$$
\pi(\bx|\by)^\kappa = \int \pi(\bx,\theta_1|\by)\,\text{d}\theta_1
\times\cdots\times\int \pi(\bx,\theta_\kappa|\by)\,\text{d}\theta_\kappa \eqno{(8.10)} %Ugly, so-un-latex-like...
$$
where $\theta_i=(\beta_i,\bmu_i,\sigma^2_i)$ $(1\le i\le\kappa)$. Deduce from this representation an
optimization scheme that slowly increases $\kappa$ over iterations and that runs a Gibbs sampler for the
integrand of (8.10) at each iteration.
\end{exoset}

The representation (8.10) is obvious since
\begin{align*}
\left( \int \pi(\bx,\theta|\by)\,\text{d}\theta \right)^\kappa &= \int \pi(\bx,\theta|\by)\,\text{d}\theta
\times\cdots\times\int \pi(\bx,\theta|\by)\,\text{d}\theta \\
&= \int \pi(\bx,\theta_1|\by)\,\text{d}\theta_1
\times\cdots\times\int \pi(\bx,\theta_\kappa|\by)\,\text{d}\theta_\kappa
\end{align*}
given that the symbols $\theta_i$ within the integrals are dummies.

This is however the basis for the so-called SAME algorithm of Doucet, Godsill and Robert (2001), 
described in detail in Robert and Casella (2004).

\end{document}